\documentclass[aps,prb,twocolumn,groupedaddress,showpacs,letterpaper,floatfix,10pt]{revtex4-1}
\usepackage{amsmath,amssymb,graphicx,hyperref}
\usepackage{wrapfig}
\usepackage{multirow}
\usepackage[toc,page]{appendix}
\usepackage{listings}
\usepackage{color}

\hypersetup{pdfnewwindow=true, colorlinks=true, linkcolor=blue, anchorcolor=blue,
  citecolor=blue, filecolor=blue, menucolor=blue, urlcolor=blue}

\DeclareMathOperator{\tr}{tr}

\usepackage{braket}

\allowdisplaybreaks

\newcommand{\R}{\mathcal{R}}

\newcommand{\G}{\mathcal{G}}

\newcommand{\D}{\mathcal{D}}

\newcommand{\be}{\begin{equation}}
\newcommand{\ee}{\end{equation}}
\newcommand{\bea}{\begin{eqnarray}}
\newcommand{\eea}{\end{eqnarray}}
\newcommand{\ba}{\begin{eqnarray*}}
\newcommand{\ea}{\end{eqnarray*}}
\newcommand{\dagga}{{\phantom{\dagger}}}

\newcommand{\Tr}{\mathrm{Tr}}

\begin{document}

\title{Correlated Materials Design: Prospects and Challenges}

\author{Ran Adler$^1$}
\author{Chang-Jong Kang$^1$}
\author{Chuck-Hou Yee$^1$}
\author{Gabriel Kotliar$^{1,2}$}
\affiliation{
$^1$Dept. of Physics \& Astronomy, Rutgers, The State University of New Jersey,
Piscataway, NJ 08854, USA\\
$^2$Condensed Matter Physics and Materials Science Department,
Brookhaven National Laboratory, Upton, New York 11973, USA
}

\date{\today}

\begin{abstract}
  The design of correlated materials challenges researchers to combine the
  maturing, high throughput framework of DFT-based materials design with the
  rapidly-developing first-principles theory for correlated electron systems.
  We review  the field of correlated materials,   distinguishing  two  broad classes of correlation effects, static and dynamics,
  and describe methodologies to take them into account.  We  introduce a  material design workflow,   and  illustrate it
  via examples in several materials classes, including
  superconductors, charge ordering  materials and  systems near an electronically driven metal to insulator
  transition, highlighting the
  interplay between theory and experiment  with a view towards finding new
  materials. We  review the statistical formulation of the errors of currently available methods to estimate formation
  energies. We formulate an approach for estimating a lower-bound for the probability of a new compound to form.
  Correlation effects have to be considered in all the material design steps. 
  These include bridging between structure and property,   obtaining  the correct  structure and    predicting material stability. We introduce a post-processing strategy to take them into account.   
  
\end{abstract}

\maketitle

\section{Introduction}
\label{sec:intro}

The ability to design new materials with desired properties is crucial to the
development of new technology. The design of Silicon and Lithium-ion based
materials are well known examples which led to the proliferation of consumer
hand-held devices today.   Materials discovery has historically
proceeded via trial and error, with a mixture of serendipity and intuition
being the most fruitful path. For example, all major classes of
superconductors--from elemental Mercury in 1911, to the heavy Fermions,
Cuprates and most recently, the iron-based superconductors--have been
discovered largely by chance~\cite{Greene_2012}. 

The dream of materials design is to create an effective workflow for
discovering new materials by combining our theories of electronic structure,
chemistry and computation. It is an inverse problem: start with the materials properties
desired, and work to back out the chemical compositions and crystal structures
which would lead to desirable properties.
It requires a conceptual framework for thinking about the physical properties of materials, and sufficiently
accurate methods for computing them. In addition it requires algorithms for predicting
crystal structures and testing them for stability against decomposition, efficient codes implementing them and broadly accessible databases of known materials and their properties.

For weakly correlated materials, systems for which band theory works, significant progress in all these fronts has been made.  Fermi liquid theory justifies our thinking of the excitations of a solid in terms
of quasiparticles.
Kohn Sham  density functional theory  (DFT)  is a good tool
for computing total energies and a good starting point for
computing those quasiparticle properties in perturbation theory in the screened Coulomb interactions.
Practical implementations of DFT such as LDA and GGA   have become the underlying workhorse of the
scientific community. Extensive benchmarks of software
implementations~\cite{Lejaeghere_2016} have shown that DFT reliably produces
the total energy of a given configuration of atoms, enabling comparisons of
stability between different chemical polymorphs. The maturity of DFT, combined
with searchable repositories of experimental and calculated data (Materials Project~\cite{Jain_2013}, OQMD~\cite{Kirklin2015}, AFLOWlib~\cite{Setyawan2011} and NIMS~\cite{Neugebauer2012}), has
fostered the growth of databases of computed materials properties
to the point where one can successfully design materials  (see for example Refs.~\onlinecite{Fennie_2008,
 Gautier_2015, Fredeman_2011}).

Indeed these advances are beginning to pan out.
The search for new topological materials such as topological insulators or Weyl semimetals is now  greatly aided  by electronic structure
calculations (for a recent review see Ref.~\onlinecite{Armitage2017}).
Another clear example of this
coming of age is the recent prediction of superconductivity in H$_3$S under
high pressure near 190~K~\cite{Duan_2014}. Subsequently, hydrogen sulfide was
observed to superconduct near 200~K, the highest temperature superconductor
discovered so far~\cite{Drozdov_2015}.

The situation is different for strongly correlated materials.
Many aspects of  the physics of correlated electron materials are still not well understood.
Correlated systems exhibit novel phenomena not observed
in weakly-correlated materials: metal-insulator transitions, magnetic order and
unconventional superconductivity are salient examples. Designing and
optimizing materials with these properties would advance both technology and
our understanding of the underlying physics.

Furthermore, material specific predictive theory for this class of materials is not fully developed,
so even the direct problem of predicting properties of correlated materials with known atomic coordinates is very challenging. It requires going beyond  perturbative approaches, and we currently lack methods for reliably modeling materials properties which scale up to the massive number of calculations necessary for material design purposes.
In this article, we  examine the challenges  of material design projects involving strongly correlated electron systems. Our goal is to present the state of the art in the field stressing the outstanding challenges as it pertains to correlated materials, and propose strategies to solve them. We begin by providing a clear definition of correlations (Sec.~\ref{sec:correlations}), distinguishing two important  types, static and dynamic, and some available tools to treat them.  Next we  introduce the material
design workflow (Sec.~\ref{sec:workflow}). Then we give five examples of materials
design in correlated systems to illustrate the application of our ideas
(Sec.~\ref{sec:tuning}-\ref{sec:bacoso}) and conclude with a brief outlook.

\section{What are correlated materials. Static and Dynamic Correlations}

\label{sec:correlations}

The standard model of periodic solids views the electrons in a crystal
as freely propagating waves with well defined quantum numbers, crystal
momentum and band index. Dating back to Sommerfeld and Bloch, it now
has a firm foundation based on the Fermi liquid theory and the renormalization
group, which explains why the effects of Coulomb interactions disappear
or ``renormalize away'' at low energies, and provides an exact description
of the excitation spectra in terms of quasiparticles. Another route
to the band theory of solid, is provided by the density functional
theory in the Kohn-Sham implementation, where a system of non-interacting
quasiparticles is designed so as to provide the exact density of a
solid. While this wave picture of a solid has been extraordinarily
successful and is the foundation for the description of numerous materials,
it fails dramatically for a class of materials, which we will denote
strongly correlated electron systems. 

The basic feature of correlated materials is their electrons cannot
be described as non-interacting particles. Since the constituent electrons
are strongly coupled to one another, studying the behavior of individual
particles generally provides little insight into the macroscopic properties
of a correlated material. Often, correlated materials arise when electrons
are subjected to two competing tendencies: the kinetic energy of hopping
between atomic orbitals promotes band behavior, while the potential
energy of electron-electron repulsion prefers atomic behavior. When
a system is tuned so that the two energy scales are comparable, neither
the itinerant nor atomic viewpoint is sufficient to capture the physics.
The most interesting phases generally occur in this correlated and
difficult-to-describe regime, as we shall see in subsequent sections.

\textcolor{black}{These ideas have to be sharpened in order to quantify
correlation strength,}\textcolor{red}{{} }as there is no sharp boundary
between weakly and strongly correlated materials. Ultimately one would
like to have a methodology which can explain the properties of any
solid and which seeks to make predictions for comparison with experimental
observations. To arrive at an operational definition of a correlated
material, we examine DFT and how it relates to the observed electronic
spectra.

The key idea behind DFT is that the free energy of a solid can be
expressed as a functional of the electron density $\rho(\vec{r})$.
Extremizing the free energy functional one obtains the electronic
density of the solid, and the value of the functional at the extremum
gives the total free energy of the material. The functional has the
form $\Gamma[\rho]=\Gamma_{univ}[\rho]+\int d^{3}rV_{cryst}(r)\rho(r)$
where $\Gamma_{univ}[\rho]$ is the same for all materials, and the
material-specific information is contained in the second term through
the crystalline potential. The universal functional is written as
a sum of $T[\rho]$ the kinetic energy, $E_{H}$, a Hartree Coulomb
energy and a rest which is denoted as $F_{xc}$ the exchange correlation
free energy. This term needs to be approximated since it is not exactly
known, and the simplest approximation is to use the free energy of
the electron gas at a given density. This is called the Local Density
Approximation (LDA).

The extremization of the functional was recast by Kohn and Sham~\cite{Kohn_1965}
in the form of a single particle Schr{\"o}dinger equation with the Hartree
atomic units~$\left(m_{e}=e=\hbar=\frac{1}{4\pi\varepsilon_{0}}=1\right)$

\begin{equation}
\left[-\frac{1}{2}\nabla^{2}+V_{KS}\left(\vec{r}\right)\right]\psi_{\vec{k}j}\left(\vec{r}\right)=\epsilon_{\vec{k}j}\psi_{\vec{k}j}\left(\vec{r}\right).\label{Kohn-Sham}
\end{equation}
\begin{equation}
\sum_{\vec{k}j}|\psi_{\vec{k}j}(\vec{r})|^{2}f(\epsilon_{\vec{k}j})=\rho(\vec{r})\label{KS2}
\end{equation}
reproduces the density of the solid. It is useful to divide the Kohn-Sham
potential into several parts: $V_{KS}=V_{H}+V_{cryst}+V_{xc}$, where
one lumps into $V_{xc}$ exchange and correlation effects beyond Hartree.
In practice, the exchange-correlation term is difficult to capture,
and is generally modeled by approximations known as the local density
approximation (LDA) or generalized gradient approximation (GGA). Density
functional calculations using the LDA/GGA approximation have become
very precise so that the uncertainties are almost entirely systematic.
To get a feel for the numbers, convergence criteria of $10^{-1}$
to $10^{-4}$~meV/atom are routinely used whereas differences between
experimental and theoretical heats of formation routinely differ by
over 100~meV/atom~\cite{Stevanovic_2012}. 

The eigenvalues $\epsilon_{\vec{k}j}$ of the solution of the self-consistent
set of Eqs.~(\ref{Kohn-Sham}) and~(\ref{KS2}) are not to be interpreted
as excitation energies. Konh Sham excitations are \textit{not} Landau
quasiparticle excitations. The latter represent the excitation spectra,
which are the experimental observable in angle resolved photoemission
and inverse photoemission experiments and should be extracted from
the poles of the one particle Green's function: 

\begin{equation}
G\left(\vec{r},\vec{r'},\omega\right)=\frac{1}{\omega+\frac{1}{2}\nabla^{2}+\mu-V_{H}-V_{cryst}-\Sigma\left(\vec{r},\vec{r',}\omega\right)}.\label{eq:gwr}
\end{equation}
Here $\mu$ is the chemical potential and we have singled out in Eq.~(\ref{eq:gwr})
the Hartree potential $V_{H}(\vec{r})=\int\frac{\rho\left(\vec{r}'\right)}{|\vec{r}-\vec{r}'|}d^{3}r'$
expressed in terms of the exact density, $V_{cryst}$ is the crystal
potential and we lumped the rest of the effects of the correlation
in the self-energy operator $\Sigma\left(\vec{r},\vec{r',}\omega\right)$
which depends on frequency as well as on two space variables.

\begin{figure}
\includegraphics[width=0.5\textwidth]{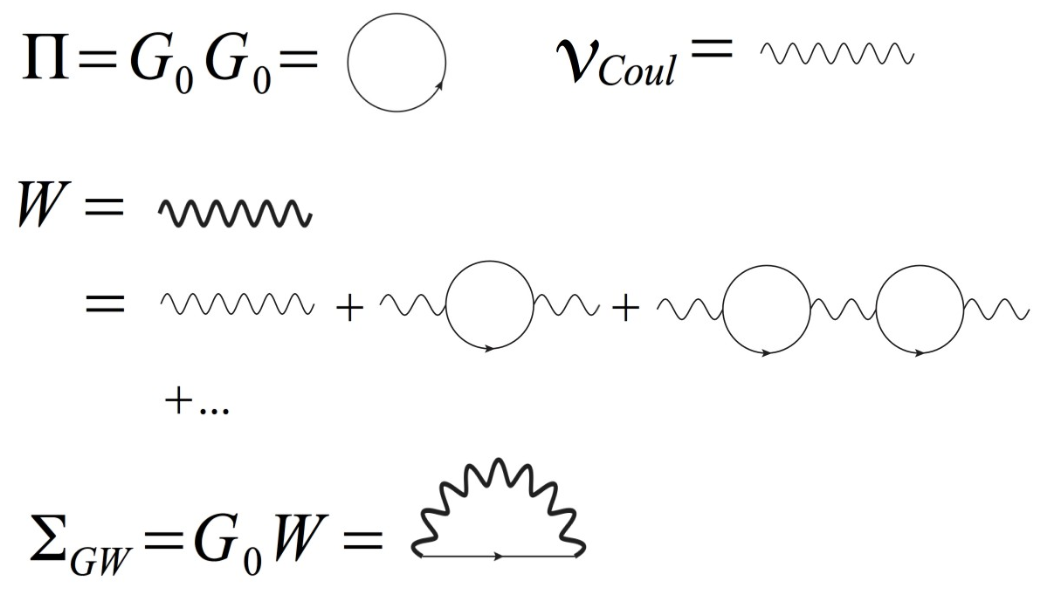}\caption{\label{fig:GW} Schematic diagrams for the GW method. Starting from
some $G_{0}$ a polarization bubble is constructed, which is used
to screen the Coulomb interactions resulting in an interaction W.
This W is then used to compute a self-energy $\Sigma_{GW}$ using
W and $G_{0}$ . To obtain the full Green's function G in Eq. ~(\ref{eq:gwr}),
one goes from $\Sigma_{GW}$ to $\Sigma$ by subtracting the necessary
single particle potential and uses the Dyson equation $G^{-1}={G_{0}}^{-1}-\Sigma$
as discussed in the text. Adapted from \onlinecite{DMFT_at_25}.}
\end{figure}

Since taking $\Sigma=V_{xc}$ generates the Kohn-Sham spectra, we
define weakly correlated materials as ones where
\begin{equation}
\left|\Sigma(\omega)-V_{\text{xc}}\right|\label{eq:deviation}
\end{equation}
is small for low energies, so our definition of weakly correlated
materials are those for which the Kohn-Sham spectra is sufficiently
close to experimental results.

We can refine this definition, by taking into account first order
perturbation theory in the screened Coulomb interactions, taking LDA
as a starting point. This is the $G_{0}W_{0}$ method, which we now
describe using diagrams in Fig.~\ref{fig:GW}. This figure first
describes the evaluation of the polarization bubble $\Pi$
\begin{equation}
\Pi\left(t,t'\right)=G_{0}\left(t,t'\right)G_{0}\left(t',t\right)\label{eq:bubble}
\end{equation}
Next, the screened Coulomb potential $W$ in the random phase approximation
(RPA) which is the infinite sum of diagrams depicted and represent
the expression 

\begin{equation}
W^{-1}=v_{Coul}^{-1}-\Pi\label{eq:Q}
\end{equation}
 where $v_{Coul}$ is the bare Coulomb potential. Then one proceeds
to the evaluation of a self-energy

\begin{equation}
\Sigma_{GW}=G_{0}W\label{eq:self_GW}
\end{equation}
which represents the lowest order contribution in perturbation theory
in W (given in real space by Fig.~\ref{fig:GW}), and then $G^{-1}=G_{0}^{-1}-\Sigma$
using Dyson's equation.

$G_{0}$ above is just a Green's function of non-interacting particles,
and it can thus be defined in various ways, leading to different variants
of the GW method. In the ``one-shot'' (that is, a method with no
self-consistency loop) GW method (aka $G_{0}W_{0})$ one uses the
LDA Kohn-Sham Green's function. 
\begin{equation}
G_{0}\left(i\omega\right)^{-1}=i\omega+\mu+\frac{1}{2}\nabla^{2}-V_{H}-V_{cryst}-V_{xc}^{LDA}.\label{eq:GW_G0}
\end{equation}
and the self-energy is thus taken to be $\Sigma=\Sigma_{GW}-V_{xc}^{LDA}$.
Through this paper we use a matrix notation loosely and view operators
as matrices. For example, in the Dyson equations $W,v_{Coul},\Pi$
are operators (matrices) with matrix elements $\bra{r}W(\omega)\ket{r'}=W(\omega,r,r')$,
$\bra{r}\Pi(\omega)\ket{r'}=\Pi(\omega,r,r')$, $\bra{r}v_{Coul}\ket{r'}=\frac{1}{|r-r'|}$. 

\begin{figure*}
\includegraphics[scale=0.8]{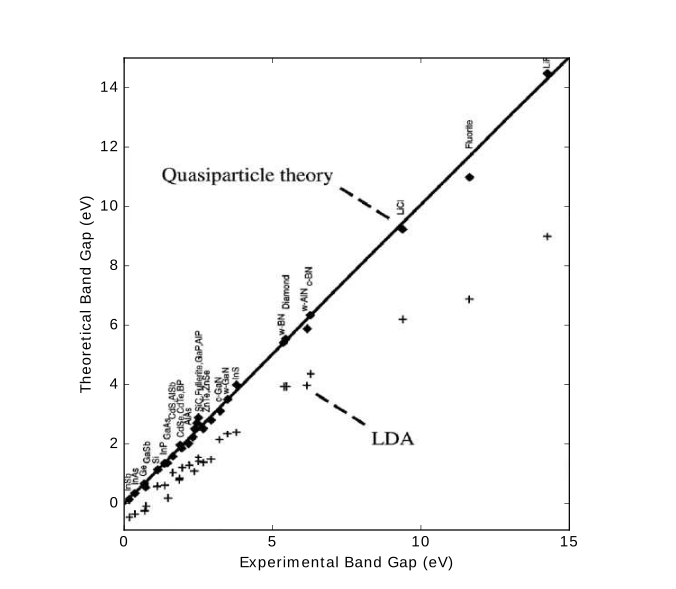}\caption{Theoretically-determined semiconductor gap in a one shot LDA $G_{0}W_{0}$
calculation versus experiment (data complied by E. Shirley). Adapted
from Chapter III. ``First-principles theory of electron excitation
energies in solids, surface, and defects'' (article author: Steven
G. Louie) in Topics in Computational Materials Science, edited by
C. Y. Fong (World Scientific, 1998) {[}\onlinecite{louie1998first}{]}.
Diamonds are the $G_{0}W_{0}$ excitation gap, while the crosses are
the LDA value.\label{fig:louie}}
\end{figure*}

This \textbf{$G_{0}W_{0}$ method}, introduced by Hybertsen and Louie
\cite{Hybertsen1985}, systematically improves the gaps of all semiconducting
materials. We show this in Fig.~\ref{fig:louie}. The success of
this $G_{0}W_{0}$ method implies that in this kind of materials the
Kohn-Sham references system is sufficiently close to the exact self-energy
that the first order perturbation theory correction $\Sigma_{G_{0}W_{0}}(\omega)-V_{\text{xc}}^{LDA/GGA}$
brings us close enough to the experimental results.

However, there are many materials (usually containing atoms with open
$d$ or $f$ shells) where the photoemission spectra (and many other
physical properties) are not so well described by this method. A successful
many body theory of the solid state aims to describe all these systems.
For the most widely used DFT starting points, LDA and GGA, what is
the physical basis for the deviations in Eq.~(\ref{eq:deviation})?
It is useful to think about two limiting cases, one in which the self-energy
$\Sigma$ is a strong function of frequency, in which case we talk
about dynamical effects, and a case where $\Sigma$ is strongly momentum-dependent,
or in real space - highly non-local, but weakly frequency dependent,
and we talk about static correlations. In materials with strong dynamical
correlations the spectral function displays additional peaks, which
are not present in the band theory, and reflect the atomic multiplets
of the material. 

Electron correlation is customarily divided into dynamical and non-dynamical,
but there is no strict definition of these terms. In the context of
Quantum Chemistry calculations, these terms are mainly used to describe
the ability of different methods to capture significant correlation
effects, and the type of wave function which would approximate the
exact solution of the Schr{\"o}dinger equation. Non-dynamical or static
correlations in the chemistry context means that energetically-close
/ degenerate electronic configurations are appreciably present in
the wave function. This requires multiple Slater determinants of low
lying configurations, and multi-reference methods to describe them,
such as the multi-reference Hartree Fock method, or multi-reference
coupled cluster methods. Dynamical correlation refers to a situation
where a single Slater determinant, such as a closed shell configuration
of some orbitals, is a good reference system - which then needs to
be dressed by including double (or higher) excitations from strongly
occupied core shells to empty orbitals. In addition, other virtual
processes can modify the orbitals of the original slater determinant.
This situation is well described in the standard coupled cluster method
which is considered the gold standard in Quantum Chemistry~\cite{shavitt2009many}. 

Confusingly, the chemist's delocalization error corresponds to our
definition of $k$ dependent self-energy, which we denote as static
correlation (since it does not involve frequency dependence of the
self-energy), while the chemist's static correlation corresponds to
what we call dynamical correlation as it requires a strongly frequency
dependent self-energy in condensed matter physics. We use the solid
state physicist convention in this article. 

Another useful way to classify the correlations is by the level of
locality of the self-energy. Introducing a complete basis set of localized
wave functions labeled by site and orbital index we can expand the
self-energy as 
\begin{equation}
\Sigma\left(\vec{r},\vec{r}',\omega\right)=\sum_{\alpha\vec{R},\beta\vec{R}'}\chi_{\alpha\vec{R}}^{*}(\vec{r})\Sigma{}_{\alpha\vec{R},\beta\vec{R}'}\left(\omega\right)\chi_{\beta\vec{R}'}\left(\vec{r}'\right).\label{eq:sigma_basis}
\end{equation}

The self-energy is approximately local when the on-site term $R=R'$
in Eq.~(\ref{eq:sigma_basis}) is much larger than the rest. Notice
that the notion of locality is defined with reference to a basis set
of orbitals.

Equation~(\ref{eq:sigma_basis}) allows us to introduce an approximation
to the self-energy \cite{tomczak} involving a sum of a non-local
but frequency independent term plus a frequency dependent but local
self-energy:

\begin{equation}
\Sigma(\vec{k},\omega)\simeq\Sigma(\vec{k})+\sum_{\vec{R},\alpha\beta\in L}|\vec{R}\alpha\rangle\Sigma_{\alpha\vec{R,\beta}\vec{R}}(\omega)\langle\vec{R}\beta|\label{eq:sigma_ansatz}
\end{equation}
\textcolor{black}{This ansatz was first introduced by Sadovskii }\textit{\textcolor{black}{et
al}}\textcolor{red}{\cite{Sadovskii2005}.} It is useful when the
sum over orbitals in Eq.~(\ref{eq:sigma_ansatz}) runs over a small
set $L$ (much smaller than the size of the basis set), for example
over a single shell of $d$ or $f$ orbitals. This form captures both
static and dynamical correlations and is also amenable to computation
using Dynamical Mean Field Methods to be introduced in section \ref{sec:How-to-treat}.

\section{How to treat correlations\label{sec:How-to-treat}}

\begin{figure}
\includegraphics[viewport=15bp 512bp 330bp 760bp,clip,scale=0.7]{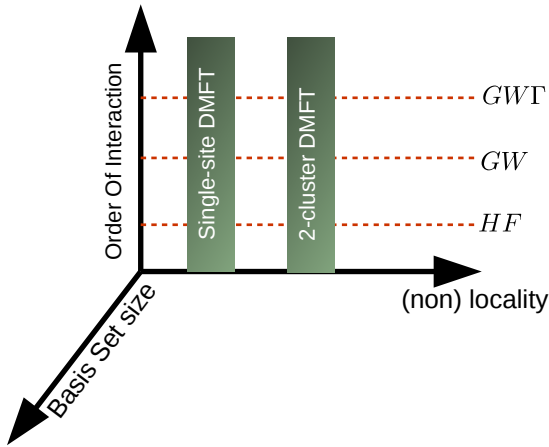}

\caption{Two complementary approaches to the treatment of correlations. One
axis represents the systematic perturbative expansion in powers of
an interaction (for example the screened Coulomb interaction W). The
second axis, sums perturbation theory to all orders, but at the local
level. When locality is just a lattice site, we have the single site
DMFT, improvements involve larger clusters. In addition when one goes
beyond model Hamiltonians towards the realistic treatment of solids
we need to introduce a basis set and estimate that the results are
converged as function of the size of the basis set.\label{fig:axes}}
\end{figure}

Having defined correlations as a departure of the Green's function
from the results of lowest order perturbation theory around LDA (i.e.
$G_{0}W_{0}$), we now review various ways to correlations into account.
One should keep in mind that different materials may require stronger
momentum or frequency dependence in their self-energy, and may exhibit
different degrees of locality. This section lays out several complementary
approaches to treat correlations beyond $G_{0}W_{0}$. They represent
different compromises between speed and accuracy, and can target different
levels of locality and different correlations strengths. A schematic
view of the grand challenge posed by the treatment of correlations
in the solid state is presented in Fig. \ref{fig:axes}, which explains
the need to converge the calculations along multiple axis.

\textbf{Linearized Self Consistent Quasiparticle GW}. We begin our
treatment with the GW approximation, which was introduced in the previous
section. One obvious flaw of the $G_{0}W_{0}$ method is its dependence
on the LDA input. This makes the method increasingly inaccurate as
the strength of the correlations increase. One way to eliminate this
dependence, is to introduce some level of self consistency. Hedin\cite{Hedin65}
proposed a full self-consistent $GW$ scheme, namely to use $G_{0}=G$
is in Eq.~(\ref{eq:self_GW}). We can think of this as setting $V_{xc}=0$,
so it is not used in intermediate steps. There are numerous advantages,
however, to using a non-interacting form for $G_{0}$ in the algorithm,
and in practice the spectra in self-consistent GW turned out to be
consistently worse in solids than the non-self-consistent approach
for spectral properties\cite{Holm98}. Nevertheless, GW can be reasonably
accurate for total energy calculations, as they can be obtained as
stationary points of a functional\cite{Stan06,Kutepov09}.

To improve on the spectra relative to $G_{0}W_{0}$ while retaining
some level of self consistency so as not to depend on the starting
point, the self-consistent quasi-particle GW (QPGW)~\cite{mark_V}was
proposed. Here one uses the ``best'' non-interacting Green's function
$G_{0}$, which is defined in terms of an ``exchange and correlation
potential'' $V_{xc}^{QPGW}$ chosen to reproduce the spectra of the
full G as closely as possible:

\begin{equation}
G_{0}^{QPGW}\left(i\omega\right)^{-1}=i\omega+\mu+\frac{1}{2}\nabla^{2}-V_{H}-V_{cryst}-V_{xc}^{QPGW}.
\end{equation}

To determine $V_{xc}^{QPGW}$ (which once again we view as a matrix
with matrix elements $\bra{r}V_{xc}^{QPGW}\ket{r'}=V_{xc}^{QPGW}(r)\delta(r-r')$),
it was proposed to approximate the spectra and the eigenvectors of
G by those of $G_{0}^{QPGW}$- by solving a set of non-linear equations
on the real axis\cite{mark_V}. An alternative approach that works
on the imaginary axis is to linearize the GW self-energy at each iteration.
Namely, after the evaluation of the self-energy in Eq. (\ref{eq:self_GW}),
this quantity is Taylor expanded around zero frequency (hence the
name ``linearized''): 

\[
\Sigma_{lin}(\vec{k},i\omega)=i\omega(1-Z(\vec{k})^{-1})+\Sigma(\vec{k},0)
\]
and $G_{0}^{QPGW}\left(i\omega\right)$ is obtained by solving the
usual Dyson equation with the linearized self-energy, and multiplying
the result by the quasiparticle residue, $Z$, to obtain a properly
normalized quasiparticle Green's function. 
\begin{align}
 & G_{0}^{\textrm{QPGW}}=\label{eq:linearized_QSGW}\\
 & \,\sqrt{Z(\vec{k})}[i\omega+\mu+\frac{1}{2}\nabla^{2}-V_{H}-V_{cryst}-\Sigma_{lin}]^{-1}\sqrt{Z(\vec{k})}\nonumber 
\end{align}

Note that this defines the exchange correlation potential of the self-consistent
QPGW method. This method, the linearized self-consistent quasiparticle
GW, was introduced in Ref.~\onlinecite{Kutepov12} and an open source
code to implement this type of calculation in the linearized augmented
plane wave (LAPW) basis set is available in Ref.~\onlinecite{Kutepov17}. 

The GW or RPA method captures an important physical effect. Electrons
are charged objects which interact via the long range Coulomb interactions.
\textcolor{black}{Quasiparticles, on the other hand, interact through
the screened Coulomb interaction. }They are composed of electrons
surrounded by screening charges, thus reducing the strength and the
range of their interaction. For this reason, in many model Hamiltonians
describing metals, only the short range repulsion is kept. On the
other hand, it is well known that the RPA fails in describing the
pair correlation function at short distances. One can say that the
GW method captures the long range of the screening effects of the
long range Coulomb interactions and produce a self-energy which is
non-local in space, but with a weak frequency dependence (indeed the
self-energy is linear in a broad range of energies). It turns out
that this method is not able to capture the effects of the short range
part of the Coulomb interactions which in turn induces strong frequency
dependence (i.e. strong non-locality in time), but in turn is much
more local in space.

\begin{figure}
\includegraphics[width=0.4\textwidth]{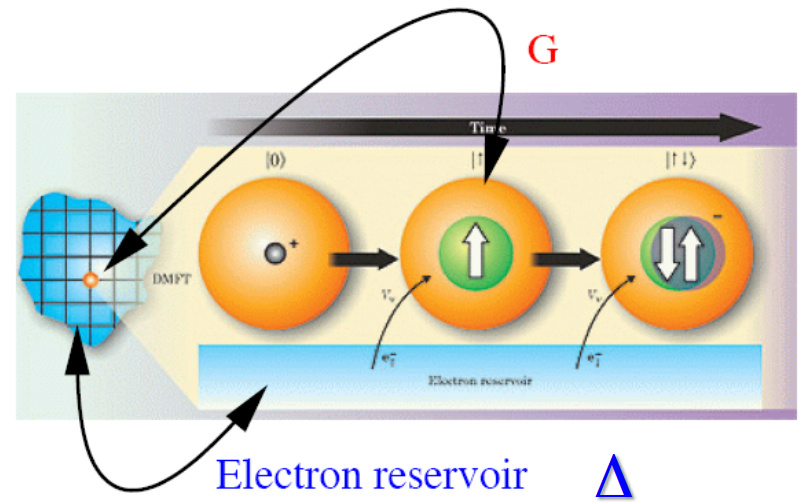} \caption{\label{fig:dmft_mapping} Dynamical Mean Field Theory (DMFT) maps
(or \textbf{truncates}) a lattice model to a single site embedded
in a medium (impurity model) with a hybridization strength which is
determined self consistently. Adapted from Ref.~\onlinecite{physics_today}. }
\end{figure}

\textbf{Dynamical Mean Field Theory (DMFT)}. To capture dynamical
local correlations one uses Dynamical mean field theory\cite{georges_kotliar_PRB},
which is the natural extension of the Weiss mean field theory of spin
systems to treat quantum mechanical model Hamiltonians. Dynamical
Mean Field Theory becomes exact in the limit of infinite dimensions,
which was introduced by Metzner and Vollhardt \cite{dieter_walter}.
With suitable extensions it plays an important role in realistically
describing strongly correlated electron materials. Here we describe
the main intuitive DMFT ideas as a quantum embedding, starting from
the example of a one-band Hubbard model (describing $s$ electrons),
in which the relevant atomic configurations are $\ket{0},\ket{\uparrow},\ket{\downarrow},\ket{\uparrow\downarrow}$
as described in Fig.~\ref{fig:dmft_mapping}. It involves two steps.
The first step, focuses on a single lattice site, and describes the
rest of the sites by a medium with which an electron at this site
hybridizes. This \textbf{truncation\index{truncation}} to a single
site problem is common to all mean field theories. In the Weiss mean
field theory one selects a spin at a given site, and replaces the
rest of the lattice by an effective magnetic field or Weiss field.
In the dynamical mean field theory, the local Hamiltonian at a given
site is kept, and the kinetic energy is replaced by a hybridization
term with a bath of non-interacting electrons, which allows the atom
at the selected site to change its configuration. This is depicted
in Fig.~\ref{fig:dmft_mapping} where we apply the method to the
one-band Hubbard model. The system consist of one band of $s$ electrons.
The Fourier transform of the hopping integral is given by $t(\overrightarrow{k})$.

It is used in the second step, which involves the reconstruction of
lattice observables by \textbf{embedding\index{embedding}} the local
impurity self-energy into a correlation function of the lattice, 
\[
G_{latt}(\vec{k},i\omega)^{-1}=i\omega+\mu-t(\vec{k})-\Sigma_{imp}(i\omega).
\]
Here $\Sigma_{imp}(i\omega)$ are viewed as functionals of the Weiss
field. The requirement that $\sum_{k}G_{latt}=G_{loc}$ determines
the Weiss field. Table~\ref{tab:dmft_wm} summarizes the analogies
between Weiss mean field theory and dynamical mean field theory.

\begin{table}[h]
\begin{tabular}{l|l}
\hline 
Weiss Mean Field Theory & Dynamical Mean Field Theory\tabularnewline[2mm]
\hline 
Ising Model $\rightarrow$ Single Spin & Hubbard Model $\rightarrow$ \tabularnewline
in effective Weiss Field & Impurity in effective bath\tabularnewline
\hline 
Weiss field: $h_{eff}$ & effective bath: $\Delta(\imath\omega_{n})$\tabularnewline
\hline 
Local observable: $m=<s_{i}>$ & Local Observable: $G_{loc}(\imath\omega_{n})$\tabularnewline
\hline 
Self-consistent condition: & Self-consistent condition:\tabularnewline
$\tanh\left(\beta\sum_{j}J_{ij}s_{j}\right)=m$ & $i\omega_{n}-E_{imp}-\Delta\left(i\omega_{n}\right)$\tabularnewline
 & ~~$-\Sigma\left(i\omega_{n}\right)=\left[\sum_{\vec{k}}G_{\vec{k}}\left(i\omega_{n}\right)\right]^{-1}$\tabularnewline
\hline 
\end{tabular}

\vspace{3bp}

\caption{\label{tab:dmft_wm}Corresponding quantities in Dynamical MFT (right)
and Weiss or static MFT in statistical mechanics (left). }
\end{table}

\begin{figure}
\includegraphics[width=0.48\textwidth]{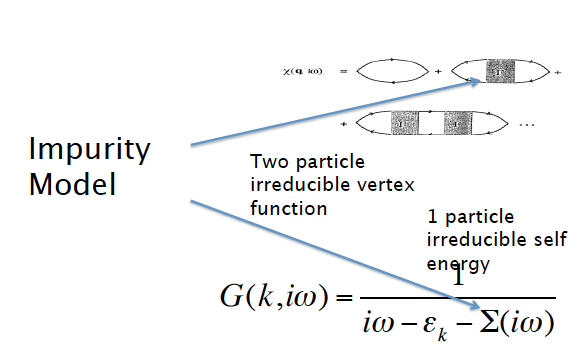} \caption{\label{fig:dmft_mapping2} The DMFT impurity model is used to generate
irreducible quantities such as self-energies and one particle vertices.
These are then \textbf{embedded} in the lattice model to generate
momentum dependent lattice quantities such as spectral functions,
or spin susceptibilities. Adapted from \onlinecite{DMFT_at_25}.}
\end{figure}

The DMFT mapping of a lattice model into an impurity model gives a
local picture of the solid, in terms of an impurity model, which can
then be used to generate lattice quantities such as the electron Green's
function and the magnetic susceptibility by computing the corresponding
irreducible quantities. This is illustrated in Fig.~\ref{fig:dmft_mapping2}. 

The self-consistent loop of DMFT is summarized in the following iterative
cycle

\medskip{}

\begin{tabular}{|c|c|c|c|c|}
\cline{1-1} \cline{3-3} \cline{5-5} 
 &  &  &  & \tabularnewline
{\scriptsize{}$E_{imp},$ $\Delta\left(i\omega_{n}\right)$} & {\scriptsize{}$\rightarrow$} & {\scriptsize{}Impurity Solver} & {\scriptsize{}$\rightarrow$} & {\scriptsize{}$\Sigma_{imp}\left(i\omega_{n}\right),$ $G_{loc}\left(i\omega_{n}\right)$}\tabularnewline
 &  &  &  & \tabularnewline
\cline{1-1} \cline{3-3} \cline{5-5} 
\multicolumn{1}{c}{{\scriptsize{}$\uparrow$}} & \multicolumn{1}{c}{} & \multicolumn{1}{c}{} & \multicolumn{1}{c}{} & \multicolumn{1}{c}{{\scriptsize{}$\downarrow$}}\tabularnewline
\cline{1-1} \cline{3-3} \cline{5-5} 
 &  & {\scriptsize{}$G_{\vec{k}}\left(i\omega_{n}\right)=$} &  & \tabularnewline
{\scriptsize{}Truncation} & {\scriptsize{}$\leftarrow$} & {\scriptsize{}$\frac{1}{i\omega_{n}+\mu-t\left(\vec{k}\right)-\Sigma\left(i\omega_{n}\right)}$} & {\scriptsize{}$\leftarrow$} & {\scriptsize{}Embedding}\tabularnewline
 &  &  &  & \tabularnewline
\cline{1-1} \cline{3-3} \cline{5-5} 
\end{tabular}

\medskip{}

From the simplest model, the one-band Hubbard model, one can proceed
to more realistic descriptions of correlated materials by replacing
$t(\vec{k})$ by a tight-binding model Hamiltonian matrix. The DMFT
equations can be derived from a functional 
\begin{align}
 & \Gamma_{DMFT_{model}}\left[G_{\alpha\beta,\vec{R}},\Sigma_{\alpha\beta,\vec{R}}\right]=\label{eq:DMFT_model}\\
 & -\Tr\,\ln\left[i\omega_{n}-H(\vec{k})-\Sigma_{\alpha\beta\vec{R}}(i\omega_{n})\right]\nonumber \\
 & -\sum_{n}Tr\left[\Sigma\left(i\omega_{n}\right)G\left(i\omega_{n}\right)\right]+\sum_{\vec{R}}\Phi\left[G_{\alpha\beta,\vec{R}},U\right]\nonumber 
\end{align}
where $\Phi\left[G_{\alpha\beta,\vec{R}},U\right]$ is the Baym-Kadanoff
functional - the sum of all two particle irreducible diagrams in terms
of the full Green's function $G$, and the Hubbard interaction $U$,
which denotes a four rank tensor $U_{\alpha\beta\gamma\delta}$. It
can also be evaluated from the Anderson impurity model expressed in
terms of the full local Green's function of the impurity $G$. The
impurity model is the engine of a DMFT calculation. Multiple approaches
have been used for its solution, and full reviews have been written
on the topic. The introduction of continuous time Monte Carlo method
for impurity models~\cite{gull} have provided numerically exact
solutions reducing the computational cost relative to the Hirsch-Fye
algorithm that was used in earlier DMFT studies. 

\textbf{DFT+DMFT method.} This is the next step towards a more realistic
description of solids. It was introduced in Refs.~\onlinecite{anisimov1997,lichtenstein}.
In these early implementations, it consisted of replacing the Hamiltonian
$H(\vec{k})$ by the Kohn-Sham matrix in Eq.~(\ref{eq:DMFT_model})
with a correction to subtract the correlation energy that is contained
in the Kohn-Sham Hamiltonian (double counting correction). The original
DFT calculations were carried out with an LDA exchange and correlation
potential, but they could be done with GGA and other functionals.
Furthermore the exchange and correlation potential in the Dyson equation
for the Green's function can be replaced by another static mean field
theory like hybrid DFT or QPGW, but in the following we will use the
terminology LDA+DMFT.

Starting from the Anderson model Hamiltonian point of view, one divides
the orbitals into two sets. The first set contains the large majority
of the electrons are properly described by the LDA Kohn-Sham matrix.
The second set contains the more localized orbitals ($d$-electrons
in transition metals and $f$-electrons in rare earths and actinides)
which require the addition of DMFT corrections. A subtraction (called
the double counting corrections) takes into account that the Hartree
and exchange correlation has been included in that orbital twice,
since it was treated both in LDA and in DMFT. The early LDA+DMFT calculations,
proceeded in two steps (one-shot LDA+DMFT). First an LDA calculation
was performed for a given material. Then a model Hamiltonian was constructed
from the resulting Kohn-Sham matrix corrected by $E_{dc}$ written
in a localized basis set. The values of a Coulomb matrix for the correlated
orbitals were estimated or used as fitting parameters. Finally DMFT
calculation were performed to improve on the one particle Green's
function of the solid.

In reality, the charge is also corrected by the DMFT self-energy,
which in turn changes the exchange and correlation potential away
from its LDA value. Therefore charge self-consistent LDA+DMFT is needed.
This was first implemented in Refs.~\onlinecite{savrasov2004,Savrasov_2001}.

For this purpose it is useful to notice that the LDA+DMFT equations
can be derived as stationary points of an LDA+DMFT functional, which
can be viewed as a functional of the density and local Green's function
of correlated orbitals. This is a spectral density functional\index{spectral density functional theory}.
Evaluating the functional at the stationary point gives the free energy
of the solid, and the stationary Green's functions gives us the spectral
function of the material. We can arrive at the DFT+DMFT functional
by performing the substitutions $-\frac{1}{2}\nabla^{2}+V_{KS}(\vec{r})$
for $H(\vec{k})$ in the model DMFT functional Eq.~(\ref{eq:DMFT_model})
and then adding terms arising from the density functional theory,
namely:
\begin{align}
 & \Gamma_{DFT+DMFT}\left[\rho\left(\vec{r}\right),G_{\alpha\beta,\vec{R}},V_{KS}\left(\vec{r}\right),\Sigma_{\alpha\beta,\vec{R}}\right]\nonumber \\
 & =\Gamma_{DMFT_{model}}[H(\vec{k})\rightarrow-\frac{1}{2}\nabla^{2}+V_{KS}(\vec{r})]\nonumber \\
 & +\Gamma_{2}[V_{KS}\left(\vec{r}\right),\rho\left(\vec{r}\right)]-\Phi_{DC}
\end{align}
where
\begin{align}
\Gamma_{2}[V_{KS}\left(\vec{r}\right),\rho\left(\vec{r}\right)] & =-\int V_{KS}\left(\vec{r}\right)\rho\left(\vec{r}\right)d^{3}r\nonumber \\
 & +\int V_{ext}\left(\vec{r}\right)\rho\left(\vec{r}\right)d^{3}r\nonumber \\
 & +\frac{1}{2}\int\frac{\rho\left(\vec{r}\right)\rho\left(\vec{r}'\right)}{|\vec{r}-\vec{r}'|}d^{3}rd^{3}r'+E_{xc}^{DFT}\left[\rho\right]
\end{align}

We then arrive at the DFT+DMFT functional which we write in full below.

\begin{widetext}

\begin{eqnarray}
 &  & \Gamma_{DFT+DMFT}\left[\rho\left(\vec{r}\right),G_{\alpha\beta,\vec{R}},V_{KS}\left(\vec{r}\right),\Sigma_{\alpha\beta,\vec{R}}\right]=\nonumber \\
 &  & -\Tr\ln\left[i\omega_{n}+\mu+\frac{\nabla^{2}}{2}-V_{KS}-\sum_{R,\alpha\beta\in L}\chi_{\alpha\vec{R}}^{*}\left(\vec{r}\right)\Sigma_{\alpha\beta\vec{R}}(i\omega_{n})\chi_{\beta\vec{R}}\left(\vec{r}'\right)\right]\nonumber \\
 &  & -\int V_{KS}\left(\vec{r}\right)\rho\left(\vec{r}\right)d^{3}r-\sum_{n}\Tr\left[\Sigma\left(i\omega_{n}\right)G\left(i\omega_{n}\right)\right]+\int V_{cryst}\left(\vec{r}\right)\rho\left(\vec{r}\right)d^{3}r\nonumber \\
 &  & +\frac{1}{2}\int\frac{\rho\left(\vec{r}\right)\rho\left(\vec{r}'\right)}{|\vec{r}-\vec{r}'|}d^{3}rd^{3}r'+E_{xc}^{DFT}\left[\rho\right]+\sum_{\vec{R}}\Phi\left[G_{\alpha\beta,\vec{R}},U\right]-\Phi_{DC}.\label{eq:LDA+DMFT}
\end{eqnarray}

\end{widetext}

$\Phi$ is the sum of two-particle irreducible diagrams written in
terms of $G$ and $U$. It was written down first in Ref.~\onlinecite{savrasov2004},
building on the earlier work of Chitra and Kotliar ~\cite{chitra_2001,Chitra2000}.
It is essential for total energy calculations which require the implementation
of charge self-consistency in the LDA+DMFT method. The first implementation
of charge self-consistent LDA+DMFT was carried out in a full potential
linear muffin-tin orbital (FP-LMTO) basis set~\cite{savrasov2004}.
It was used to compute total energy and phonons of $\delta$-plutonium\cite{Savrasov_2001,dai_phonons}.

Alternatively, one can include the hybridization function $\Delta$
or the Weiss field $\G$ as an independent variable in the functional
in order to see explicitly the free energy of the Anderson Impurity
Model, $\G_{\alpha\beta,\overrightarrow{R}}^{-1}=G_{atom_{\alpha,\beta,\overrightarrow{R}}}^{-1}-\Delta_{\alpha\beta,\overrightarrow{R}}$:
\[
F_{imp}\left[{\cal {G}}_{\alpha\beta,\vec{R}}^{-1}\right]=-\ln\int D[c^{\dagger}c]e^{-S_{imp}[c^{\dagger},c]}
\]
with 
\begin{align*}
S_{imp}[{\cal {G}}_{\alpha\beta,\vec{R}}^{-1}] & =-\sum_{\alpha\beta}\int d\tau d\tau^{\prime}c_{\alpha}^{\dagger}(\tau){\cal {G}}_{\alpha\beta,\vec{R}}^{-1}(\tau,\tau^{\prime})c_{\beta}(\tau^{\prime})\\
 & \,\,\,\,\,\,\,\,\,\,\,\,\,\,\,+U_{\alpha\beta\gamma\delta}\int d\tau c_{\alpha}^{\dagger}(\tau)c_{\beta}^{\dagger}(\tau)c_{\delta}(\tau)c_{\gamma}(\tau)
\end{align*}
 So that:

\begin{widetext}

\begin{eqnarray}
 &  & \Gamma_{DFT+DMFT}\left[\rho\left(\vec{r}\right),G_{\alpha\beta,\vec{R}},V_{KS}\left(\vec{r}\right),\Sigma_{\alpha\beta,\vec{R}},{}_{\alpha\beta,\vec{R}},{\cal {G}}_{\alpha\beta,\vec{R}}\right]=\nonumber \\
 &  & -\Tr\ln\left[i\omega_{n}+\mu+\frac{\nabla^{2}}{2}-V_{KS}-\sum_{R,\alpha\beta\in L}\chi_{\alpha\vec{R}}^{*}\left(\vec{r}\right)\Sigma_{\alpha\beta\vec{R}}(i\omega_{n})\chi_{\beta\vec{R}}\left(\vec{r}'\right)\right]\nonumber \\
 &  & -\int V_{KS}\left(\vec{r}\right)\rho\left(\vec{r}\right)d^{3}r-\sum\Tr\left[({\cal {G}}^{-1}-\Sigma\left(i\omega_{n}\right)-G^{-1})G\left(i\omega_{n}\right)\right]+\int V_{cryst}\left(\vec{r}\right)\rho\left(\vec{r}\right)d^{3}r\nonumber \\
 &  & +\frac{1}{2}\int\frac{\rho\left(\vec{r}\right)\rho\left(\vec{r}'\right)}{|\vec{r}-\vec{r}'|}d^{3}rd^{3}r'+E_{xc}^{DFT}\left[\rho\right]+\sum_{\vec{R}}F_{imp}\left[{\cal {G}}_{\alpha\beta,\vec{R}}^{-1}\right]-\Tr\ln[G_{\alpha\beta,\vec{R}}]-\Phi_{DC}.
\end{eqnarray}

\end{widetext}

The form of the LDA+DMFT functional makes it clear that the method
is independent of the basis set used to implement the electronic structure
calculation, provided that the basis is complete enough. On the other
hand, it is clearly dependent on the parameter $U$ chosen, on the
form of the double counting correction and the choice of the projector
(i.e., the orbitals ${\chi_{\alpha}}(\vec{r})$ with $\alpha\in L$
that enter this definition) and the exchange correlation functional
$E_{xc}^{DFT}$. A projector of the form $P(r,r')=\sum_{\alpha\beta\in L}\chi_{\alpha\mathbf{\overrightarrow{R}}}^{*}(\overrightarrow{r})\chi_{\beta\mathbf{\overrightarrow{R}}}(\overrightarrow{r}')$
was used to define a truncation from $G$ to $G_{loc}$. The inverse
of $P$ is the embedding operator $E$ defined by $P\cdot E=I_{L}$
where $I_{L}$ is the identity operator in the space spanned by the
correlated orbitals. If one restricts $E\cdot P$ to the space $L$,
one also obtains the identity operator in that space. $E$ is used
to define an embedding of the self-energy $\Sigma(r,r')=\sum_{\alpha\beta}E^{\alpha,\beta}(r,r'){\Sigma^{loc}}_{\alpha,\beta}$.

However, more general projectors can be considered as long as causality
of the DMFT equations is satisfied. Ideas for choosing an optimal
projector for LDA+DMFT based on orbitals were presented in Ref.~\onlinecite{indranil}.
Choosing suitable projectors (and correspondingly a suitable value
of the $U$ matrix and a proper double counting correction) is crucial
for the accuracy of an LDA+DMFT calculation as demonstrated recently
in the context of the hydrogen molecule~\cite{H2}. DFT+DMFT is now
a widely used method. It has been successfully used across the periodic
table, and has been implemented in numerous codes~\cite{Amadon08,Amadon12,Park14,Parcollet15,amulet,LISA,LmtART,haule_DMFT,Haule-dmft,Granas12}.
Still there is ample room for advances in implementation, and on providing
a firm foundation of the method. One can view the DFT+DMFT functional
written above, as an approximation to an exact DFT+DMFT functional,
which would yield the exact density and spectra of the solid~\cite{Chitra2000}.
This viewpoint has been used recently to provide an expression for
the double counting correction $\Phi_{DC}$~\cite{Haule2015}. An
alternative perspective goes back to a fully diagrammatic many body
theory of the solid, and examines how DFT+DMFT would fit in that framework
as an approximation. We turn to this formulation next. 

\begin{figure}
\includegraphics[scale=0.37]{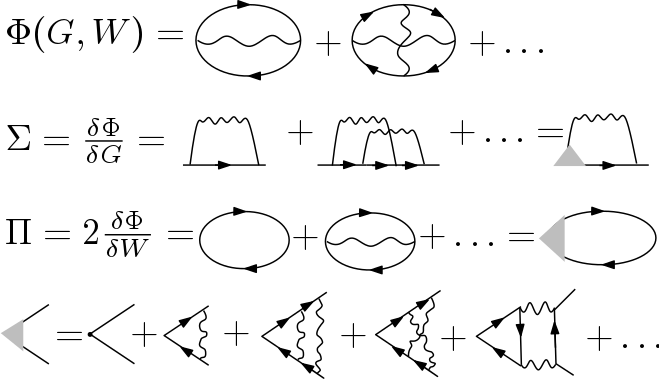}

\caption{Hedin's Equations\label{fig:Hedin's-Equations} give an exact representation
of the correlation function of the Bosonic and Fermionic correlation
in an expansion in G and W. They can be obtained by setting the functional
derivatives of the Eq. (\ref{eq:functional_GW}) to 0. $\Phi(G,W)$
(first line) is the set of 2-PI skeleton diagrams (in G and W), where
by convention the symmetry weights are omitted. The derivative by
G (line 2) shows how the self-energy $\Sigma$ is defined in terms
of a 3-legged vertex $\Gamma$. The derivative by W (line 3) equals
the polarization $\Pi.$ The bottom line shows the definition of the
vertex $\Gamma.$}
\end{figure}

\textbf{Fully Diagrammatic Methods:} The free energy of the solid
can also be expressed as a functional of $G\left(x,x'\right)$ and
$W\left(x,x'\right)$ by means of a Legendre transformation and results
in Refs.~\onlinecite{chitra_2001,Almbladh99}, where $E_{H}=\frac{1}{2}\int\frac{\rho(\overrightarrow{r})\rho(\overrightarrow{r}')}{\left|\overrightarrow{r}-\overrightarrow{r}'\right|}d^{3}rd^{3}r'$,
$\Phi$ is defined as sum of all 2-particle irreducible diagrams which
cannot be divided into two parts by cutting two Green's functions
lines (which can be either G's or W's):
\begin{eqnarray}
\Gamma\left[G,W,\Sigma,\Pi\right] & = & -\Tr\ln\left[G_{0}^{-1}-\Sigma\right]-\Tr\left[\Sigma G\right]\nonumber \\
 & + & \frac{1}{2}\Tr\ln\left[v_{Coul}^{-1}-\Pi\right]-\frac{1}{2}\Tr\left[\Pi W\right]\nonumber \\
 & + & E_{H}+\Phi\left[G,W\right],\label{eq:functional_GW}
\end{eqnarray}
This reformulation is exact and leads to the exact Hedin's equations,
shown in Fig. \ref{fig:Hedin's-Equations}. To convert this general
method into a tool of practical use, strong approximations have to
be introduced. 

\begin{figure}
\includegraphics[width=0.5\textwidth]{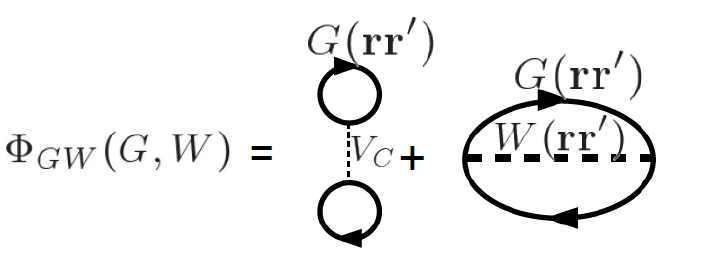} \caption{\label{fig:GW_f} Lowest order graphs in the $\Phi$-functional of
Eq.~(\ref{eq:functional_GW}). They give rise to the fully self-consistent
GW approximation, as the saddle point equations. Note that the first
term here is the Hartree energy. Adapted from \onlinecite{DMFT_at_25}.}
\end{figure}

The lowest order graphs of Eq. (\ref{eq:functional_GW}), shown in
Fig.~\ref{fig:GW_f}, reproduce the self-consistent GW approximation:
taking functional derivatives of the low order functional with respect
to the arguments produces the same equations as the GW approximation.

To summarize the discussion so far, we recall that for semiconductors,
non-local (but weakly frequency dependent) correlation effects are
needed to increase the gap from its LDA value. This admixture of exchange,
can be done within the GW method, or using hybrid density functionals.
It reflects the importance of exchange beyond the LDA, which is due
to the long-range but static part of the Coulomb interaction. These
are the \textbf{static correlation} effects. It has recently been
shown, that this type of correlation effects are important in materials
near a metal-to-insulator transition such as BaBiO$_{3}$ or HfClN~\cite{yin_kutepov}
and can have a dramatic effect in enhancing the electron-phonon interaction
relative to its LDA estimated value. In these systems, a strongly
k-dependent self-energy effect, $\Sigma(k)$, is much more important
than frequency dependence, and here GW methods work well. 

On the other hand frequency dependence, and its implied non-locality
in time, is crucial to capture Mott or Hund's physics. This physics
tends to be local in space and can be captured by DMFT. Static mean
field theories, such as the LDA, do not capture this non-locality
in time, and therefore fail to describe Mott or Hund's phenomena.
\textcolor{black}{DFT+DMFT can treat strong frequency dependency,
but has k-dependence only as inherited from the k-dependence of DFT
exchange and correlation potential, the k-dependence of the embedding
and the double counting shift. }

\textcolor{black}{In real materials both effects are present to some
degree - thus motivating physically the ansatz, Eq.~(\ref{eq:sigma_ansatz}).
Some examples discussed recently are Ce$_{2}$O$_{3}$ (using hybrid
DFT+DMFT) in Ref.~\onlinecite{jacob} and the iron pnictides and
chalcogenides in Ref.~\onlinecite{tomczak}.}

We now describe a route proposed by Chitra~\cite{chitra_2001,Chitra2000}
to embed DMFT into a many-body approach to electronic structure within
a purely diagrammatic approach formulated in the continuum.

\begin{figure}
\includegraphics[scale=0.75]{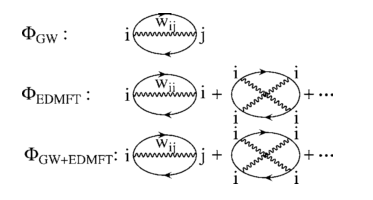}

\caption{Comparison of the functionals for the methods described in the text.
The Hartree diagram was dropped since it appears in all methods\cite{Kotliar2006a}.\label{fig:gw-dmft}}

\end{figure}

If one selects a projector, which allows us to define a local Green's
function, it was suggested in Refs.~\onlinecite{biermann,Chitra2000,chitra_2001}
that one can perform a local approximation and keep only the local
higher order graphs in a selected orbital. 
\begin{align*}
 & \Phi\left[G,W\right]\simeq\\
 & \,\,\Phi_{EDMFT}\left[G_{loc},W_{loc},G_{nonlocal}=0,W_{nonlocal}=0\right]+\\
 & \,\,\Phi_{GW}-\Phi_{GW}\left[G_{loc},W_{loc},G_{nonlocal}=0,W_{nonlocal}=0\right]
\end{align*}
Since the lowest-order graph is contained in the GW approximation,
one should start from the second order graph and higher order. This
$\Phi_{GW+DMFT}$ functional is shown in Fig. \ref{fig:gw-dmft}. 

These ideas were formulated and fully implemented in the context of
a simple extended Hubbard model~\cite{ping_sun1,ping_sun2}. An open
problem in this area, explored in Ref.~\onlinecite{ping_sun2} is
the level of self consistency that should be imposed. This important
issue is already present in the implementation of the GW method, and
the work of Ref.~\onlinecite{ping_sun2} should be revisited using
the lessons from the QPGW method~\cite{tomczak}. There has been
a large number of works exploring GW+DMFT and related extensions and
combinations, and we refer the reader to recent reviews for the most
recent references~\cite{RevModPhys90025003}. Recently, we proposed
the self consistent quasiparticle GW+DMFT~\cite{Choi2016,Tomczak2012c},
as a theory that contains both the most successful form of a GW approximation
and DMFT as limiting cases. Further understanding of this method requires
the systematic treatment of vertex corrections, an approach which
is now vigorously pursued. 

\textcolor{black}{The }\textbf{\textcolor{black}{LDA+U}}\textcolor{black}{{}
is a method was introduced by Anisimov }\textcolor{black}{\emph{et
al.}}~\textcolor{black}{\cite{Anisimov1991a}. It was made rotationally
invariant in Refs.}~\textcolor{black}{\onlinecite{Dudarev1998,Liechtenstein95}.
One can view this is as a special case of LDA+DMFT, where in the functional
(Eq.}~\textcolor{black}{(\ref{eq:LDA+DMFT})) $\Phi$ (the sum of
graphs) is restricted to the Hartree-Fock graphs: $\Phi\rightarrow\Phi_{HF}$,
and $\Sigma(\imath\omega_{n})$ is replaced by a constant matrix $\lambda$.}
Then the LDA+U functional $\Gamma_{LDA+U}$ can be written in as follows:

\begin{align}
 & \Gamma_{LDA+U}\left[\rho\left(\vec{r}\right),n_{\alpha\beta},V_{KS}(\vec{r}),\lambda_{\alpha\beta}\right]=\nonumber \\
 & \,\,-\Tr\ln\left(i\omega+\mu+\frac{\nabla^{2}}{2}-V_{KS}-\sum_{\alpha\beta\in L}\chi_{\alpha}^{*}\left(\vec{r}\right)\lambda_{\alpha\beta}\chi_{\beta}\left(\vec{r}'\right)\right)\nonumber \\
 & \,\,-\int V_{KS}\left(\vec{r}\right)\rho\left(\vec{r}\right)d^{3}r-\lambda_{\alpha\beta}n_{\alpha\beta}+\int V_{cryst}\left(\vec{r}\right)\rho\left(\vec{r}\right)d^{3}r\nonumber \\
 & \,\,+E_{H}\left[\rho\left(\vec{r}\right)\right]+E_{xc}^{LDA}\left[\rho\left(\vec{r}\right)\right]+\Phi_{HF}\left[n_{\alpha\beta}\right]-\Phi_{DC}\left[n_{\alpha\beta}\right],\label{eq:LDAU-functional}
\end{align}
where $n_{\alpha\beta}$ is the occupancy matrix. $\Phi_{HF}$ in
Eq.~(\ref{eq:LDAU-functional}) is the Hartree-Fock approximation

\begin{equation}
\Phi_{HF}\left[n_{\alpha\beta}\right]=\frac{1}{2}\sum_{\alpha\beta\gamma\delta\in L}\left(U_{\alpha\gamma\delta\beta}-U_{\alpha\gamma\beta\delta}\right)n_{\alpha\beta}n_{\gamma\delta},\label{eq:HF-functional}
\end{equation}
where indexes $\alpha,\beta,\gamma,\delta$\textcolor{black}{{} refer
to the fixed angular momentum $L$ of correlated orbitals, and the
matrix $U_{\alpha\beta\gamma\delta}$ is the on-site Coulomb interaction
matrix element.}

\textcolor{black}{Similarly to DMFT, the on-site Coulomb interaction
is already considered within LDA approximately, so it is subtracted,
hence the double-counting term $\Phi_{DC}$. One of the popular choices
is so-called ``fully-localized limit'' (FLL) whose form is}~\cite{Anisimov97}

\[
\Phi_{DC}^{FLL}\left[n_{\alpha\beta}\right]=\frac{1}{2}\bar{U}\bar{n}\left(\bar{n\!}-\!1\right)-\frac{1}{2}\bar{J}\left[\bar{n}^{\uparrow}\!\left(\bar{n}^{\uparrow\!}-\!1\right)+\bar{n}^{\downarrow}\!\left(\bar{n}^{\downarrow}\!-\!1\right)\right],
\]
where 
\begin{align*}
\bar{n}^{\sigma}\! & =\!{\displaystyle \sum_{\alpha\in L}}\!n_{\alpha\alpha}^{\sigma},\\
\bar{n}\! & =\!\bar{n}^{\uparrow}\!+\!\bar{n}^{\downarrow},\\
\bar{U} & =\frac{1}{\left(2L+1\right)^{2}}\sum_{\alpha\beta\in L}U_{\alpha\beta\beta\alpha},\\
\bar{J} & =\bar{U}-\frac{1}{2L(2L+1)}\sum_{\alpha\beta\in L}\left(\!U_{\alpha\beta\beta\alpha}\!-\!U_{\alpha\beta\alpha\beta}\!\right).
\end{align*}

The constant matrix $\lambda_{\alpha\beta}$ is determined by the
saddle point equations $\frac{\delta\Gamma_{LDA+U}}{\delta n_{\alpha\beta}}=0$:

\begin{align}
\lambda_{\alpha\beta} & =\frac{\delta\Phi_{HF}}{\delta n_{\alpha\beta}}-\frac{\delta\Phi_{DC}}{\delta n_{\alpha\beta}}\nonumber \\
 & =\sum_{\gamma\delta}\left(U_{\alpha\gamma\delta\beta}-U_{\alpha\gamma\beta\delta}\right)n_{\gamma\delta}\nonumber \\
 & \,\,\,-\delta_{\alpha\beta}\left[\bar{U}\left(\bar{n}-\frac{1}{2}\right)-\bar{J}\left(\bar{n}^{\sigma}-\frac{1}{2}\right)\right].
\end{align}

\textcolor{black}{The FLL double-counting term tends to work quite
well for strongly correlated materials with very localized orbitals.
However, for weakly correlated materials, FLL scheme describes the
excessive stabilization of occupied states and leads to quite unphysical
results such as the enhancement of the Stoner factor}~\cite{Petukhov03}\textcolor{black}{.
In order to resolve the problems, ``around mean-field'' (AMF) was
introduced in Ref.}~\onlinecite{Czyzyk94} and further developed
in Ref.~\onlinecite{Petukhov03}.

\textcolor{black}{One can say that the LDA+U method works when correlations
are static - and at the same time local. For example cases where magnetic
or orbital order are very well developed. For a review of the LDA+U
method see Ref.}~\textcolor{black}{\onlinecite{Himmetoglu2014}.}

\textbf{Slave-Boson Method.} The physics of strongly correlated electron
materials requires to take into account - on the same footing, localized
- quasi-atomic degrees of freedom, which are important at high energies,
together with strongly renormalized itinerant quasiparticles which
emerge at low energy. DMFT captures this physics via a sophisticated
quantum embedding that requires the solution of a full Anderson impurity
model in a self consistent bath. A\textcolor{red}{{} }\textcolor{black}{less
accurate but computational faster methd to solve} the strong correlation
problem, which precedes DMFT is the Gutzwiller method which has been
shown to be equivalent to the slave boson method in the saddle point
approximation. This approach starts with an exact quantum many body
problem, but one enlarges the Hilbert space so as to introduce explicitly
operators which describe the different atomic multiplet configurations,
and additional fermionic degrees of freedom which will be related
to the emergent low energy quasiparticles. The method proceeded by
writing a functional integral in the enlarged Hilbert space, supplemented
by Lagrange multipliers which enforce multiple constraints. The approach,
at zero temperature is very closely connected to the Gutzwiller method,
which appears as a saddle point solution in the functional integral
formalism \cite{kotliar_ruckenstein}. In its original formulation
this method was not manifestly rotationally invariant, but it was
extended in this respect in Refs.~\onlinecite{wolfle1,wolfle2,fresard}.
Further generalizations in the multi-orbital formulation and to capture
non-local self-energies was introduced in Ref.~\onlinecite{lechermann},
and we denote this formulation as the RISB (rotationally invariant
slave boson) method. Within the slave particle method it is possible
to go beyond mean field theory, and fluctuations around the saddle
point generate the Hubbard bands in the one particle spectra~\cite{raimondi}.
The RISB method can be used\textcolor{black}{{} to compute the energy
of} lattice models. When in conjunction with the DMFT self-consistency
condition, it gives the same results as the direct application of
the method to the lattice model \textcolor{red}{~\cite{lanata_prx15}}.
In this review, we will restrict ourselves to the RISB mean field
theory, specifically from the perspective of the free-energy functionals
that describe the free-energy of the system. We explain the physical
meaning of the variables used in this method, and summarize succinctly
the content of the mean field slave boson equations using a functional
approach. A precise operational formulation of the method was only
given recently~\cite{Lanata2017}. For pedagogical reasons we start
again with a Hubbard model with a tight binding one body Hamiltonian
\textbf{$H(\vec{k})$}.

The variables used in RISB can be motivated by noticing that the many-body
local density matrix $\rho$ (with matrix elements $\bra{\Gamma'}\rho\ket{\Gamma}$)
admits a Schmidt decomposition, which can be written in terms of the
expectation value of matrices of slave-boson operators $\phi_{Bn}$
and $\phi_{Bn}^{\dagger}$, which become $\phi_{Bn,}\,\phi_{Bn}^{*}$
when the single-particle index $\alpha$ is M-dimensional, and can
be stored as $2^{M}\times2^{M}$ matrices $\Phi,$ $[\Phi]_{An}\equiv\phi_{An}$,
$[\Phi^{\dagger}]_{nA}=\phi_{An}^{*}$, so that:

\begin{equation}
\rho=\Phi\Phi^{\dagger}.
\end{equation}

The method also introduces Fermionic operators $f_{\alpha}$ at each
site (site indices are omitted in the following) which will represent
the low energy quasiparticles at the mean field level . The physical
electron operator $\underline{d}$ is represented in the enlarged
Hilbert space by 
\begin{equation}
\underline{d}_{\alpha}=R_{\alpha\beta}[\phi]f_{\beta}\label{eq:d_alpha}
\end{equation}
where the matrix $\R$, with elements $R_{\alpha\beta}$ at the mean
field level, has the interpretation of the quasiparticle residue,
relating the physical electron to the quasiparticles. When $\R$ is
small it exhibits the strong renormalizations induced by the electronic
correlations. An important feature of the rotational invariant formalism
is that the basis that diagonalizes the quasiparticles represented
by the operators $f$ is not necessarily the same basis as that which
would diagonalize the one electron density matrix expressed in terms
of the operators $d$ and $d^{\dagger}$. Of central importance is
the expression of the matrix $\R$, in terms of the bosonic amplitudes:

\begin{align*}
 & R_{\alpha\beta}=\\
 & \sum_{\gamma}\sum_{ABnm}F_{\alpha,A,B}^{\dagger}F_{\gamma,nm}^{\dagger}[\Phi^{\dagger}]_{nA}[\Phi]_{Bm}\left[\left(\Delta^{p}(1-\Delta^{p})\right)^{-1/2}\right]_{\gamma\beta}\\
 & =\sum_{\gamma}\mathrm{Tr}\left[\Phi^{\dagger}F_{\alpha}^{\dagger}\Phi F_{\gamma}\right]\left[\left(\Delta^{p}(1-\Delta^{p})\right)^{-1/2}\right]_{\gamma\beta}.
\end{align*}

We introduced here the matrices $F$, 
\[
[F_{\alpha}]_{nm}\equiv_{f}\langle n|f_{\alpha}|m\rangle_{f}.
\]
The matrices 
\[
\Delta_{\alpha\beta}^{p}\equiv\sum_{Anms}\langle m|f_{\alpha}^{\dagger}|s\rangle\langle s|f_{\beta}|n\rangle\Phi^{\dagger}{}_{nA}\Phi_{Am}=\mathrm{Tr}\left[F_{\alpha}^{\dagger}F_{\beta}\Phi^{\dagger}\Phi\right]
\]
 have the physical interpretation of a quasiparticle density matrix:
\[
<f_{\alpha}^{\dagger}f_{\beta}>=\Delta_{\alpha,\beta}^{p}.
\]

For a multi-band Hubbard model with a tight-binding one-body Hamiltonian
$H(\vec{k})$ and interactions $\Sigma_{i}H_{i}^{loc}$, the RISB
functional, whose extremization gives the slave-boson mean field equations,
is expressed in terms of $\phi_{i},\,\phi_{i}^{\dagger}$ (the slave-boson
amplitude matrices) and the matrices $\,\lambda_{i}^{c}$, $\lambda_{i}$,
${\cal D}$. These are $N\times N$ matrices of Lagrange multipliers:
(i) $\lambda_{i}^{c}$ enforces the definition of $\Delta_{i}^{p}$
in terms of the RISB amplitudes (ii) $\lambda_{i}$ enforces the Gutzwiller
constraints and (iii) $\D_{i}$ enforces the definition of $\R_{i}$
in terms of slave boson amplitudes. Another variable, $E^{c}$ enforces\textcolor{red}{{}
}\textcolor{black}{the normalization $\mathrm{Tr}[\Phi\Phi^{\dagger}]=1$. }

\textcolor{black}{The variables ${\cal {R}}$,~$\lambda$ can be
thought as a parametrization of the self-energy. While the matrices
${\cal {\cal {D}}}$, $\lambda^{c}$ are a parametrization of a small
impurity model (the dimension of the bath Hilbert space is the same
as that of the impurity Hilbert space) , ${\cal {\cal {D}}}$ is the
hybridization function of the associated impurity model while $\lambda^{c}$
parametrizes the energy of the bath. $\Delta^{p}$ describes the quasiparticle
occupancies, which are the static analogs of the impurity quasiparticle
Green's function. }

The RISB (Gutzwiller) functional for a model Hamiltonian with a local
part which is bundled together with a local interaction term in $H^{loc}$
and a kinetic energy matrix which is non-local $H(\vec{k})^{nonloc}$,
was constructed in Ref.~\onlinecite{lanata_prx15}:

\begin{widetext}

\begin{eqnarray}
 &  & \Gamma_{\text{model}}[\phi,E^{c};\,\R,\R^{\dagger},\lambda;\,\D,\D^{\dagger},\lambda^{c};\,\Delta^{p}]=\nonumber \\[-0.6mm]
 &  & -\lim_{\mathcal{T}\rightarrow0}\frac{\mathcal{T}}{\mathcal{N}}\sum_{k}\sum_{m\in\mathbb{Z}}\Tr\,\ln\!\left(\frac{1}{i(2m+1)\pi\mathcal{T}-\R H(\vec{k})^{nonloc}\R^{\dagger}-\lambda+\mu}\right)e^{i(2m+1)\pi\mathcal{T}0^{+}}\\
 &  & +\sum_{i}\Tr\bigg[\phi_{i}^{\dagga}\phi_{i}^{\dagger}\,H_{i}^{loc}\!+\!\sum_{a\alpha}\left(\left[\D_{i}\right]_{a\alpha}\,\phi_{i}^{\dagger}\,F_{i\alpha}^{\dagger}\,\phi_{i}^{\dagga}\,F_{ia}^{\dagga}+\text{H.c.}\right)\!+\!\sum_{ab}\left[\lambda_{i}^{c}\right]_{ab}\,\phi_{i}^{\dagger}\phi_{i}^{\dagga}\,F_{ia}^{\dagger}F_{ib}^{\dagga}\bigg]\!+\!\sum_{i}E_{i}^{c}\!\left(1\!-\Tr\big[\phi_{i}^{\dagger}\phi_{i}^{\dagga}\big]\right)\nonumber \\[-0.6mm]
 &  & -\sum_{i}\bigg[\sum_{ab}\big(\left[\lambda_{i}\right]_{ab}+\left[\lambda_{i}^{c}\right]_{ab}\big)\left[\Delta_{i}^{p}\right]_{ab}+\sum_{ca\alpha}\left(\left[\D_{i}\right]_{a\alpha}\left[\R_{i}\right]_{c\alpha}\big[\Delta_{i}^{p}(1-\Delta_{i}^{p})\big]_{c\alpha}^{\frac{1}{2}}+\text{c.c.}\right)\bigg]\,.\label{Lag-SB}
\end{eqnarray}

\end{widetext}

This method can also be turned into an \textit{ab-initio} DFT+G method
(or DFT+RISB). To motivate the construction of a \textbf{DFT+G }functional
we simply follow the same path used above to go from the model DMFT
Hamiltonian to a DFT+DMFT functional. We substitute $H(\vec{k})$
for $-\frac{1}{2}\nabla^{2}+V_{KS}(\vec{r})$, which has a local and
a nonlocal part, and follow the same steps as in the DFT+DMFT section. 

\begin{align*}
 & \Gamma_{DFT+G}\left[\rho\left(\vec{r}\right),V_{KS}\left(\vec{r}\right),\phi,E^{c};\R,\R^{\dagger},\lambda;\D,\D^{\dagger},\lambda^{c};\Delta_{i}^{p}\right]=\\
 & \,\,\,\Gamma_{model}[H(\vec{k})\rightarrow-\frac{1}{2}\nabla^{2}+V_{KS}(\vec{r})]+\Gamma_{2}[V_{KS}\left(\vec{r}\right),\rho\left(\vec{r}\right)]-\\
 & \,\,\,\,\,\sum_{i}\Phi_{DC}[\Delta_{i}^{p}]
\end{align*}
where $\Gamma_{2}$ and $\Phi_{DC}$ are the same functionals defined
in the subsection on DMFT. The LDA+RISB and the LDA+G method are completely
equivalent (more precisely, the slave boson method has a gauge symmetry,
and a specific gauge needs to be chosen to correspond to the multi-orbital
Gutzwiller method introduced in Ref.~\onlinecite{Bunemann2007a}.
DFT+G was formulated in Refs.~\onlinecite{Deng2007,Ho08}. The slave
boson method in combination with DFT was first used in Ref.~\onlinecite{Savrasov2005}
in a non-rotationally-invariant framework and with full rotational
invariance  in Ref.~\onlinecite{lechermann}. For a recent review
see Ref.~\onlinecite{Piefke2017}. 

\subsubsection*{Comparing the methods, critical discussion, future directions and
outlook}

For weakly correlated systems we argued in section \ref{sec:correlations}
that once the structure is known, we have a well-defined path to compute
their properties using DFT and the $G_{0}W_{0}$ method. To go beyond
requires to move in the space illustrated in Fig \ref{fig:axes}.
This has to be done while respecting as many general properties such
as conservation laws (Refs.\onlinecite{Baym1961,PhysRev.127.1391,PhysRevB.96.075155}),
sum rules, unitarity and causality (Refs.\onlinecite{PhysRevB.69.205108,PhysRevB.94.125124,PhysRevB.90.115134})
as possible. This is a very difficult problem which is under intensive
investigation. 

This section reviewed several Green's-function-based approaches available
for studying strongly-correlated-electron materials. The reader may
wonder why we considered multiple methods. There are two reasons.
First, as stressed throughout the paper and demonstrated in the examples,
presented in the next sections there are materials where correlations
are mostly static, and others where dynamical correlations dominate
the physics. These different types of correlations require different
methods. Second, even when two methods treat the same type of correlations,
they have different accuracies and computational speeds. Finding the
correct trade-off between speed and accuracy will be important, in
particular when high throughput studies start becoming feasible for
strongly correlated systems. 

As we strive towards a fully controlled but practical solution of
the full many-body problem for solid state physics, we will need more
exact and thus slower methods to benchmark the faster but more practical
ones. Hence it is important to compare them and understand their connection.
Static correlations can be treated by GW methods, and one can view
the hybrid-functional exchange-correlation potentials as faster approximations
to the QPGW exchange / correlation potential. One can also assess
whether the GGA (or LDA) exchange / correlation potential is a good
approximation to the self-energy in a given material - by checking
how close it is to the corresponding self-consistent QPGW exchange
correlation potential. 

In the same spirit one can understand the successes of LDA+DMFT from
the \textbf{GW+DMFT }perspective. One issue is the definition of $U$
in a solid. The functional $\Phi$ can be viewed as the functional
of an Anderson impurity model which contains a frequency-dependent
interaction $U(\omega)$ obeying the self-consistency condition: 
\begin{equation}
U^{-1}=W_{loc}^{-1}+\Pi_{loc}.
\end{equation}

This provides a link between LDA+DMFT, which uses a parameter $U$,
and the GW+DMFT method, where this quantity is self-consistently determined.
An important question is thus under which circumstances one can approximate
the Hubbard $U$ by its static value. For projectors constructed on
a very broad window, $U(\omega)$ is constant on a broad range of
frequencies~\cite{Kutepov2010}. An important open question is how
one can incorporate efficiently the effects of the residual frequency
dependence of this interaction. 

Another question is the validity of the local ansatz for graphs beyond
the $GW$ approximation. This question was first addressed in Ref.~\onlinecite{zein_first},
who showed that the lowest order $GW$ graph is highly non-local in
all semiconductors, which can be understood as the exchange Fock graph
is very non-local. On the other hand, higher-order graphs in transition
metals in an LMTO basis set were shown to be essentially local.

Consider a system such as Cerium, containing light $spd$ electrons
and heavier, more correlated, $f$ electrons. We know that for very
extended systems, the GW quasiparticle band structure is a good approximation
to the LDA band structure. Therefore the self-energy of a diagrammatic
treatment of the light electrons can be approximated by the exchange-correlation
potential of the LDA (or by other improved static approximations if
more admixture of exchange is needed) . Diagrams of all orders but
in a local approximation are used for the $f$ electrons. In the full
many-body treatment $\Sigma_{ff}$ is computed using skeleton graphs
with $G_{loc}$ and $W_{loc}$. To reach the LDA+DMFT equations, one
envisions that at low energies the effects of the frequency dependent
interaction $U(\omega)$ can be taken into account by a static $U$,
which should be close to (but slightly larger than) $U(\omega=0)$.
The $ff$ block of the Green's function now approaches $\Sigma_{ff}-E_{dc}$.

We reach the LDA+DMFT equations with some additional understanding
on the origin of the approximations used to derive them from the GW+DMFT
approximation, as summarized schematically in

$\Sigma_{GW+DMFT}\left(\vec{k},\omega\right)\longrightarrow$

\[
\left(\begin{array}{cc}
0 & 0\\
0 & \Sigma_{ff}-E_{dc}
\end{array}\right)+\left(\begin{array}{cc}
V_{xc}[\vec{k}]_{spd,spd} & V_{xc}[\vec{k}]_{spd,f}\\
V_{xc}[\vec{k}]_{f,spd} & V_{xc}[\vec{k}]_{f,f}
\end{array}\right).
\]

Realistic implementations of combinations of GW and DMFT have not
yet reached the maturity of LDA+DMFT implementations, and are a subject
of current research. Recent self-consistent implementations include
Refs.~\onlinecite{PhysRevMaterials.1.043803,Choi2016}.

When strong dynamical correlations are involved, the spectra is very
far from that of free fermions. The one electron spectral function
$A(\vec{k},\omega)$ displays not only a dispersive quasiparticle
peak, but also other features commonly denoted as satellites. The
collective excitation spectra, which appear in the spin and charge
excitation spectra, does not resemble the particle-hole continuum
of the free Fermi gas with additional collective (zero sound, spin
waves) modes, produced by the residual interactions among them. Finally
the damping of the elementary excitations in many regimes does not
resemble that of a Fermi liquid. Strong dynamical correlations are
accompanied by anomalous transport properties, large transfer of optical
spectral weight, large mass renormalizations, as well as metal-insulator
transitions as a function of temperature or pressure. These can be
captured by DMFT, which combined with electronic structure, enable
the treatment of these effects in a material-specific setting, but
not by LDA+G which only provides a quasiparticle description of the
spectra. Many successful comparisons with experimental ARPES and optical
and neutron scattering data have been made over the last two decades
using LDA+DMFT which makes an excellent compromise of accuracy for
speed, and it is now the mainstay for the elucidation of structure
property relations in strongly correlated materials. LDA+G can only
describe at best the quasiparticle featurs in that spectra. 

On the other hand, as it will be stressed through examples, for total
energy evaluations - which are a central part of the material design
workflow, faster methods are currently needed. We described above
two methods, the LDA+U method, and the Gutzwiller RISB method, which
fall in this ``fast but less accurate'' category. These methods
can be viewed as approximations to the many body problem within a
DMFT perspective. As pointed out in Refs.~\onlinecite{lanata_prx15}
and~ \onlinecite{PhysRevB96235139}, the Gutzwiller RISB leads to
a DMFT-like impurity solver with a bath consisting of only one site.
LDA +U can be viewed as a limiting case of DMFT, where a static local
self-energy is considered. 

There are numerous algorithmic challenges in optimizing studies of
materials based on DMFT. While CTQMC runs for solving the Anderson
impurity model, i.e. the single orbital case, as well as 3 or 2 orbitals
($t_{2g}$ and $e_{g}$ electrons) can be completed on one CPU in
less than one day for extremely low temperatures, a full d-shell (5
electrons) requires several days, and the full f-shell is still at
the border of what can be done with current methods. All this assumes
high symmetry situations, where the hybridization function is diagonal.
Off-diagonal hybridization introduces severe minus sign problems.
Alternative exact diagonalization-based methods, such as NRG or DMRG
will be needed. This would also help with the problem of reducing
the uncertainties involved in the process of analytic continuation. 

\textcolor{black}{While the ansatz \ref{eq:sigma_ansatz} has reproduced
the photoemission spectra of many materials, there have not been high-throughput
studies which would enable us to systematically search for deviations.
This requires the improvement of computational tools, an area of active
research. What if the $\boldsymbol{k}$ and $\omega$ dependencies
cannot be disentangled? This situation may arise near a quantum critical
poin}t. Methods to incorporate the non-local correlations beyond DMFT
are an important subject of active research, which is reviewed in
Ref.~\onlinecite{RevModPhys90025003}.

Armed with an understanding of methods to treat correlations and their
physical and computational trade-offs, we proceed in section \ref{sec:workflow}
to construct a workflow for designing correlated materials. 

{\def\bibliography 
\bibliography{cormatdes,gabi}

}

\section{Materials Design Workflow\label{sec:workflow}}

\begin{figure}
\includegraphics[scale=0.5]{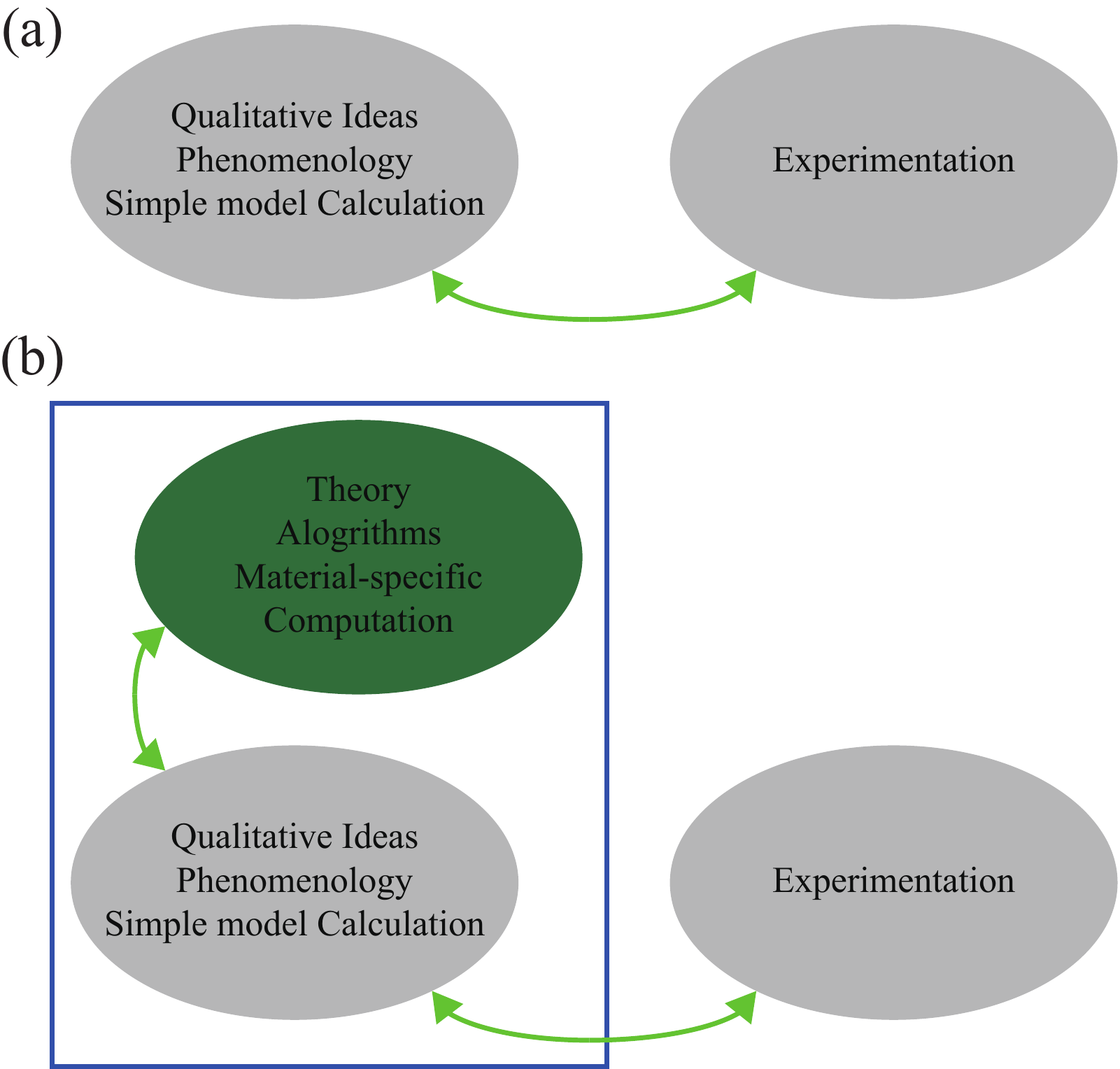}

\caption{(a) Traditional theory experiment theory interaction. This loop is
initiated by an experimental discovery for which many theories are
proposed and corrected by further experimentation. (b) The modern
material design loop augments the theoretical contribution, now this
loop can be initiated by either theory or experiment. }

\label{fig:flow}
\end{figure}

Condensed matter physics has a long standing tradition of constant
interplay between theory and experimentation. The field of strongly
correlated electron systems, has been driven by unexpected experimental
discovery a new materials, followed by a large number of theoretical
ideas which get refined as new experimental information becomes available.
This is described in Fig.~\ref{fig:flow} panel (a). Panel (b) describes
how theory, algorithms and computational power have enhanced theoretical
capabilities, which make the approach material-specific, thus enabling
theory-assisted material design. 

At the current state of development, the material design process for
strongly correlated electron compounds should begin at the intuitive
level, for example some physical idea of model that one would like
to explore or test, a property of a strongly correlated material which
could result in a useful property, a class of compounds one would
like to investigate comparatively, some ideas of chemistries of strongly
correlated materials which would enhance desirable solid state properties.
This zeroth order step can be refined with simplified quantitative
calculations using model Hamiltonians, or other computational tools
which refines our intuition. After this zeroth order step, it is natural
to divide the process of materials design into three additional steps
as summarized in Table~\ref{tbl:workflow}, which will be illustrated
by example in the following sections of this paper.

\begin{table*}
\begin{tabular}{l|c|c}
\hline 
Name  & Step  & Tools \tabularnewline
\hline 
{\footnotesize{}Motivation and analysis} & {\footnotesize{}defining directions and hypotheses to be tested, } & Heuristics, theory, simplified computations\tabularnewline
 & {\footnotesize{}refining them with simplified computations} & \tabularnewline
\hline 
\hline 
electronic structure  & structure $\to$ property  & electronic structure codes, DFT, DFT+DMFT \tabularnewline
structure prediction  & composition $\to$ structure  & evolutionary algorithms, Monte Carlo, minima hopping \tabularnewline
global stability  & chemical system $\to$ composition  & convex hull from materials databases \tabularnewline
\hline 
\end{tabular}

\vspace{3bp}
\caption{Three step workflow of materials design. Electronic structure and
structure prediction have been accessible for decades, with density
functional theory (DFT) as the key underlying method. In contrast,
checks for global stability require knowledge of all other structures
within a chemical system, which was only possible after the creation
of extensive computational materials databases.}
\label{tbl:workflow} 
\end{table*}

The first step is the quantitative calculation of the \emph{electronic
structure}, namely, how to go from a well defined structure (i.e.
atomic positions and ionic charges) to physical or chemical properties.
Given a crystal structure, we seek to compute one or several electronic
properties such as orbital occupancies, transport coefficients (resistivities,
mobilities, thermoelectric coefficients, etc.) Mott or charge transfer
gap sizes, magnetic order parameters, etc. As we have seen in section~\ref{sec:correlations},
for correlated electron materials the computational method used for
electronic structure depends on the strength and kind of correlations
and should be chosen appropriately. Sometimes, this step is divided
into two. In the first step first one derives from first principles
an effective model Hamiltonian, and then one solves the model and
explores its consequences. While the advanced functionals described
in the previous section go directly from structure to property bypassing
the model Hamiltonian, the latter can be useful for the interpretation
of the results. 

The second step is \emph{structure prediction}: predict the crystal
structure given a fixed chemical composition. A successful prediction
would take a formula, like Fe$_{2}$O$_{3}$ for example, and return
the correct crystal structure--in this case, the composition forms
in the corundum structure, called the $\alpha$ phase. For a more
complete characterization, we would seek to predict not only the ground
state structure, but nearby local minima as well, termed polymorphs.
Again taking Fe$_{2}$O$_{3}$ as an example, this composition also
forms in the spinel structure, found naturally as the mineral maghemite
and termed the $\gamma$ phase, as well as cubic $\beta$ and orthorhombic
$\epsilon$ phases. Polymorphs generally are formed in different temperature
and pressure regimes, and modeling these effects add an additional
layer of complexity. However, simply enumerating the low-energy local
minima at zero temperature can already provide a broader picture of
the structural diversity of a composition. Furthermore, if the structure
one is interested turns out not to be the ground state but a metastable
structure one can design methods for stabilizing it, either by choosing
an appropriate synthesis method or by applying external perturbations
such as stress exerted by a properly chosen substrate. 

The general procedure for structure prediction involves placing atoms
in a unit cell and using an algorithm to efficiently traverse the
space of atomic configurations and cell geometries to arrive at low
energy structures. This step requires having an accurate method for
producing the energy of a given configuration of atoms and sampling
these configurations. There are a number of structure prediction techniques
(see the review Ref. \onlinecite{Woodley08}) including the simulated
annealing approach \cite{Pannetier90,Kirkpatrick83}, evolutionary
algorithm methods \cite{Woodley99,Oganov_2006,Trimarchi07}, structure
models by analogy based on data mining and machine learning \cite{Curtarolo03,Fischer06},
metadynamics \cite{Laio02,Barducci11}, basin and minima hopping \cite{Wales97,Goedecker04},
random structure searching \cite{Pickard11,Stevanovic16}, and so
on.

The third step is testing for \emph{global stability}: given the lowest
energy structure of a fixed composition, check whether it is stable
against decomposition to all other compositions (phase separation)
in the chemical system. 

The steps involving total energies are assisted by electronic structure
methods and material databases. In particular the third step which
requires the knowledge of all other known stable compositions, their
crystal structures and total energies, is now facilitated by materials
databases containing data in standardized computable formats, such
as the Materials Project \cite{Jain_2013}, the Open Quantum Materials
Database (OQMD)\cite{Kirklin2015}, AFLOWlib\cite{Setyawan2011} and
NIMS\cite{Neugebauer2012}. With this information, the energetic convex
hull for a chemical system can be constructed and the target composition
checked for stability against decomposition (phase separation).

For weakly-correlated materials, the entire workflow can be built
around $LDA/GGA$ for total energies and $G_{0}W_{0}$ for spectral
properties. For correlated materials, GGA is a good starting point
for computing total energy differences (such as reaction energies
or structural energy differences). In this review we highlight some
failures of the LDA/GGA energies in structural prediction and phase
stability, and ways to introduce corrections to account for the correlation
effects.

The need to correct DFT total energies for materials design projects,
is broadly recognized in the context of all the material databases
where the GGA/LDA results are corrected using semi-empirical schemes.
There are three broadly used schemes in literature, and they have
associated databases, the\textit{ fitted elemental-phase reference
energy} (FERE) scheme~\cite{Stevanovic_2012}, Materials Project
\cite{Jain_2013}, and Open Quantum Materials Database (OQMD)\cite{Kirklin2015}.

While the details of the implementation are different, they have two
key elements in common. First, instead of using GGA, the total energies
are computed within the GGA+U method, with some U values empirically
assigned to each element. Second, the experimental formation energies
$\Delta H^{\text{exp}}$ are used to determine best fits for elemental
energies $E^{\textrm{Fitted}}(A)$, where A is an element, for a training
set of compounds by solving the linear least-squares problem.

\begin{align}
 & \Delta H^{\text{exp}}(\text{A}_{m}\text{B}_{n})\thickapprox\mathcal{E}_{\textrm{corrected}}^{\text{GGA+U}}(\text{A}_{m}\text{B}_{n})\label{eq:reqction-fit-1}\\
 & =E^{\text{GGA+U}}(\text{A}_{m}\text{B}_{n})-mE^{\textrm{Fitted}}(A)-nE^{\textrm{Fitted}}(B)\nonumber 
\end{align}

We note that all elemental energies are fitted for all elements in
FERE (whithin the set of relevant elements) while only selected ones
are fitted in MP and OQMD (especially in the ``fit-partial'' scheme
in OQMD).

We describe the different correction methods in Appendix \ref{sec:Empirical-corrections}.
In this article we will use the Materials Project database for analysis
of phase stability and estimate probabilities based on their data.
We will show in section \ref{sec:bacoso} that careful consideration
of correlations is essential not only for calculation of phase stability,
but also for structure prediction.

Notice that the theoretical workflow, outlined in Table~\ref{tbl:workflow}
progresses in an order different from experimental solid state synthesis.
There, elements and simple compounds in a chemical system are combined
and subjected to heating/cooling cycles to provide the kinetic energy
necessary for atomic rearrangement to form new stoichiometries (of
which there may be more than one). Finally the stoichiometries crystallize
to form structures which are then isolated for the study of their
properties.

\section{\emph{Statistical Interpretation}}

\label{sec:Probability-Estimation}

DFT has reached a high degree of stability and scalability, enabling
software packages such as USPEX~\cite{Oganov_2006,Lyakhov_2013}
to implement genetic algorithms on top of DFT to successfully predict
never before observed structures. As discussed in the previous sections,
correlations in the form of U and empirical corrections are now available
in several databases. Since these methods are not exact methods and
suffer from systematic errors, compounds predicted to be stable will
not necessarily be found in experiment, and vice versa.

The main question we address in this section is the interpretation
of the above-hull/below-hull energies that we compute within LDA/GGA
(with or without the empirical corrections). There are two related
questions: (1) what is the likely error in the computed energy, (2)
how likely is the compound to be synthesized given its energy relative
to the convex hull. Namely, how likely is one to find the target compound?\textcolor{red}{{}
}\textcolor{black}{This assessment serves as a background for the
conclusion section \ref{sec:conclusion}, where we evaluate the results
of various material design projects. }

\begin{figure}[h]
\includegraphics[scale=0.8]{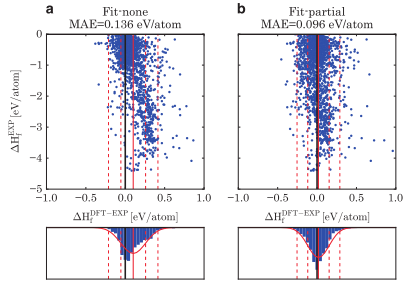}

\caption{Comparison of ``Fit-None'' to ``Fit-Partial'' in OQMD (from Ref.
\onlinecite{Kirklin2015}). Left side shows raw data (with U added
to some correlated elements in certain compounds) - there is an evident
skewness and drift. On the right the data is corrected by a small
set of chemical potentials. The skewness and drift are eliminated.\label{fig:OQMD-fits}}
\end{figure}

Reference \onlinecite{hautier_prb12} modeled the computational error
- the difference between computed and experimental formation energies
- as a random variable with normal distribution. A normal-distribution
was also used by Ref. \onlinecite{Kirklin2015}, as seen in Fig. \ref{fig:OQMD-fits}.
We follow a similar statistical approach to the question. 

Denote by $\mathcal{{E}}_{exp}$ the experimental heat or enthalpy
per atom of the reaction A+B$\,\rightarrow$~AB at low temperature,
and denote by $\mathcal{{E}}_{calc}$ the same quantity computed using
an approximate method, like GGA, or the empirically-corrected value
of this computation. We treat \textbf{$\boldsymbol{\mathcal{{E}}}_{\boldsymbol{calc}}$}
and $\boldsymbol{\mathcal{{E}}}_{\boldsymbol{exp}}$ as real random
variables, and analyze the distribution of the variable $d=\boldsymbol{\mathcal{{E}}}_{\boldsymbol{calc}}-\boldsymbol{\mathcal{{E}}}_{\boldsymbol{exp}}$:

\begin{equation}
P(\boldsymbol{\mathcal{{E}}}_{\boldsymbol{calc}}-\boldsymbol{\mathcal{{E}}}_{\boldsymbol{exp}}=d)=F_{\alpha\beta\mu}(d)\label{eqn:pm}
\end{equation}
where $F_{\alpha\beta\mu}(d)$ is some probability distribution function
(PDF) with center $\mu$ and scale $\alpha$, as well as some shape
parameter $\beta$. Since GGA (or one of its correction schemes) is
reasonably accurate, we expect $F_{\alpha\beta\mu}(d)$ to be concentrated
around the center.

In order to study this distribution, we observe that the same quantity
in Eq.~(\ref{eqn:pm}) describes computational errors in \textit{energies
above-the-hull} as well as computational errors in \textit{formation}
energies. This holds because the distribution applies to \textit{computational
error in reaction energies in general}. The crucial point that makes
this possible is that the number of atoms is balanced on the left
and on the right (so as to cancel core energies). Therefore we can
train a statistical model on the distribution of computational error
for formation energies, for which there exists a reasonably-sized
experimental data set, and then make predictions based on above-hull
energies.

We expect a stronger statistical-correlation when all the systems
A, B, and AB are weakly correlated than when the correlations are
strong, hence the parameters $\alpha,\,\beta,\,\mu$ should be taken
in a well-defined space of materials defined from the outset. Since
we will use this model for prediction, it is necessary to fit the
parameters using a large-enough sample of representative materials. 

\begin{figure*}
\includegraphics{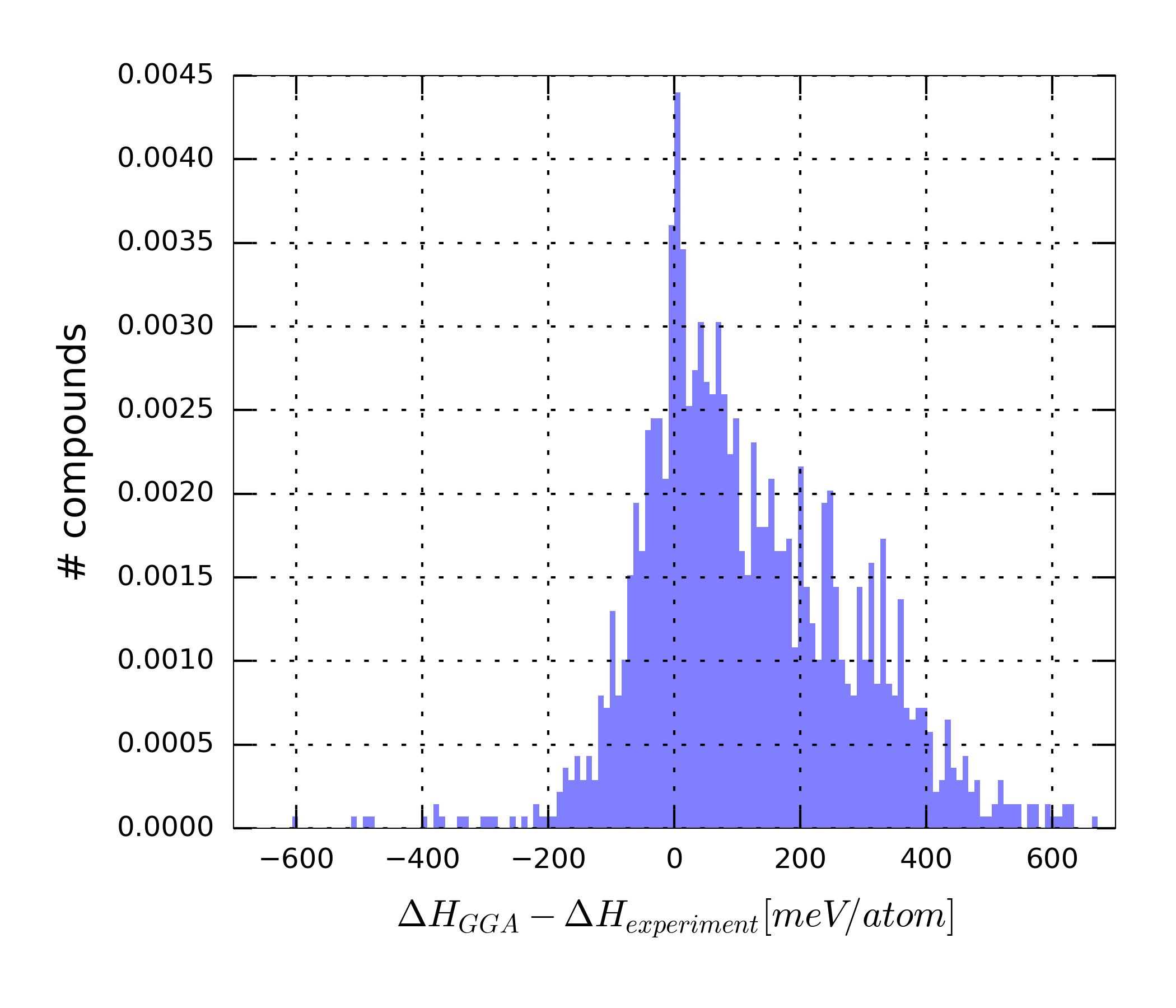}\includegraphics{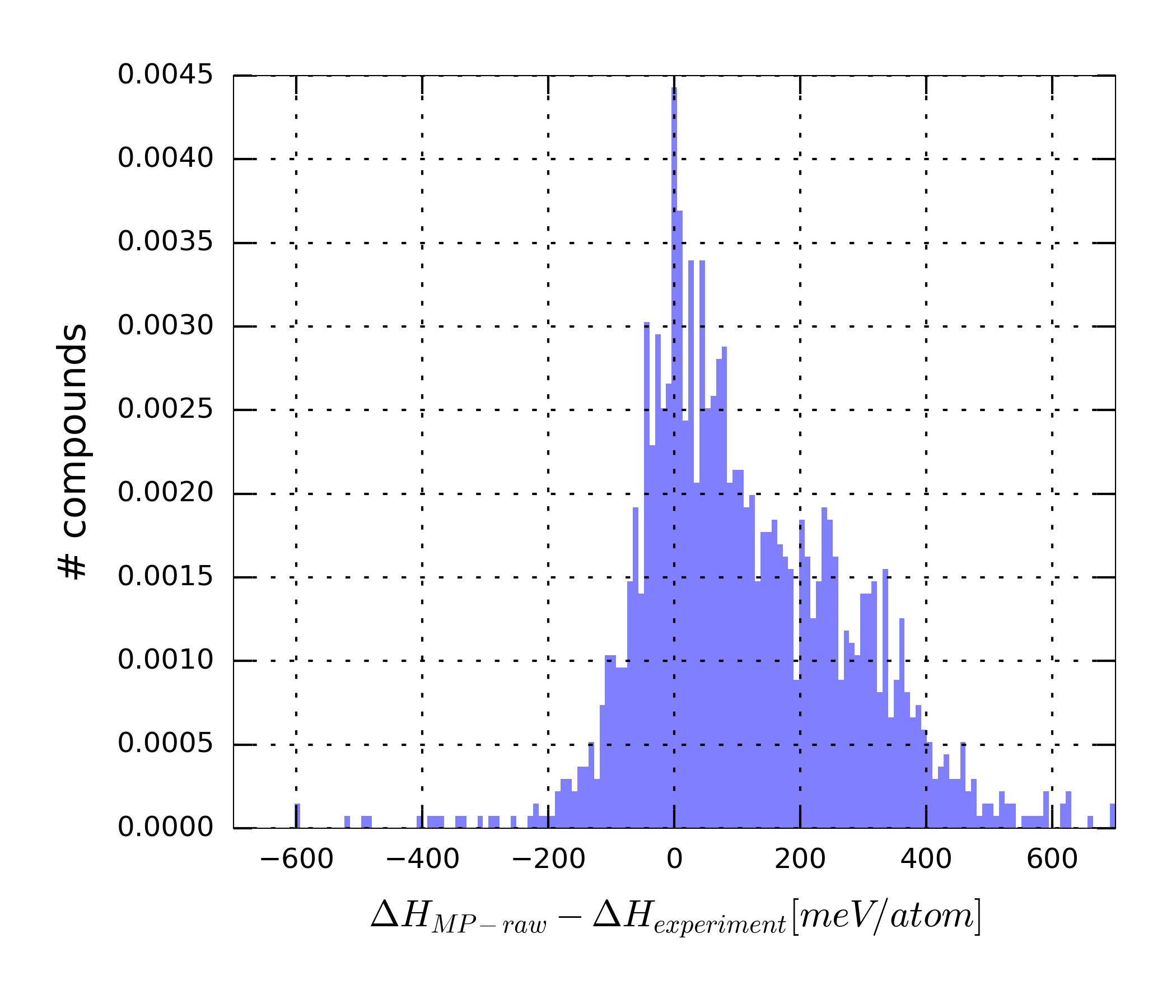}

\includegraphics{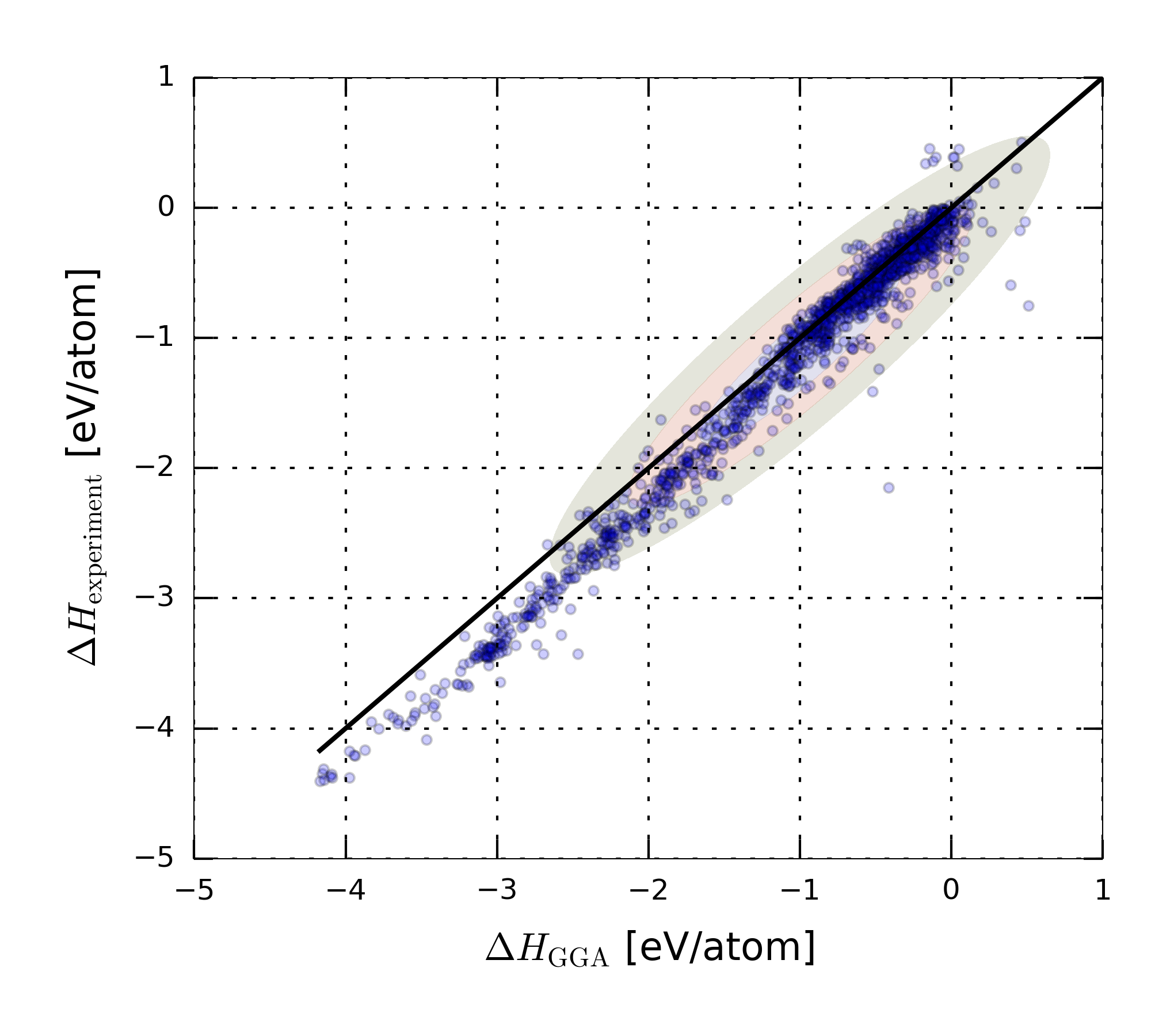}\includegraphics{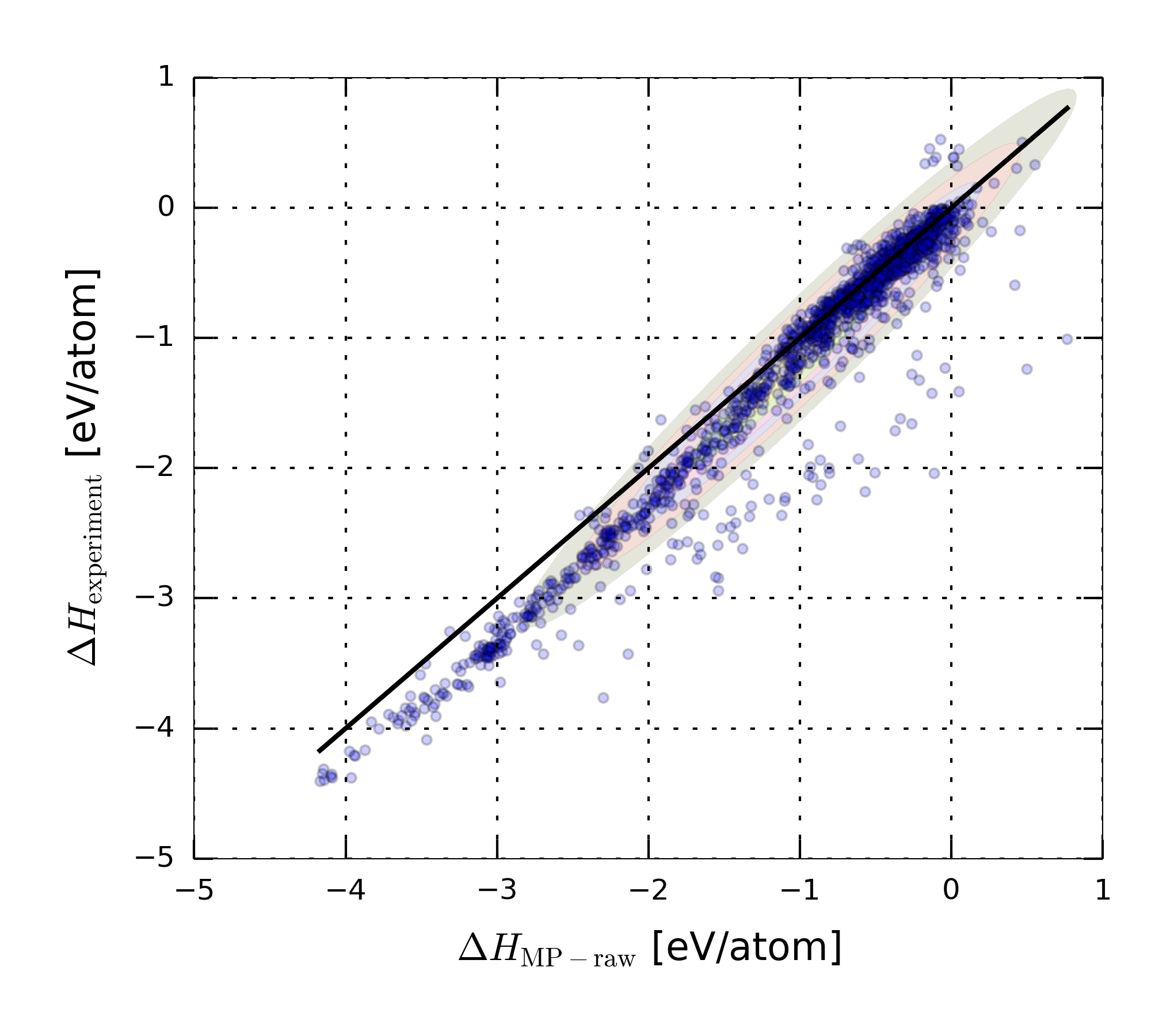}

\caption{\textbf{Left}: Distribution of computational error reproduced with
VASP GGA(PBE) for compounds collected in the data set. The data set
includes 1,500 substances from OQMD and the materials listed in table
II in \onlinecite{Stevanovic_2012}. To get the experimental formation
energies for the corresponding compounds in the data set, we use the
OQMD database, where the experimental data is available as well as
the crystal structure. The experimental data is also merged with table
II in \onlinecite{Stevanovic_2012}. The pure GGA(PBE) formation energies
are calculated for the compounds in the data set and compared with
the corresponding experimental formation energies. \textbf{Right:}
plots for Materials Project's raw data, which adds nonzero U for some
elements in certain compounds. As can be seen, addition U is not enough
to make the distribution $F_{\alpha\beta\mu}(d)$ un-skewed or fix
the drift to the bottom (this is also evident in the Fit-none distribution
from OQMD~\cite{Kirklin2015}, shown in Fig. \ref{fig:OQMD-fits}).
Also shown are the ellipses of the bivariate-normal estimators, demonstrating
the drift.}

\label{fig:RAW_GGA}
\end{figure*}

Figure \ref{fig:RAW_GGA} shows the distribution of computational
error $\boldsymbol{\mathcal{{E}}}_{\boldsymbol{calc}}-\boldsymbol{\mathcal{{E}}}_{\boldsymbol{exp}}$
for formation energies of compounds collected in the data set. The
data set includes 1,500 substances from the OQMD database and the
materials listed in table II in {[}14{]}. It is noteworthy that the
experimental formation energies are available as well as crystal structures
of the compounds in the OQMD database (query \footnote{The SQL query sent to OQMDv1.1 was: 

select

d.composition\_id, '\#', 

d.delta\_e as formation\_energy,

e.delta\_e as dft\_formation\_energy,

d.source as expt\_source,

e.source as dft\_source 

from

expt\_formation\_energies as d,

expt\_formation\_energies as e 

where

d.dft =0 and e.dft=1 and 

e.composition\_id = d.composition\_id} was used to collect these pairs). The experimental data originates
from 2 sources: the SGTE Solid SUBstance (SSUB) database\cite{Kong2015},
and the thermodynamic database at the Thermal Processing Technology
Center at the Illinois Institute of Technology (IIT)\cite{IIT-database},
as described in Ref.~\onlinecite{Kirklin2015}. The experimental
formation energies were also collected from table II in \onlinecite{Stevanovic_2012}
and were merged with the data set. In Fig. \ref{fig:RAW_GGA} the
left-side plots show the computational error for pure GGA calculations
(reproduced with our own VASP calculations for the compounds in the
data set), whereas the right hand side includes +U for some of the
compounds, as in the from the Materials Project's recipe (see appendix
for details). We observe that the distribution is skewed to the right,
and that application of the U correction is not enough to undo the
skewness, although it reduces $\sigma$ from 427 meV to 266 meV (with
an increase of the mean error from 155 meV to 172 meV). This data
is summarized in Table \ref{tbl:stat-params}. 

\begin{figure*}
\includegraphics{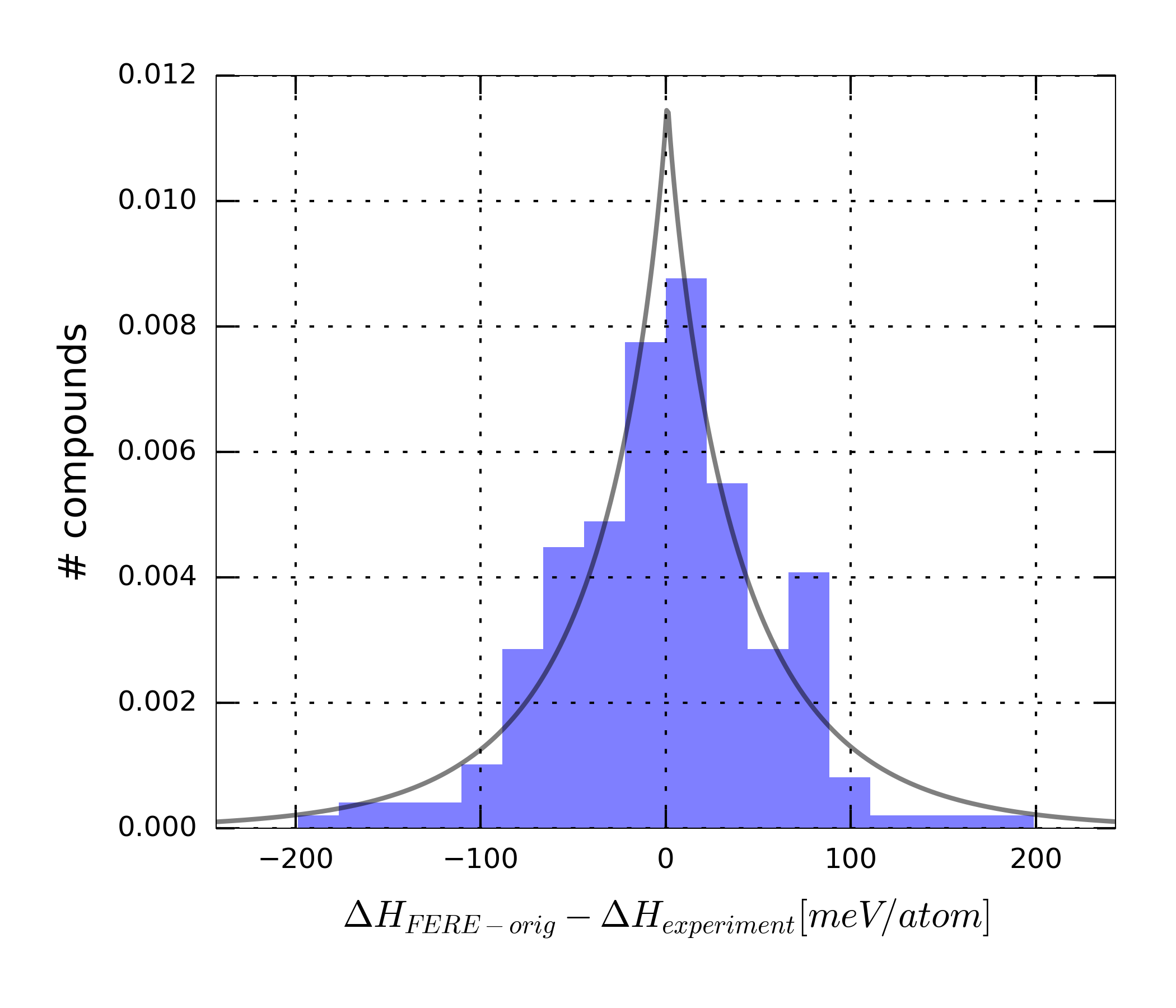}\includegraphics{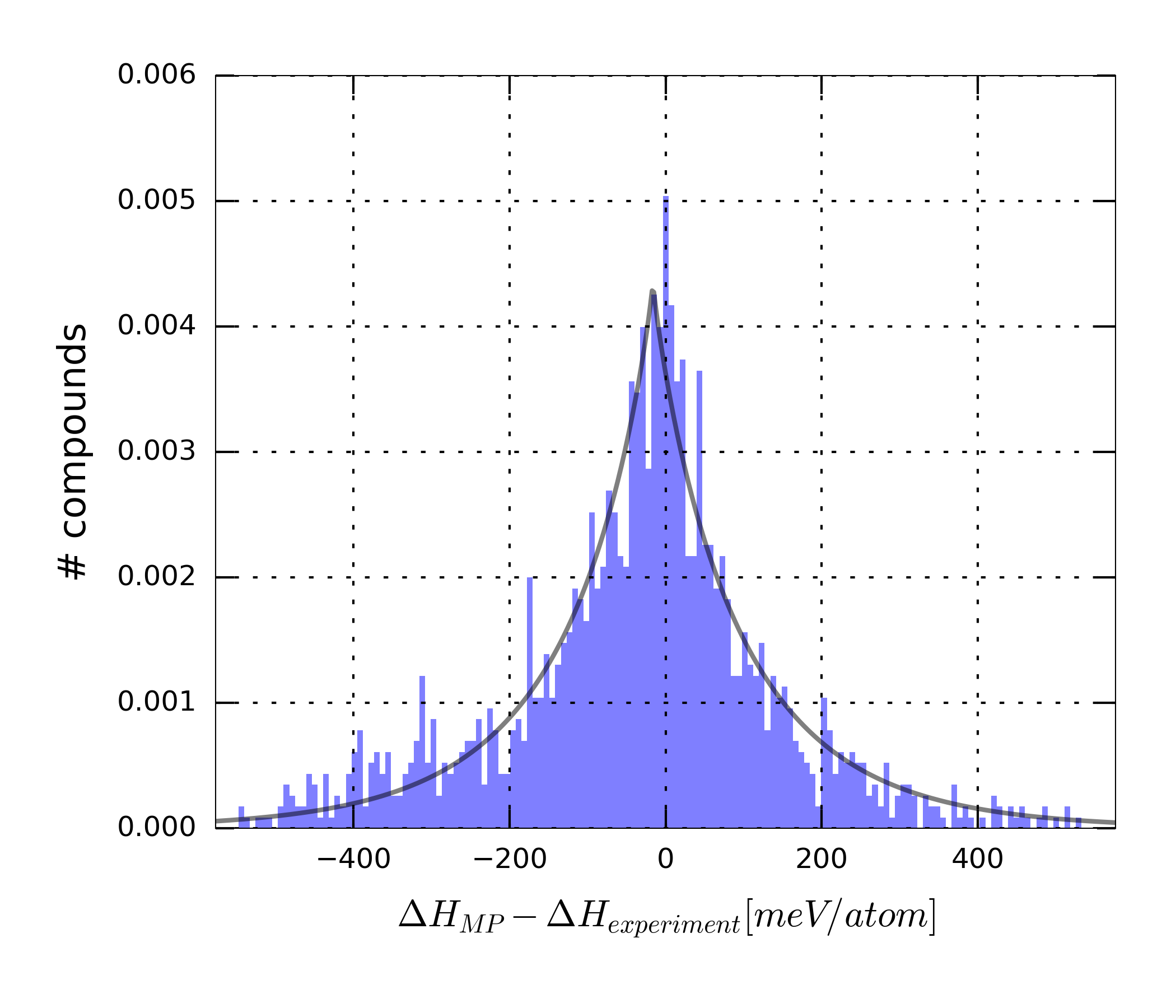}

\includegraphics{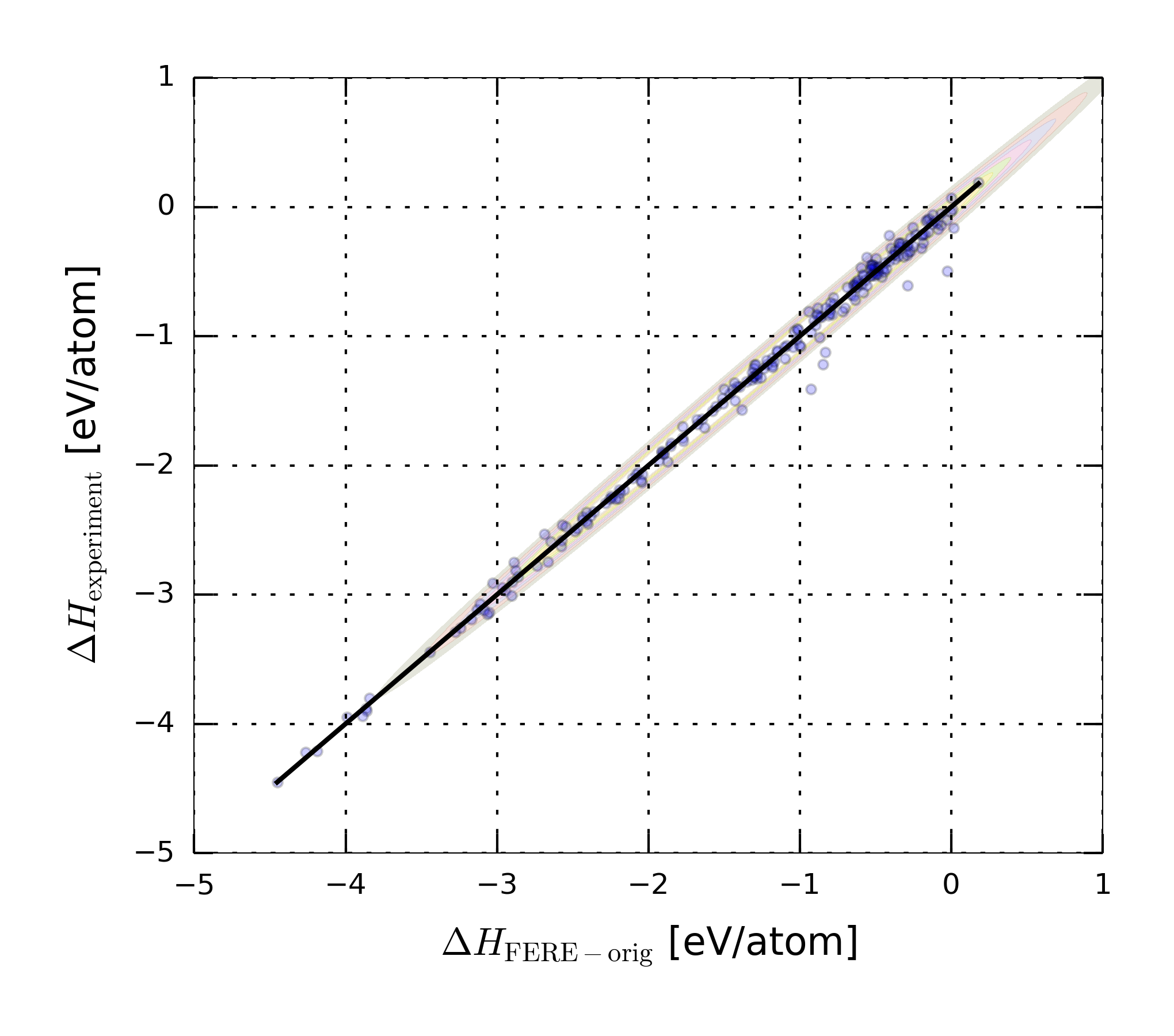}\includegraphics{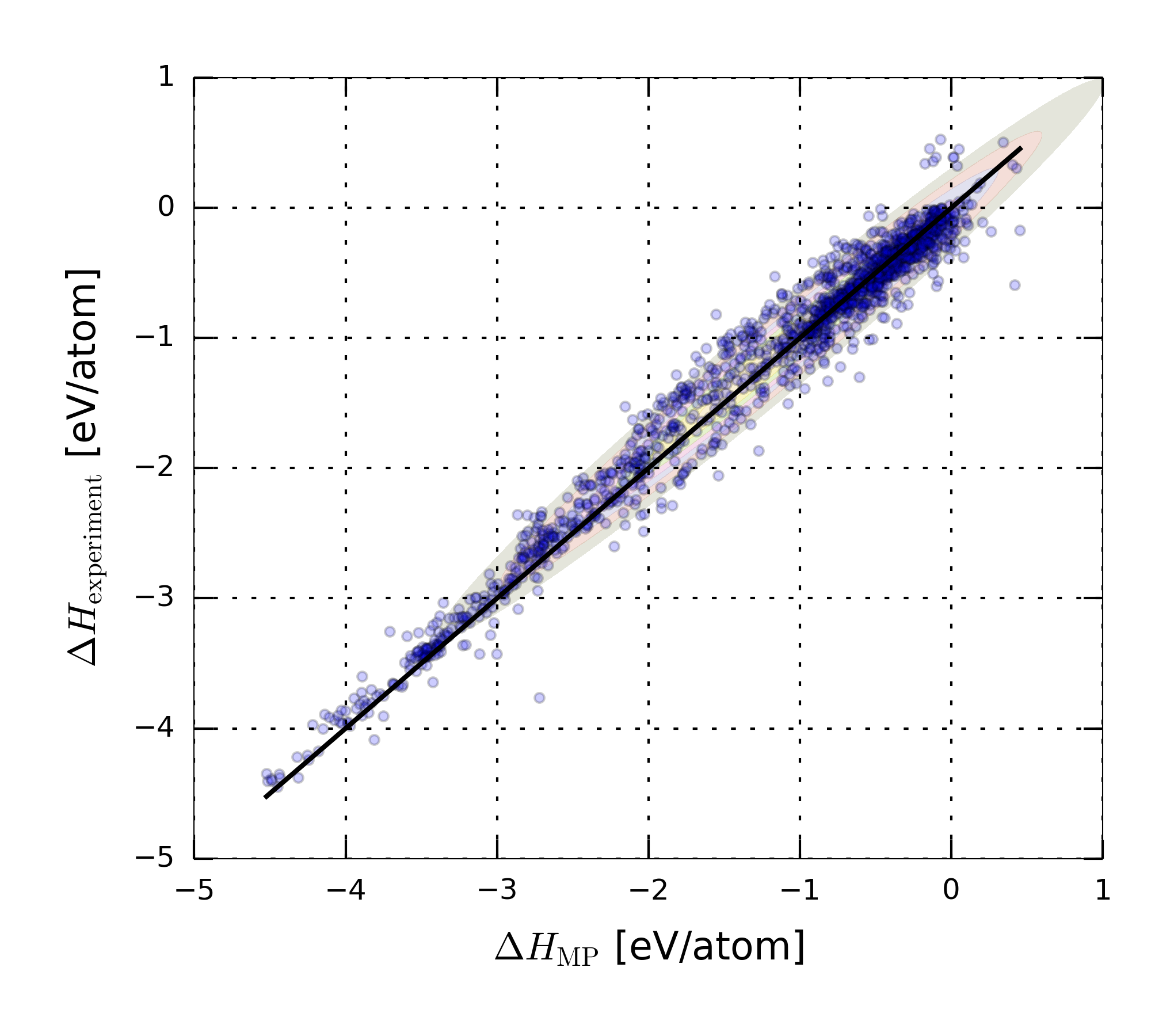}

\caption{Distribution of computational error for empirical correction schemes:
FERE (left) and Materials Project corrected scheme (right). For FERE,
the data comes from table II in Ref.~\onlinecite{Stevanovic_2012}.
For the Material Project, the corresponding corrected GGA values were
downloaded using their Python interface. Fitting of the distribution
was done using the maximum likelihood method \texttt{\textcolor{black}{fit()}}(from
Scipy\cite{SCIPY}) for the generalized-normal distribution (\texttt{\textcolor{black}{scipy.stats.gennorm}}).
The distributions are summarized in the third and fourth columns of
table \ref{tbl:stat-params}. }

\label{fig:CORRECTED}
\end{figure*}

As discussed above, various authors have corrected the GGA/GGA+U values
for formation energies by shifting the chemical potentials of elements.
This is usually achieved by fitting the set of experimental results
on the equations for GGA/GGA+U formation energies as in Eq. (\ref{eq:reqction-fit-1}).
The effect of this procedure on the distribution $F_{\alpha\beta\mu}(d)$
is to make it centered, and to eliminate the drift of the bivariate
distribution. This can be seen in the OQMD fit (Fig. \ref{fig:OQMD-fits},
as well as in FERE and Materials Project (Fig.\ref{fig:CORRECTED}). 

\begin{figure}[h]
\includegraphics{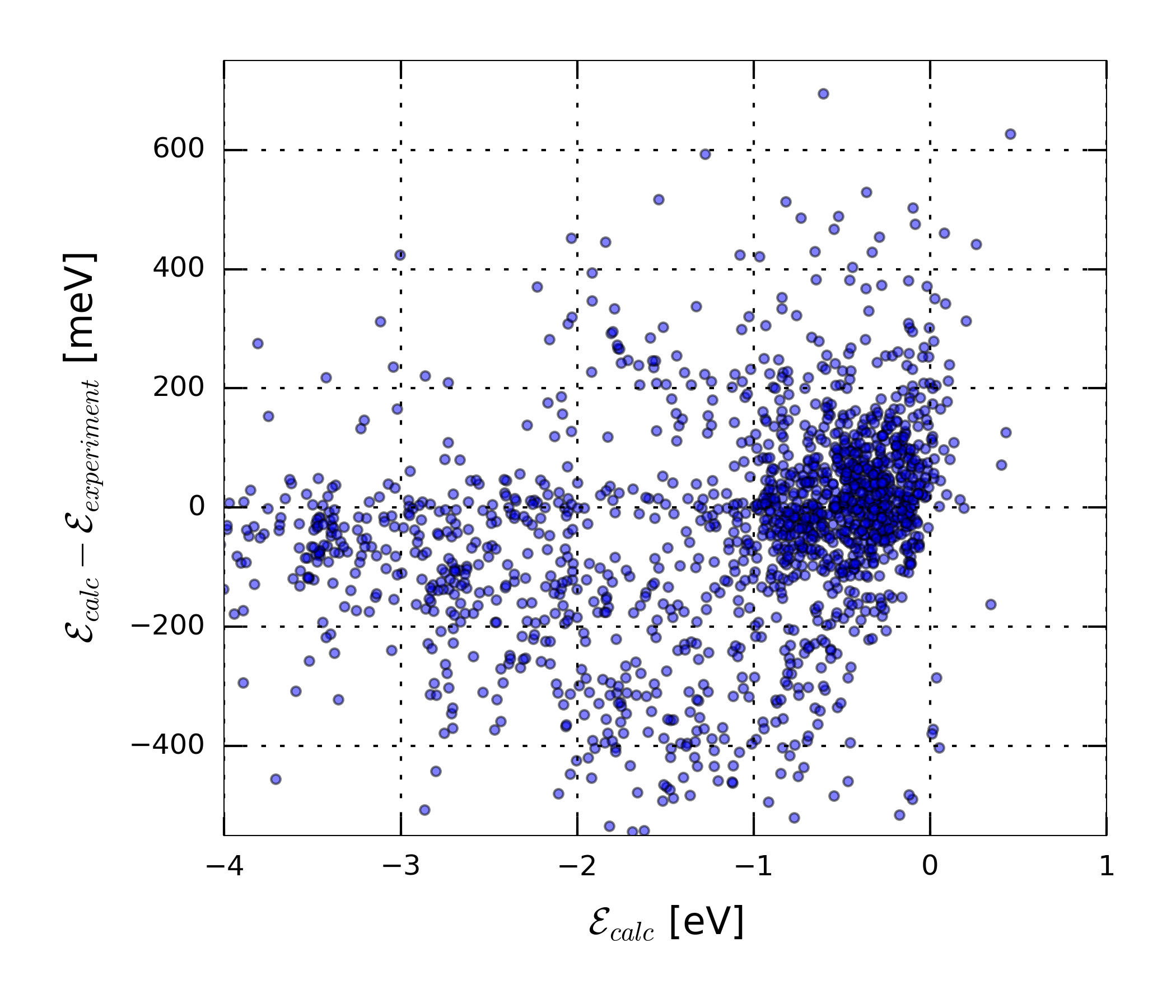}

\caption{Plot of the computational error in Materials Project formation energies
against the calculated formation energy, demonstrating little or no
correlation between $\mathcal{E}_{exp}-\mathcal{E}_{calc}$ and $\mathcal{E}_{calc}$.
Each data point corresponds to a pair of calculated and experimental
formation energy for a material in the data set. \label{fig:independence}}
\end{figure}

\begin{figure}[h]
\includegraphics[scale=0.38]{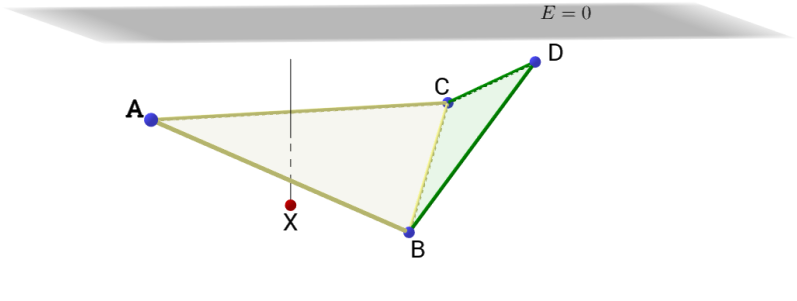}

\caption{\textcolor{black}{The determinant reaction for a new material X -
X lies either above or below the convex hull of reactions. In the
illustration above it is well below the ABC triangle, although its
energy is above the B compound (note that in this case X still forms
out of A, B, and C). X's computed distance below the hull, which we
denoted in the text by $\mathcal{{E}}_{X}$, is the quantity that
determines its formation probability in our analysis. In this diagram
it is equal to X's vertical distance from the triangle ABC, and therefore
the reaction ABC$\rightarrow$X is the determinant reaction in this
hull. }\label{fig:det-reaction}}
\end{figure}

Another important property of the corrected distributions is that
the error can be seen as approximately independent of the value of
the computed energy:

\begin{align*}
P(\boldsymbol{\mathcal{E}}_{\boldsymbol{calc}}-\boldsymbol{\mathcal{{E}}_{exp}} & =d|\boldsymbol{\mathcal{{E}}}_{\boldsymbol{calc}}=x)\approx P(\boldsymbol{\mathcal{E}}_{\boldsymbol{calc}}-\boldsymbol{\mathcal{{E}}_{exp}}=d)\\
= & F_{\alpha\beta\mu}(d)
\end{align*}

This (approximate) independence is evident in Fig. \ref{fig:independence}.
One can reason that as long as the computational error is small, it
should not be correlated with the value of the computed energy. Finally,
with just a few 1000's of points, there is not enough data to split
the domain into sub-ranges and make meaningful statistical analysis.
With more data one could refine the distribution parameters on ranges
of $\boldsymbol{\mathcal{{E}}}_{\boldsymbol{calc}}=x$, or possibly
other variables that we did not consider here. 

\begin{figure}[!h]
\includegraphics{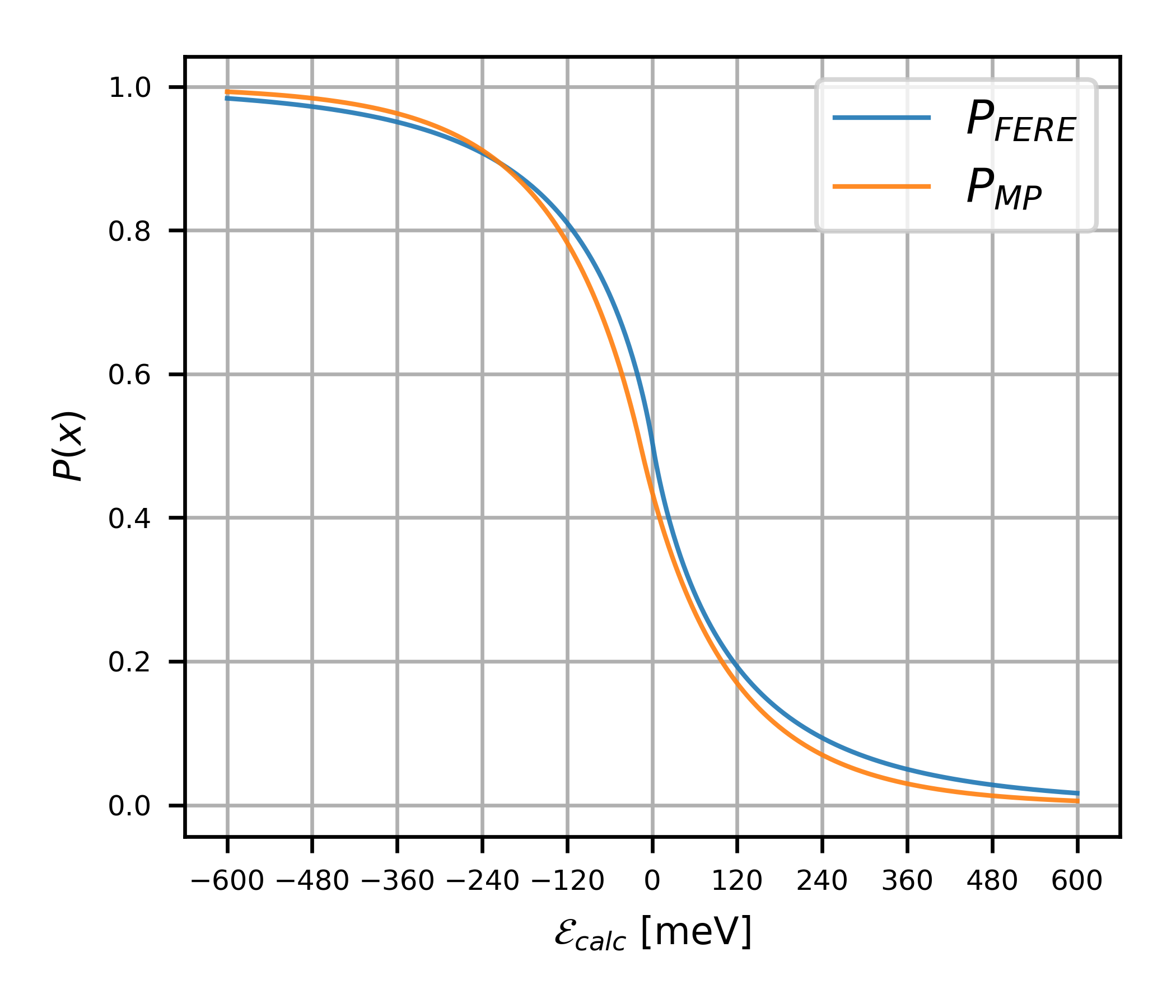}\caption{$\mathcal{{P}}(\mathcal{E}_{calc})$, probability for a compound to
exist, given the computed formation energy $\mathcal{E}_{calc}$ with
the two correction schemes summarized in Table \ref{tbl:stat-params}.\label{fig:probabilities}}
\end{figure}

For prediction, the\textit{ probability that AB forms as the ground
state}, when the computed energy is at a distance $x$ above the Hull
is given by 
\begin{eqnarray}
\mathcal{{P}}(x) & = & \int_{-\infty}^{0}P(\boldsymbol{\mathcal{{E}}}_{\boldsymbol{exp}}=y|\boldsymbol{\mathcal{{E}}}_{\boldsymbol{calc}}=x)dy\label{eq:CDF}\\
 & = & \int_{-\infty}^{0}P(\boldsymbol{\mathcal{E}}_{\boldsymbol{calc}}-\boldsymbol{\mathcal{{E}}_{exp}}=x-y|\boldsymbol{\mathcal{{E}}}_{\boldsymbol{calc}}=x)dy\nonumber \\
 & \approx & \int_{-\infty}^{0}F_{\alpha\beta\mu}(x-y)dy=1-{\cal {\cal {\cal F}}}_{\alpha\beta\mu}(d\leqslant x)\nonumber 
\end{eqnarray}
where $\mathrm{{\cal F}_{\alpha\beta\mu}(d\leqslant x)}$ is the cumulative
distribution function corresponding to $F_{\alpha\beta\mu}$. This
expression can be evaluated numerically by estimating the number of
data points in the distribution tail (where \# denotes the number
of elements in the set):

\[
\mathcal{{P}}(x)\approx\frac{\#\{(\boldsymbol{\mathcal{E}}_{\boldsymbol{calc}},\boldsymbol{\mathcal{{E}}_{exp}})\,|\mathcal{\,\boldsymbol{E}}_{\boldsymbol{calc}}>x\}}{\#\{(\boldsymbol{\mathcal{E}}_{\boldsymbol{calc}},\boldsymbol{\mathcal{{E}}_{exp}})\}},
\]

however, it is convenient to have an analytic form. For the corrected
distributions, we postulate that $F_{\alpha\beta\mu}$ is Normal or
generalized-Normal distribution - which includes also Normal $(\beta=2)$,
exponential $(\beta=1)$, as well as uniform $(\beta=\infty)$ distributions:

\begin{equation}
F_{\alpha\beta\mu}(d)={\frac{\beta}{2\alpha\Gamma(1/\beta)}}\;e^{-(|d-\mu|/\alpha)^{\beta}}\label{eq:PDF}
\end{equation}
 Using the experimental data (see above), we calculated the numbers
listed in Table~\ref{tbl:stat-params}. We used the maximum likelihood
method to estimate $\alpha,\beta$ and $\mu$. The \textcolor{black}{first}
column summarizes the distribution of raw GGA calculations, which
were reproduced with our own GGA runs for the compounds in the data
set. The \textcolor{black}{second} column corresponds to Materials
Project's GGA or GGA+U (GGA for most compounds and GGA+U only for
certain correlated materials specified in appendix), which was evaluated
using the raw data from Materials Project. These distributions are
depicted in Fig.~\ref{fig:RAW_GGA}. We did not include model estimates
for the non-corrected distributions, since they do not comply with
some of the assumptions (as explained above). The \textcolor{black}{third}
column in the table is fitted for $\Delta H_{FERE}$ from Table II
in Ref. \onlinecite{Stevanovic_2012}. As expected, the parameters
for FERE show a smaller standard error (for a smaller dataset). The
\textcolor{black}{fourth} column corresponds to corrected formation-energies
from the Materials Project. Again, the parameters show a smaller standard
error compared to bare GGA. These distributions are depicted in Fig.~\ref{fig:CORRECTED}.
Figure~\ref{fig:probabilities} shows the calculated probabilities
in each one of the schemes for ${\cal E}_{calc}$. 

\begin{table}[h]
\begin{tabular}{c|c|c||c|c}
\hline 
\multirow{2}{*}{} & \multicolumn{2}{c||}{Before corrections} & \multicolumn{2}{c}{After correction}\tabularnewline
\cline{2-5} \cline{3-5} \cline{4-5} \cline{5-5} 
 & GGA (U=0) & MP (+U)  & ~FERE~ & MP\tabularnewline
\hline 
Mean  & 136 & 151 & 6 & -31\tabularnewline
\hline 
MAE  & 155 & 172 & 51 & 130\tabularnewline
\hline 
$\sigma$  & 427 & 266 & 81 & 192\tabularnewline
\hline 
\hline 
center $\mu$  & -- & -- & 1 & -16\tabularnewline
\hline 
shape $\beta$  & -- & -- & 0.85 & 0.90\tabularnewline
\hline 
scale $\alpha$  & -- & -- & 39 & 109\tabularnewline
\hline 
\# data points & 1500 & 1598 & 227 & 1598\tabularnewline
\hline 
\end{tabular}

\vspace{3bp}

\caption{Statistics for the distribution of error $\mathcal{F}$. Maximum likelihood
values for the parameters of the distribution function in Eq. (\ref{eq:PDF})
\label{tab:CDF} are given for the correction schemes. All units are
meV except for $\beta$ (unitless) and the number of points. MP (+U)
stands for Materials Project's raw data, which only includes the U
correction for certain correlated compounds, as described in appendix.}
\label{tbl:stat-params} 
\end{table}

Interestingly, the mean values for pure GGA and Materials Project's
raw data (table \ref{tbl:stat-params}) are positive, meaning that
GGA/GGA+U tend to over-estimate formation energies. We also observe
that the corrected distributions are more exponential-like then normal.
In fact, they all get a very low score on statistical Shapiro testing
for Normalcy, but pass the Kolmogorov test for generalized-Normal
distribution. 

\textcolor{black}{In general we would like to apply the same kind
of statistical error analysis to the systematic computational error
made by }\textbf{\textcolor{black}{correlated}}\textcolor{black}{{}
electronic structure calculations, such as DMFT, or Gutzwiller. However,
since there exists no database of energies for these methods, at this
point in time it is impossible to estimate the parameters of the model
Eq.~(\ref{eq:PDF}) for correlated methods. Still, since we expect
these methods to be more accurate than GGA, we expect their standard
error $\sigma$ to be smaller, and their mean $\mu$ to be closer
to 0 than what we found for GGA. Since our probability estimate depends
only on the distance from the convex hull, it is determined only by
one triangle in the diagram, which we refer to as the determinant
reaction, see Fig.}~\textcolor{black}{\ref{fig:det-reaction}. Only
the energies of the relevant compounds (in the diagram, these are
A, B, C, and X) need to be re-estimated using the computationally
expensive correlated method, and thus one can improve the probability
estimate for X, without requiring too much extra computation. Since
the cumulative distribution function (Eq.}~\textcolor{black}{(\ref{eq:CDF}))
for the correlated computation will be sharper, if the energy of X
falls below the hull, its probability will clearly be larger than
0.5, whereas if the energy is well above the hull we argue that it
is unlikely to exist.}

Narayan \textit{et al.}\cite{Narayan_2016} constructed a related
statistical model to address theoretical predictions of new materials
using formation energies computed with the Materials Project corrections.
They proposed an optimal cutoff energy $\epsilon_{0}=0.1\,eV$, to
minimize the rates of false-positives and false-negatives. They suggested
that the synthesis of materials with $\epsilon<\epsilon_{0}$ should
be attempted. Our analysis can be seen as a refinement of the idea
of a cutoff.

From an experimental standpoint, the question to answer is whether
an experimental search should be pursued or not. This depends not
only on the function $\mathcal{{P}}(x)$ but also on how interesting
is the estimated property that we are seeking. Also, notice that we
are asking the question of whether we will obtain a material close
to its ground state. It is well known that many known compounds are
metastable~\cite{Sun_2016}. Therefore, when looking for compounds
which have not been synthesized before, we are obtaining a lower bound
for the probability of finding a new material, which is in greater
than $\mathcal{{P}}(x)$. Third, notice that stability analyses based
on convex hulls can only rule out the stability of a compound, rather
than definitively prove its stability, since technically there are
an infinite number of compositions and structures to consider in the
phase diagram.

Theory assisted material discovery has already had notable successes,
where the predicted compounds were successfully synthesized. It has
lead to new multiferroic materials (Refs.~\onlinecite{Ryan2013,Fennie_2008}).
The material design strategy was used to discover and synthesize 18
new ternary semiconductors\cite{Gautier_2015}. It has also been used
to find two new structures in the Ce-Ir-In system~\cite{Fredeman_2011}.
There are also reports of broad theoretical searches where experimental
synthesis did not find the theoretical approach predictive\cite{Narayan_2016}.
As an example from that work, we examine KScS$_{2}$ which has $\mathcal{E}_{calc}=-0.136\,eV$
relative to the convex hull. According to our analysis it has probability
0.8 to exist and indeed it has been synthesized successfully in Ref.~\onlinecite{Havlak2015a}.

{\def\bibliography 
\bibliography{cormatdes,gabi}

}

\section{Case Studies}
\label{sec:exhibits}

In the following, we illustrate the concepts of
the previous section providing   examples of the design of correlated
materials.
We limited ourselves to a few projects we are familiar with and
omitted important areas, as for example the design of
nuclear fuels~\cite{Deng2013a,Yin2011,Lanata2017,Butler2004,Soderlind2010,Yun2011}, where strong
correlations in the solid state play a  major role.

The examples  encompass materials displaying 
superconductivity \ref{sec:tuning}, \ref{sec:cuprates2}, \ref{sec:fe112},
charge disproportionation \ref{sec:tlcscl3},
and metal-insulator transitions \ref{sec:bacoso}.

The motivation for material design projects in the field of strongly correlated electron materials is not only to find new compounds , but to use the process to check our understanding of the theory, validating correct ideas and approaches and discarding wrong ones.
In each case the material design project begins with some intuitive idea that one would like to test.
We then move through the procedure outlined in section \ref{sec:workflow} , describing the steps to go from structure to property, from composition to structure and from components to composition.  We conclude each section  with the lessons learned from the project.
We
highlight the tools currently in use and the role that correlations play in
each stage of the workflow.

The field of material design of strongly correlated electron systems is in its
infancy and the examples span the development of the workflow itself---in some
cases  in the examples, only part of the full workflow was applied in the design process. Throughout the presentation we stress the
importance   of
qualitative ideas and chemical principles  and how they can be supported and enhanced with modern computational techniques.

\subsection{Tuning the charge-transfer energy}

\subsubsection*{\label{sec:tuning} Background }

The cuprate superconductors are classic examples of correlated materials.
Since this family of compounds exhibits the highest known superconducting
transition temperatures at ambient pressure (surpassed only by H$_{3}$S
at high pressures), they have been intensely studied since their discovery
in the mid-1980s. They have motivated an immense body of work in condensed
matter physics, both in new theories and huge leaps in experiment. 

Structurally, all the cuprate families have in common CuO$_{2}$ planes
which support superconductivity. They are described by the chemical
formula XS$_{n-1}$(CuO$_{2}$)$_{n}$, where $n$ CuO$_{2}$ planes
are interleaved with $n-1$ spacer layers S to form a multi-layer.
These multi-layers are then stacked along the c-axis, separated by
a different spacer layer X. It is known that the critical temperature
is a strong function of doping, and for each family there is an optimal
doping where the maximum superconducting $T_{\text{c}}$ ($T_{\text{c,max}}$)
occurs. Empirically, it is known that $T_{\text{c,max}}$ is strongly
materials-dependent, ranging from 40 K in La$_{2}$CuO$_{4}$ to 138
K in HgBa$_{2}$Ca$_{2}$Cu$_{3}$O$_{8}$. 

It is now agreed that superconductivity arises from doping a parent
compound which is a charge transfer antiferromagnetic insulator\textcolor{red}{\footnotesize{}~\cite{Zaanen85}},
and that the symmetry of the superconducting state in these materials
is $d$-wave, as it was predicted theoretically\textcolor{red}{\footnotesize{}~}\cite{doi:10.1142/S0217979287001079,Kotliar1988a,Zhang1988b}.
One should stress however, that there is no consensus on the \textit{mechanism}
that control the superconducting critical temperature and as a consequence
what leads to the correlation of $T_{\text{c,max}}$ with the apical
oxygen distance. Even if one accepts a mechanism, such as the proximity
to magnetism and the presence of on-site Hubbard correlations, controversies
persist, as both weak coupling and strong coupling approaches predict
the correct nature of the superconductivity. 

\emph{Setting the Targets \& Framing the Questions} -- Could materials
design help us verify or falsify theories of the high-temperature
cuprate superconductors? Or at least narrow down the set of competing
theories and sharpen our theoretical understanding while suggesting
new compounds in this important class of materials?

We chose the $T$-type layered perovskite La$_{2}$CuO$_{4}$ as the
starting point. Our intuition led us to propose the site substitution
of the apical oxygen with sulfur. Due to the larger ionic radius of
sulfur as compared to oxygen we expect the LaS charge reservoir layer
to be crowded. To compensate, we explored the effect of substituting
the large La ion with smaller trivalent ions $R$, selected from the
lanthanide-like elements. The compositions we considered were $R_{2}$CuS$_{2}$O$_{2}$
(shown in Fig.~\ref{fig:rcuso}) and $R_{2}$CuSO$_{3}$. We include
the monosulfide in hopes that the configurational entropy of only
replacing a quarter of the apical oxygens with sulfur would help stabilize
the target phase.

\subsubsection*{Structure to Property }

Cluster DMFT can describe the competition and synergy between antiferromagnetism
and $d$-wave superconductivity~\cite{Capone2006}.

\begin{figure}
\includegraphics[scale=0.5]{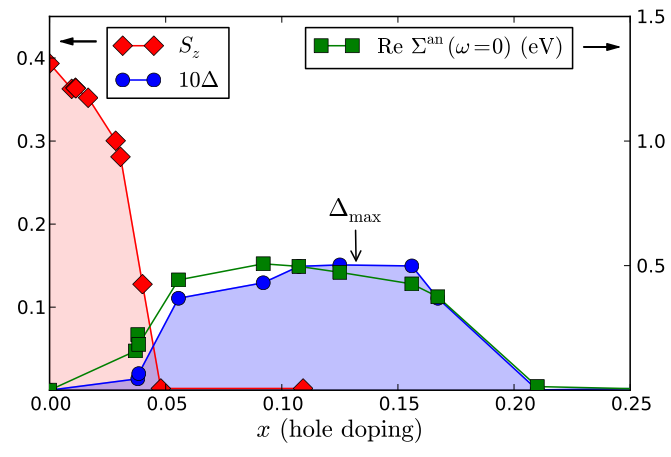}

\caption{Cluster DMFT investigation of a model of La$_{2-x}$Sr$_{x}$CuO$_{4}$
(Weber \textit{et al.} \cite{Weber2012}). Zero temperature phase
diagram connects structure to physical properties (superconductivity
and antiferromagnetism). }

\label{fig:weber2} 
\end{figure}

Weber \textit{et al.} \cite{Weber2012} used Cluster LDA+DMFT with
an exact diagonalization to bridge between the structure of La$_{2-x}$Sr$_{x}$CuO$_{4}$
(LSCO) and the physical properties of magnetism and superconductivity.
As shown in Fig. \ref{fig:weber2}, the method describes well the
zero temperature phase diagram of this compound, including the antiferromagnetic
phase (characterized by the magnetic moment $S_{z}$) and the $d$-wave
superconductor characterized by the anomalous self energy and the
superconducting gap $\Delta$. 

\begin{figure*}
\includegraphics[scale=0.85]{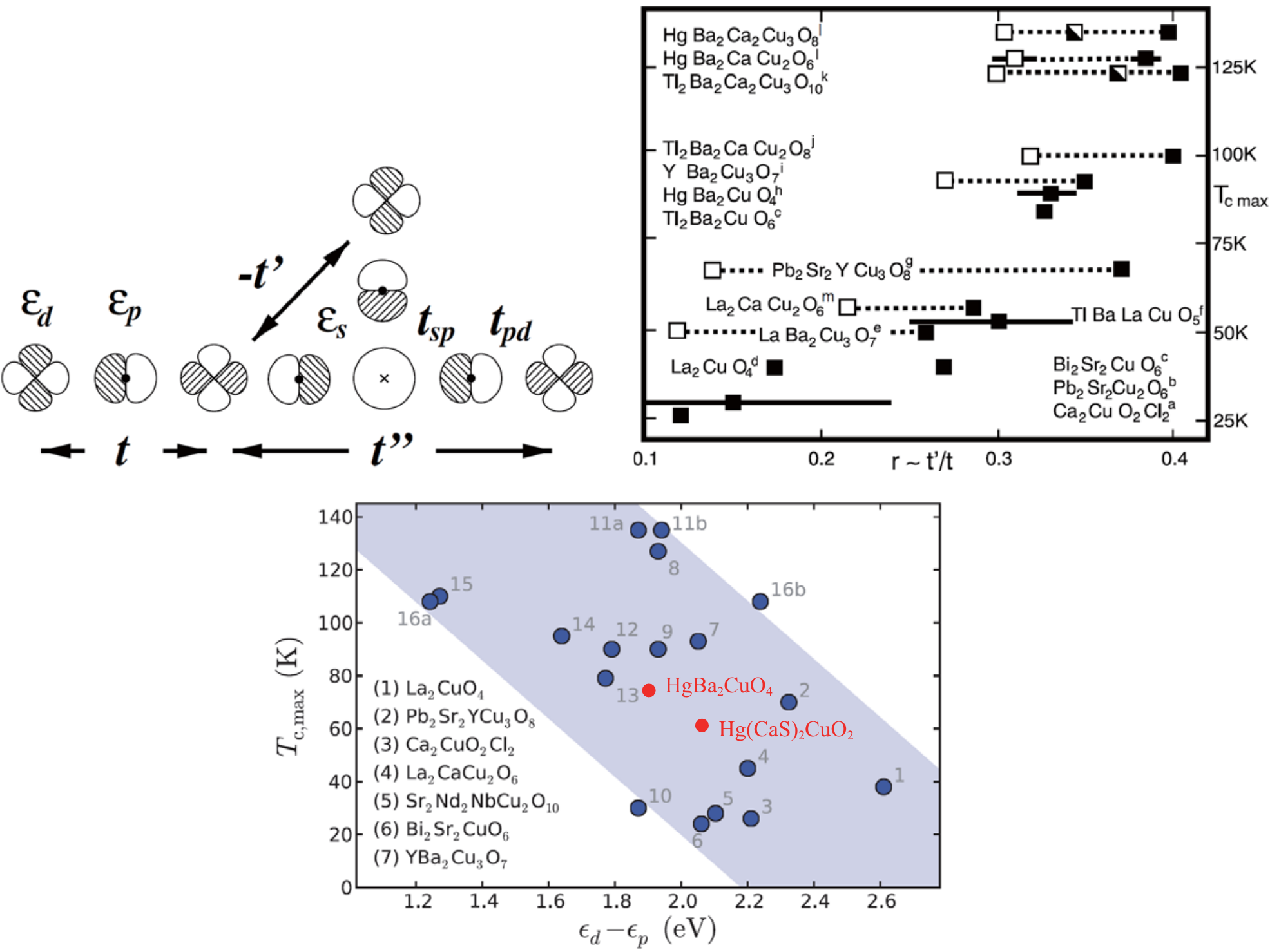}

\caption{The empirical correlation of $T_{\text{c,max}}$ with the apical oxygen
can be due to variation of many parameters in the low energy Hamiltonian
describing different processes (top left), $t'/t$ as originally suggested
by Pavarini \textit{et al.} \cite{Mishonov2004}(top right), as well
as the charge transfer gap as suggested by Weber \textit{et al.} \cite{Weber2012}
(bottom) . We have updated the bottom figure to include Hg-based cuprates.}
\label{fig:weber1} 
\end{figure*}

\begin{figure*}
\includegraphics[scale=0.5]{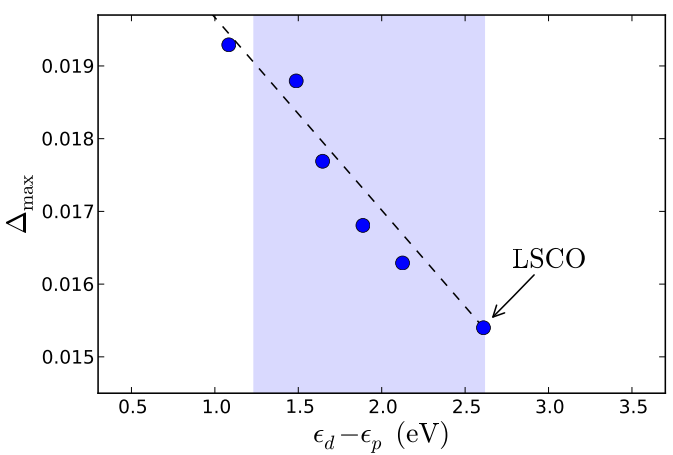}\includegraphics[scale=0.5]{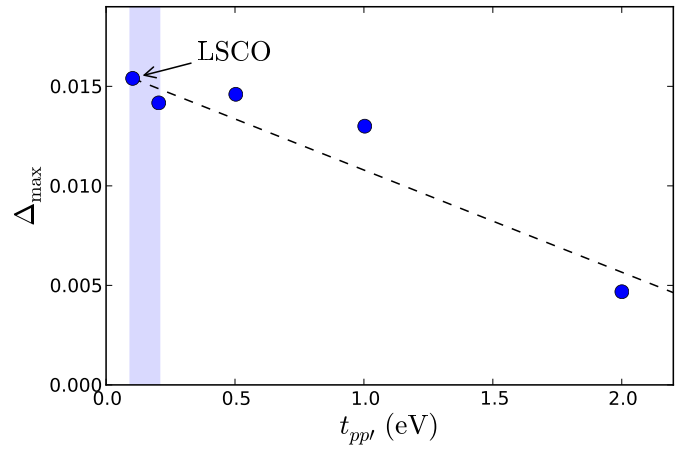}\caption{Cluster DMFT investigation of a model of the copper oxide planes (Weber
\textit{et al.} \cite{Weber2012}). After showing that the model displays
the correct phase diagram (top), starting with La$_{2-x}$Sr$_{x}$CuO$_{4}$
(LSCO) one varies independently the parameters that control the charge
transfer energy $\epsilon_{p}-\epsilon_{d}$ (bottom left) and the
oxygen oxygen overlap $t_{pp'}$ (bottom right). The figure displays
the dependence of the maximum of the superconducting order parameter
(which serves as a proxy for $T_{c,max}$). We see that the maximum
superconducting gap $\Delta_{max}$ is very sensitive to the charge
transfer gap, which thus controls (in this strong coupling theory)
the critical temperature and not with $t'$ or $t'_{pp}$. }

\label{fig:variation} 
\end{figure*}

Multiple studies concluded that $T_{\text{c,max}}$ is an increasing
function of the apical oxygen distance \cite{Ohta1991,Mishonov2004,Weber2012,Raimondi1996,Sakakibara2010}.
Changing the apical oxygen position, however, affects many different
parameters of the electronic structure of the copper oxygen planes,
ranging from the hoppings to the charge transfer energy and to the
relative positions of the $d_{x^{2}-y^{2}}$ and $d_{z^{2}}$ orbitals.
For a working design process, we need to elucidate the connection
between these parameters and $T_{\text{c,max}}$. 

To \textcolor{black}{investigate} the link between $T_{\text{c,max}}$
and the various parameters describing the electronic structure, one
needs to vary them independently, which is easy to do in-silico, based
on a combination of first principles calculations and dynamical mean-field
theory to directly model the superconducting state. Specifically,
La$_{2-x}$Sr$_{x}$CuO$_{4}$ has the largest charge transfer energy
and a small $T_{\text{c,max}}$, we vary the charge transfer energy
and the hoppings independently and the results are shown in Fig.~\ref{fig:variation}.
It is clear that (within LDA+DMFT) the superconductivity increases
rapidly when we \textcolor{black}{decrease} the charge transfer gap,
while it is very weakly sensitive (and in fact decreases) when we
increase the oxygen oxygen overlap and the hopping integral $t^{\prime}$
\textcolor{black}{(depicted in Fig.}\textcolor{red}{~\ref{fig:weber1}}\textcolor{black}{)}
on the square lattice (not shown).

Inspired by these results and the earlier heuristic considerations
of connecting structure to property, we aimed to design new cuprates
with reduced charge transfer gaps (and thus higher $T_{c}$) via sulfur
(S) substitution. To check that S substitution helps control the charge
transfer gap, and to see which other variables it affects, we used
the maximally localized Wannier function{\footnotesize{}~}\cite{Marzari97,Souza01}
implemented in the WANNIER90 code{\footnotesize{}~}\cite{Arash08}.

\begin{figure}
\includegraphics[width=1\columnwidth]{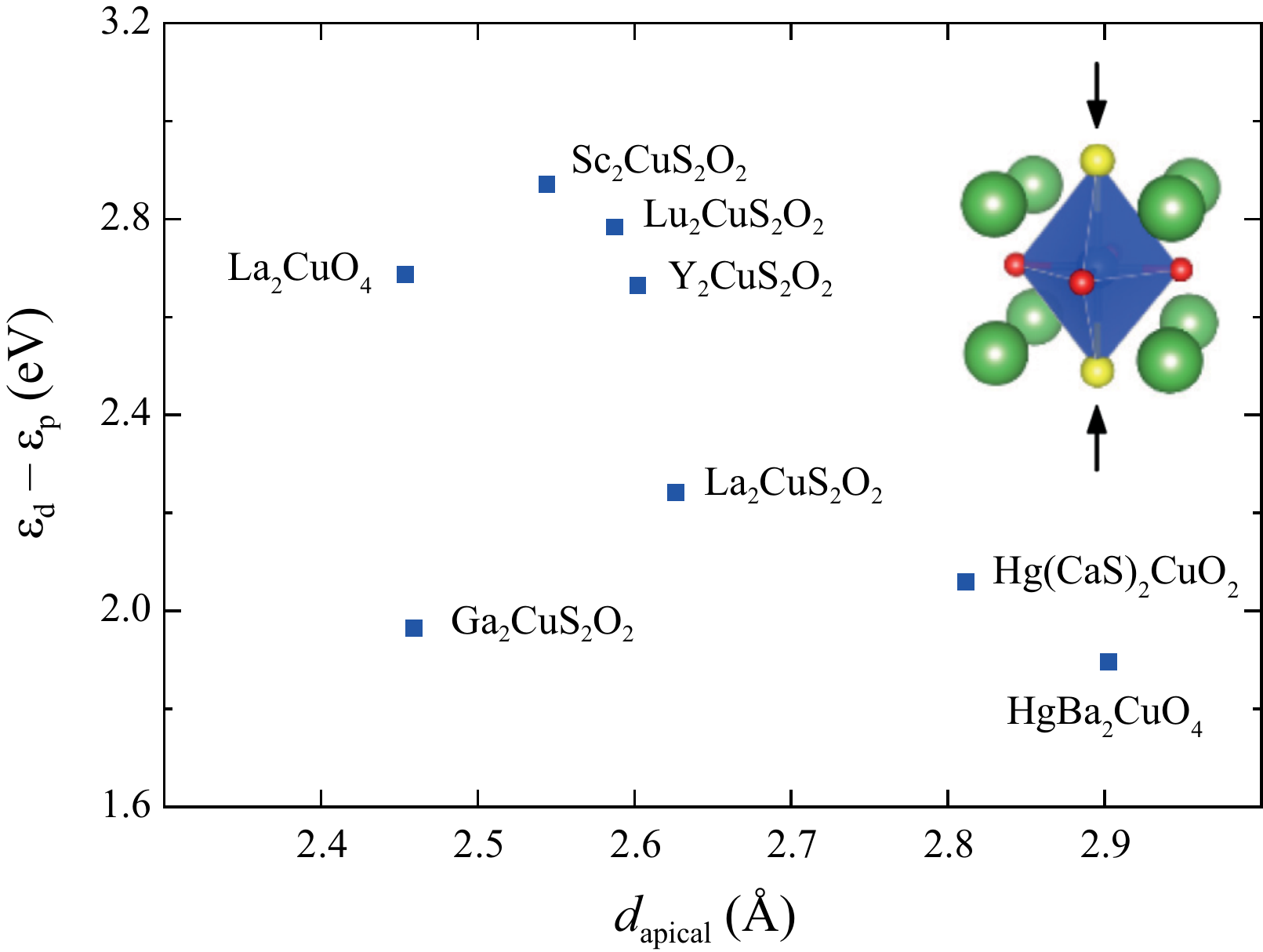} \caption{Explored compositions, generated by substituting the apical oxygens
(arrows) and rare earth ion (green spheres) in La$_{2}$CuO$_{4}$
to form the family of compounds $R_{2}$CuS$_{2}$O$_{2}$. Two mercury
compounds HgBa$_{2}$CuO$_{4}$ and Hg(CaS)$_{2}$CuO$_{2}$ are also
added for comparison. Plotted are the extracted charge-transfer energies
vs. the apical distance of sulfur atom in the proposed compounds after
structural relaxation in LDA. Note that charge-transfer energies in
known cuprates superconductors are observed in the range from $\sim1.2$
to $\sim2.6$ eV. Adapted from Ref. \onlinecite{Yee2014}.}
\label{fig:rcuso} 
\end{figure}

The results for the GGA estimates of the charge transfer energies
are displayed in Fig.~\ref{fig:rcuso} and the second column of Table~\ref{tbl:parameters}
together with the reference system La$_{2}$CuO$_{4}$. Both figure
and table suggest that Sc$_{2}$CuS$_{2}$O$_{2}$ having the highest
charge transfer energy would have the smallest $T_{\text{c,max}}$
among proposed cuprates $R_{2}$CuS$_{2}$O$_{2}$ and La$_{2}$CuS$_{2}$O$_{2}$
having smaller charge transfer energy than La$_{2}$CuO$_{4}$ would
have higher $T_{\text{c,max}}$ than that of La$_{2}$CuO$_{4}$ \textcolor{black}{-
based on the charge-transfer theory in} Ref. \onlinecite{Weber2012}.

\begin{figure}
\includegraphics[width=1\columnwidth]{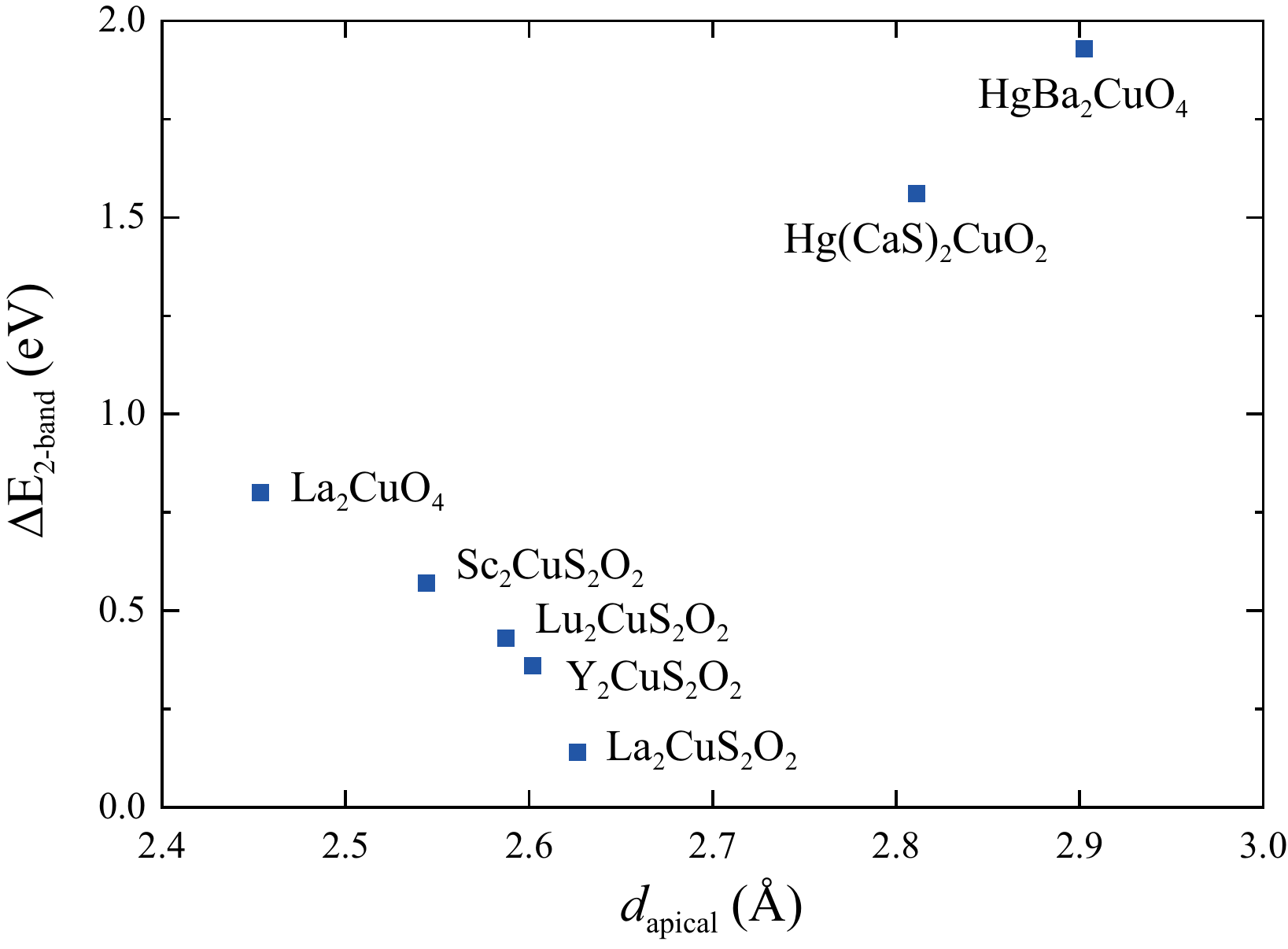} \caption{The crystal field splitting $\triangle E_{2-band}=\epsilon_{d(x^{2}-y^{2})}-\epsilon_{dz^{2}}$
computed by downfolding into the effective two-band model. La$_{2}$CuO$_{4}$
and the copper oxysulfide family are shwon with two mercury compounds
HgBa$_{2}$CuO$_{4}$ and Hg(CaS)$_{2}$CuO$_{2}$ for comparison.
$\triangle E_{2-band}$ in the oxysulfides is generally smaller, implying
stronger mixing between $d_{z^{2}}$ and $d_{x^{2}-y^{2}}$ orbitals.
Thus sulfur substitution would suppress $T_{\text{c,max}}$ according
to the two-band theory~\cite{Sakakibara2010}, while $T_{\text{c,max}}$
would be enhanced according to the charge-transfer theory~\cite{Weber2012}.
Adapted from Ref. \onlinecite{Yee2014}.}
\label{fig:crystal_field_splitting} 
\end{figure}

\textcolor{black}{Sakakibara }\textit{\textcolor{black}{et al.}}\textcolor{black}{{}
\cite{Sakakibara2010} observed a positive correlation between $T_{\text{c,max}}$
and the crystal field splitting $\Delta E=\epsilon_{d(x^{2}-y^{2})}-\epsilon_{dz^{2}}$
based on a two-band model incorporating both $d_{x^{2}-y^{2}}$ and
$d_{z^{2}}$ orbitals. They found that large crystal field splitting
$\Delta E$, which is also related to the reduction of $d_{z^{2}}$
contribution to the Fermi surface, enhances $T_{\text{c,max}}$. In
order to test their theory, we examine $\Delta E$ in Fig.~\ref{fig:crystal_field_splitting}
and the last column of Table~\ref{tbl:parameters}. The crystal field
splitting $\Delta E$ in the oxysulfides family $R_{2}$CuS$_{2}$O$_{2}$
is generally smaller than that in La$_{2}$CuO$_{4}$, implying that
sulfur substitution would suppress $T_{\text{c,max}}$. La$_{2}$CuS$_{2}$O$_{2}$
having the smallest crystal field splitting $\Delta E$ among the
oxysulfides family, for example, has the smallest $T_{\text{c,max}}$
based on the theory, while the charge-transfer theory gives the opposite
result. Therefore a detailed study of La$_{2}$CuS$_{2}$O$_{2}$
is necessary to test the validity of these two theories.}

\begin{figure}
\includegraphics[scale=0.4]{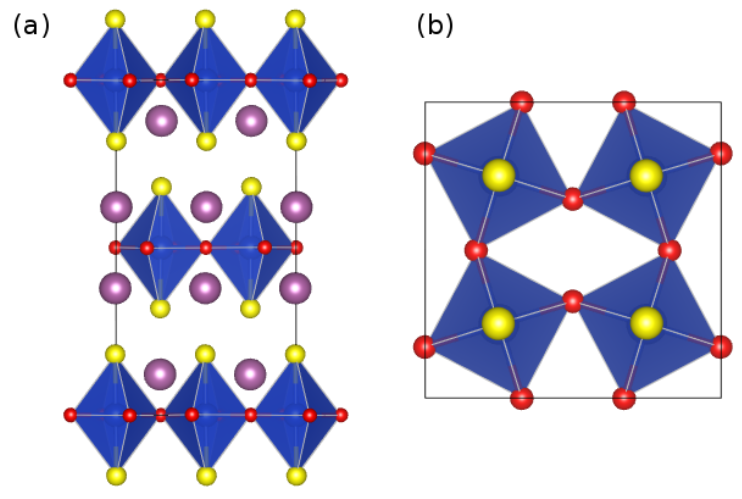}\caption{Octahedral rotations in Sc$_{2}$CuS$_{2}$O$_{2}$ as shown by a
section of the CuO$_{2}$ plane with four atoms. Adapted from Ref.
\onlinecite{Yee2014}.}

\label{fig:cuprates-distortion} 
\end{figure}

\emph{Structure Prediction }-- To check for local stability we used
a $2\times2\times1$ unit cell, we performed full structural relaxation
to check if the structure would be unstable towards distortion to
the $T'$-type layered perovskite, knowing that substitution of the
large La ion for the smaller Pr and Nd led to a rearrangement of the
charge reservoir layer into the fluorite structure. We found that
the $T$-type structure was indeed stable and there was no out-of-plane
buckling, although the CuO$_{6}$ octahedra favored axial rotations
($a^{0}a^{0}c_{p}^{-}$ in Glazer notation). See Fig.~\ref{fig:cuprates-distortion}.

\emph{Global stability} -- We checked the thermodynamic stability
of the proposed compounds against competing phases by selecting commonly
known reactants and computing the formation enthalpies of the synthesis
pathways as shown in Table~\ref{tbl:pathways}. We computed the total
energies of formation $\Delta E=E_{\text{products}}-E_{\text{reactants}}$,
and find that all differentials are positive, indicating the reactions
target phases are unfavorable. However, it is known that many functional
materials are metastable, protected from decay by large energetic
barriers. The parent cuprate La$_{2}$CuO$_{4}$ is an example: as
shown on the last line of Table~\ref{tbl:pathways}, La$_{2}$CuO$_{4}$
is actually unstable by 65 meV/atom. We also examined the volume differentials
$\Delta V=V_{\text{products}}-V_{\text{reactants}}$ with the knowledge
that often high pressure synthesis allows otherwise unstable compounds
to form. Notice that $\Delta V$ are overwhelming negative, meaning
the application of high pressure may allow the formation of the target
phases.

\begin{table}
\begin{tabular}{r|r|cl}
\hline 
$\Delta E$  & $\Delta V$  & \multicolumn{2}{c}{Synthesis pathway}\tabularnewline
\hline 
\hline 
232  & -1.92  & La$_{2}$SO$_{2}$ + CuS  & $\rightarrow$ La$_{2}$CuS$_{2}$O$_{2}$\tabularnewline
362  & -1.45  & Y$_{2}$SO$_{2}$ + CuS  & $\rightarrow$ Y$_{2}$CuS$_{2}$O$_{2}$\tabularnewline
415  & -1.36  & Lu$_{2}$SO$_{2}$ + CuS  & $\rightarrow$ Lu$_{2}$CuS$_{2}$O$_{2}$\tabularnewline
542  & -1.04  & Sc$_{2}$SO$_{2}$ + CuS  & $\rightarrow$ Sc$_{2}$CuS$_{2}$O$_{2}$\tabularnewline
\hline 
304  & -2.00  & La$_{2}$O$_{3}$ + CuS  & $\rightarrow$ La$_{2}$CuSO$_{3}$ \tabularnewline
780  & -1.11  & Sc$_{2}$O$_{3}$ + CuS  & $\rightarrow$ Sc$_{2}$CuSO$_{3}$ \tabularnewline
217  & -1.28  & La$_{2}$SO$_{2}$ + CuO  & $\rightarrow$ La$_{2}$CuSO$_{3}$ \tabularnewline
519  & -0.54  & Sc$_{2}$SO$_{2}$ + CuO  & $\rightarrow$ Sc$_{2}$CuSO$_{3}$\tabularnewline
\hline 
65  & -2.71  & LaCuO$_{2}$ + 2 La$_{2}$O$_{3}$ + La(CuO$_{2}$)$_{2}$ & $\rightarrow$ 3 La$_{2}$CuO$_{4}$ \tabularnewline
\hline 
\end{tabular}\vspace{3bp}
\caption{Synthesis pathways for various cuprate oxysulfides based on substitution
of sulfur for both (top block) or only one (middle block) of the apical
oxygens in $R_{2}$CuO$_{4}$. Energies in meV/atom and volumes in
$\AA^{3}$/atom. Since the energies of formation ($\Delta E=E_{\text{products}}-E_{\text{reactants}}$)
are positive, none of these pathways appear favorable at ambient conditions.
However, high-pressure synthesis will help stabilize these pathways,
since the majority of volume differentials ($\Delta V=V_{\text{products}}-V_{\text{reactants}}$)
are negative. We benchmark our method against the standard synthesis
pathway for La$_{2}$CuO$_{4}$, shown on the last line. Surprisingly,
$\Delta E$ is +65 meV/atom, so either DFT systemmatically overestimates
enthalpies (which means the actual enthalpies for our hypothetical
compounds are \emph{smaller}, in our favor), or we must add a bi-directional
uncertainty of $\pm70$~meV/atom to the computed enthalpies. Additionally,
positional entropy of the apical $S$ in the half-substituted $R_{2}$CuSO$_{3}$
compounds should also assist in synthesis.}
\label{tbl:pathways} 
\end{table}

\begin{figure}
\includegraphics[width=1\columnwidth]{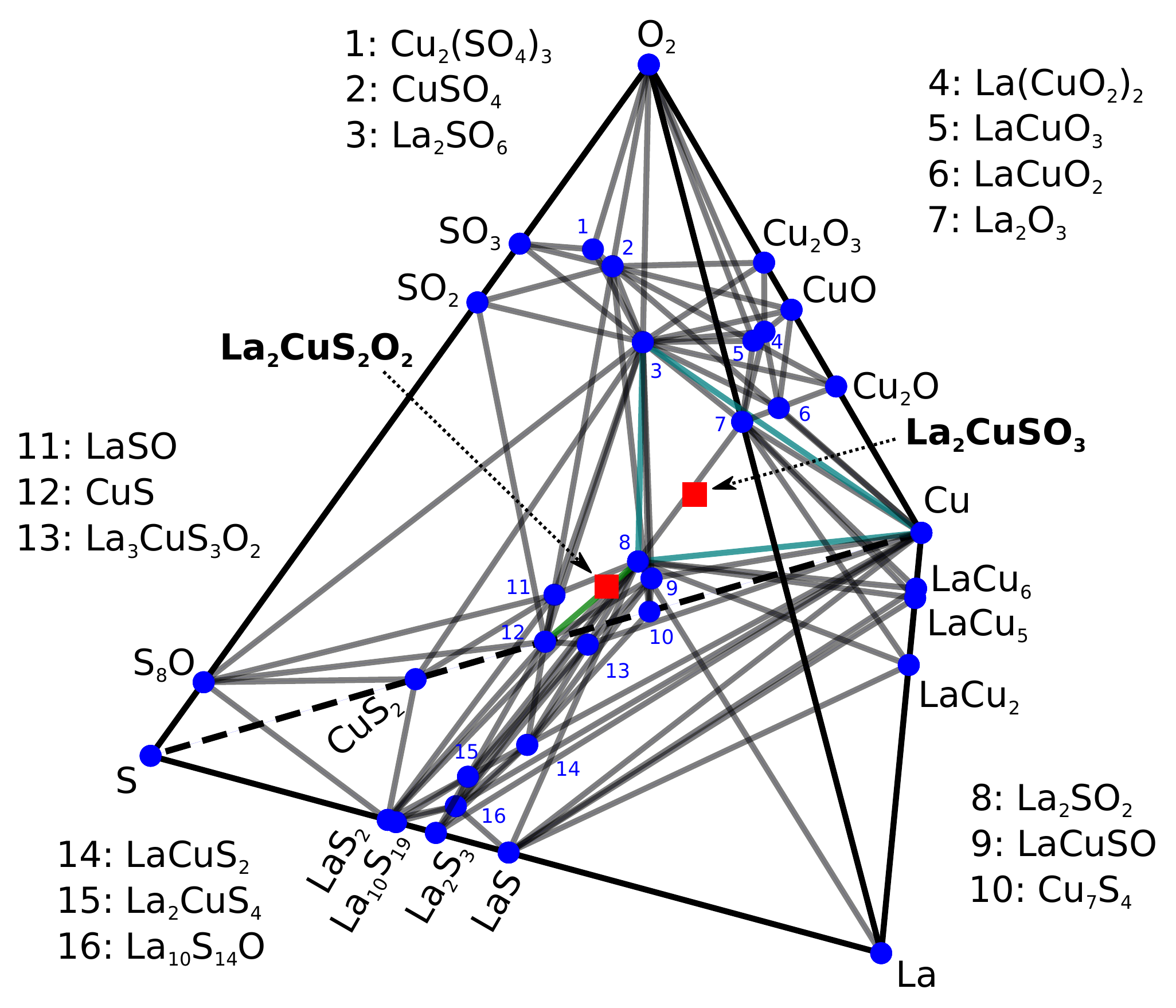} \caption{Gibbs phase diagram of La-Cu-S-O system. The two proposed compositions
are marked by red squares. They are both found to be unstable: La$_{2}$CuS$_{2}$O$_{2}$
lies on the line between CuS (12) and La$_{2}$SO$_{2}$ (8), and
La$_{2}$CuSO$_{3}$ lies on the triangular facet spanned by Cu, La$_{2}$SO$_{2}$
(8), and La$_{2}$SO$_{6}$ (6). In experiment, we find that the quaternary
phase LaCuSO (9) is preferred, likely because copper prefers the $1+$
oxidation state.}
\label{fig:lcso-hull} 
\end{figure}

To look at the question of global stability of La$_{2}$CuS$_{2}$O$_{2}$
and La$_{2}$CuSO$_{3}$, we use the modern materials databases, and
reanalyze the entire La-Cu-S-O system to construct the convex hull
(plotted in Fig.~\ref{fig:lcso-hull}) and globally investigate stability. 

\begin{figure}
\includegraphics[width=1\columnwidth]{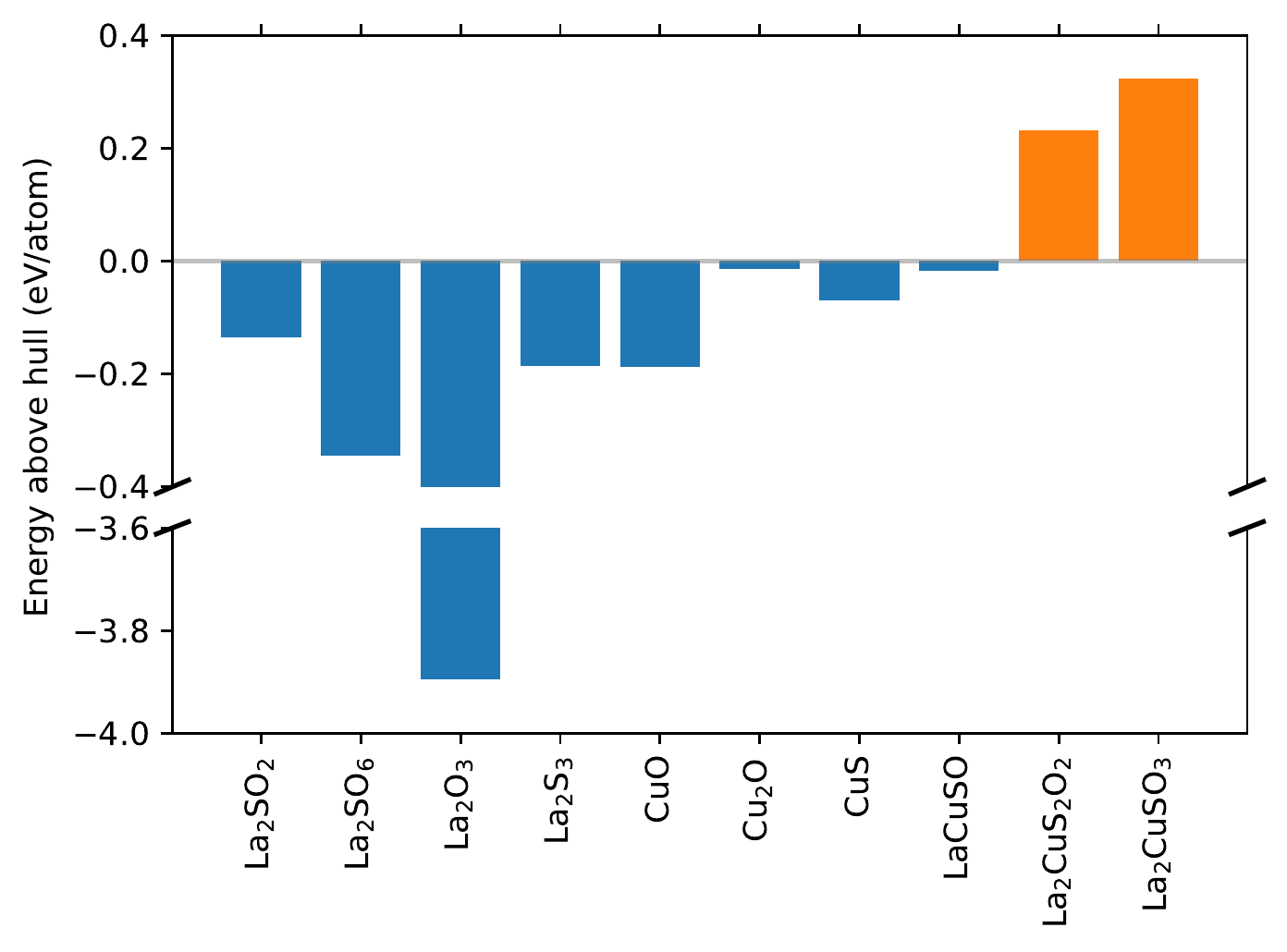} \caption{Energies relative to the convex hull for reactants and products observed
in experiment in the La-Cu-S-O system. Negative energies indicate
stable compounds, while unstable compounds have positive energies.
Both La$_{2}$CuSO$_{3}$ and La$_{2}$CuS$_{2}$O$_{2}$ are highly
unstable, lying over 200~meV/atom above the hull. The vertical axis
is broken to display the large stability energy of La$_{2}$O$_{3}$.}
\label{fig:lcso-ehull} 
\end{figure}

Using the convex hull, we can assess the stability of the reactants
and products reported in experiment. In Fig.~\ref{fig:lcso-ehull},
we plot the energies relative to the convex hull for all reported
compounds. Negative values are stability energies against decomposition.
We find that La$_{2}$CuS$_{2}$O$_{2}$ and La$_{2}$CuSO$_{3}$
are unstable at 232 and 324~meV/atom above the hull, respectively,
which is slightly more unstable than the earlier estimates which only
tested a few reactions, and we gain additional information as we learn
these compounds in equilibrium would decompose into: 
\begin{align*}
\text{La}_{2}\text{CuS}_{2}\text{O}_{2} & \rightarrow\text{La}_{2}\text{SO}_{2}+\text{CuS}\\
4\text{La}_{2}\text{CuS}\text{O}_{3} & \rightarrow3\text{La}_{2}\text{SO}_{2}+4\text{Cu}+\text{La}_{2}\text{SO}_{6}
\end{align*}

Additionally, LaCuSO lies 17 meV/atom below the hull, so it is likely
to be stable according to the analysis of section \ref{sec:Probability-Estimation}.

\textcolor{black}{Following our proposal in section \ref{sec:Probability-Estimation},
we obtain sharper probability estimates using a correlated method
- GGA(PBE) + Gutzwiller. As outlined, we only need to examine the
energies of the determinant reaction. The on-site Coulomb interaction
$U$ = 8 eV and Hund's coupling constant $J$ = 0.88 eV are used in
the Gutzwiller calculations~\cite{Anisimov04_cuprates}. We find
that La$_{2}$CuS$_{2}$O$_{2}$ and La$_{2}$CuSO$_{3}$ are still
unstable at 214 and 286 meV/atom above the hull, respectively - consistent
with a recent experiment~\cite{Hua_2016}.}

\emph{Conclusion} -- Experimental support for the idea that the reduction
of the charge transfer gap results in an increase in $T_{\text{c}}$
were recently provided by STM studies which were used to compare the
charge transfer gaps of Ca$_{n+1}$Cu$_{n}$O$_{2n}$Cl$_{2}$ and
Bi$_{2}$Sr$_{2}$Ca$_{n-1}$Cu$_{n}$O$_{2n+4}$~\cite{Ruan_2016},
hence the ideas proposed in Ref.~\onlinecite{Weber2012} are definitely
worth pursuing. It is not known however, where exactly the maximum
of $T_{\text{c}}$ is obtained, as it is clear that a very small charge
transfer energy would result in an uncorrelated material with very
low $T_{\text{c}}$. So, while this problem remains open, we can already
say that the work of Ref.~\onlinecite{Weber2012} has served to provide
context for later experimental work. 

\begin{table*}
\begin{tabular}{|c|c|c|c|c|c|}
\hline 
 & $\Delta_{pd}$ (eV) & $\varDelta E$ (eV) & $E_{g}$ (eV) & $S_{z}$ ($\mu_{B}$/Cu) & $\varDelta E_{2-band}$ (eV)\tabularnewline
\hline 
\hline 
La$_{2}$CuO$_{4}$  & 2.69 & 0.03 & 2.01 & 0.71 & 0.80\tabularnewline
\hline 
La$_{2}$CuS$_{2}$O$_{2}$  & 2.24 & 0.09 & 1.47 & 0.72 & 0.14\tabularnewline
\hline 
Sc$_{2}$CuS$_{2}$O$_{2}$  & 2.87 & 0.35 & 0.99 & 0.70 & 0.57\tabularnewline
\hline 
Lu$_{2}$CuS$_{2}$O$_{2}$ & 2.78 & 0.23 & 1.67 & 0.71 & 0.43\tabularnewline
\hline 
Y$_{2}$CuS$_{2}$O$_{2}$ & 2.66 & 0.19 & 1.48 & 0.71 & 0.36\tabularnewline
\hline 
HgBa$_{2}$CuO$_{4}$  & 1.90 & 0.29 & 1.12 & 0.68 & 1.93\tabularnewline
\hline 
Hg(CaS)$_{2}$CuO$_{2}$  & 2.06 & 0.24 & 1.73 & 0.69 & 1.56\tabularnewline
\hline 
\end{tabular}\vspace{3bp}
\caption{\textcolor{black}{Physical parameters for cuprate materials. The charge
transfer energy $\Delta_{pd}=\epsilon_{d}-\epsilon_{p}$ and crystal
field splitting $\varDelta E=\epsilon_{d(x^{2}-y^{2})}-\epsilon_{dz^{2}}$
are obtained from the Wannier method applied to GGA(PBE) nonmagnetic
calculations. All of the Cu $d$, O $p$, and S $p$ orbitals are
considered for the Wannier method. The band gap $E_{g}$ and magnetic
moment $S_{z}$ (per Cu atom) are caluclated within GGA(PBE) + U antiferromagnetic
calculations. The on-site Coulomb interaction $U=8$ eV and Hund's
coupling constant $J=0.88$ eV are used in the GGA+U calculations~\cite{Anisimov04_cuprates}.
The last column of the table describes the results for the crystal
field splitting downfolding to a two band model describing the $d_{x^{2}-y^{2}}$
and the $d_{z^{2}}$ in order to test the theory of Ref.~\onlinecite{Sakakibara2010}.}}
\label{tbl:parameters} 
\end{table*}

Synthesis performed at synchrotron light sources now allows for \textit{in
situ} tracking of intermediate products and provides a deep understanding
of the chemical reaction pathways by which a compound forms at different
temperatures~\cite{Molinder12,Kanatzidis14}. The group of M. Aronson
used this technique to study the system described in this section~\cite{Hua_2016}
and investigated the potential synthesis of La$_{2}$CuS$_{2}$O$_{2}$
and La$_{2}$CuSO$_{3}$. Their work confirmed that the two compounds
were unstable (at least at high temperatures). From the convex hull
(Fig.~\ref{fig:lcso-hull}), the target compounds are thermodynamically
unstable with respect to CuS, La$_{2}$SO$_{2}$, Cu, and La$_{2}$SO$_{6}$
and the phase with composition LaCuSO seemed to be quite stable at
high temperatures. In agreement with theory, the end products La$_{2}$SO$_{2}$
and La$_{2}$SO$_{6}$ were observed, in addition to LaCuSO which
was the preferred quaternary composition in almost all the experimentally
analyzed reactions. This experimental study thus validates in detail
the current material design workflow and its probabilistic interpretation. 

Sulfate apical substitution is not easily realized but the work of
Ref.~\onlinecite{Weber2012} has served to provide context for later
experimental work~\cite{Hua_2016}, which highlight the difficulties
in forming the desired oxysulfides. The more flexible valence of sulfur
which can now reduce copper from its starting 2+ oxidation state to
bring it to its 1+ state, and thus forms LaCuSO instead of our target
compounds, which like La$_{2}$CuO$_{4}$, requires a 2+ copper state~\cite{Hua_2016}.
With this insight we can now go back to the first step of the design
loop to search for better routes to reduce the charge transfer gap
while keeping the valence of Cu close to 2+ or turn to other routes
to vary the charge transfer energy as described in the next section.

{\def\bibliography 
\bibliography{cormatdes,gabi}

}

\subsection{Hg-based Cuprates}
\label{sec:cuprates2}

With the experience of the previous section
we revisit the question of how to design novel cuprates following Ref.~\onlinecite{Yee15}.
The challenge in the design process is finding
a new chemical composition that forms in the desired cuprate structure - with  slightly different parameters, thus
enabling the elucidation of the mechanism of superconductivity.
We focus on the charge transfer energy~\cite{Weber2012}.  We also consider
the $d_{z^2}$ admixture which is the key variable in the orbital distillation theory~\cite{Sakakibara2010}.  This theory
is supported by weak coupling calculations that show that  a reduction in the
content of  $d_{z^2}$  (which correlates with an increase of the crystal field splitting $\Delta E$) increases the superconducting critical temperature.

Design of novel cuprates is challenging because the phase space is
well-explored: most simple point substitutions likely have been attempted. On
the other hand, an exhaustive search consisting of structure prediction over
all possible compositions containing copper would be prohibitively costly.
Rather than designing novel cuprates from scratch, we took an intermediate
route: begin with the family with the highest known transition temperatures,
the Hg-based cuprates, and modulate its spacer layers~\cite{Yee15}.

\begin{figure*}
  \includegraphics[width=15cm]{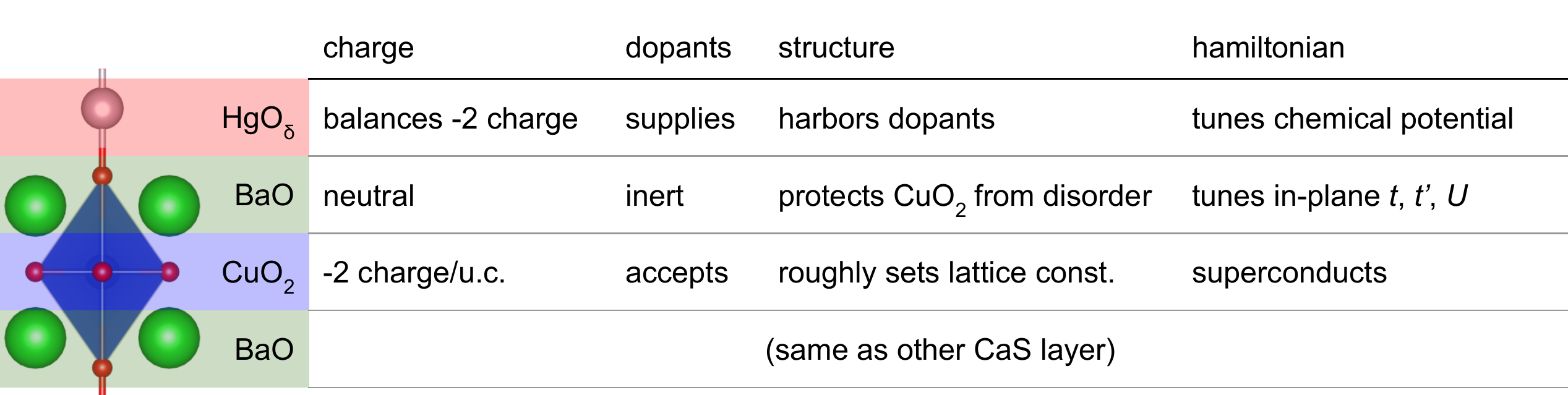}
  \caption{The cuprates are a heterostructure of functional layers, each
    performing a specific structural and electronic role in tuning the
    superconducting hamiltonian. Here, we show the example of
    HgBa$_2$CuO$_{4+\delta}$, the single layer cuprate with the highest
    transition temperature. In our work, we focus on tuning the chemistry of
    the layers immediately adjacent to the CuO$_2$ plane, as these will most
    strongly affect the in-plane Hamiltonian. Adapted from Ref. \onlinecite{Yee15}.}
  \label{fig:stack}
\end{figure*}

\emph{Setting the Targets, Framing the Questions,  Heuristic Considerations} --
Taking the Hg-based cuprates as an example
(Fig.~\ref{fig:stack}), we view the cuprates as a stack of functional layers,
with the composition of each layer chosen to play a specific role. The central
copper oxide (CuO$_2$) plane supports superconductivity and roughly constrains
the in-plane lattice constant. The remaining layers must tune the chemical
potential of the CuO$_2$ layer without rumpling the plane or introducing
disorder, and isolate each CuO$_2$ plane to create a 2D system.

Our goal is to tune the in-plane  charge transfer energy  (effective $U$),  so the relevant layers to focus
on are the BaO layers immediately adjacent to the CuO$_2$ plane. Due to their
spatial proximity, the BaO layers tune the hoppings and interaction strengths
of the in-plane Hamiltonian.  We also pay attention to the energy of the $d_{z^2}$
orbital, as it plays a key role in the orbital distillation theory~\cite{Sakakibara2010}.

Designing compounds with novel adjacent layers
provides a mechanism for controlling superconductivity. However,
cycling through all roughly $100 \times 100$ elemental substitutions for BaO in
the periodic table using structural prediction is clearly too naive and
computationally expensive.
To select plausible compositions, we noted that the BaO layers form a rock salt
structure. Using materials databases, we selected all naturally occurring rock
salt compounds AX, composed of a cation A and an anion X, starting with 333 in
total. We then quickly pre-screened candidates by discarding compositions with
(1) large lattice mismatches relative to the in-plane Cu-Cu distance, which we
took to be 3.82~\AA, and (2) anions less electronegative than Cu, as these
anions would capture dopants intended for the superconducting plane, producing
additional Fermi surfaces.

\emph{Electronic structure} --  To evaluate the prospects of
superconductivity, we follow the previous section, and  we focus on the charge transfer gap.  This is summarized in  additional entries in Table~\ref{tbl:parameters}.
We find the
charge-transfer energy of Hg(CaS)$_2$CuO$_2$ (HCSCO) to be 2.06 eV.
It is slightly larger than other Hg-based cuprate, HgBa$_{2}$CuO$_{4}$ (HBCO),
however is smaller than La$_{2}$CuO$_{4}$ and the proposed cuprates $R_2$CuO$_{2}$S$_{2}$ in Table~\ref{tbl:parameters}.
Hence we expect $T_{\text{c,max}}$ of HCSCO is smaller than that of HBCO,
but is larger than those of La$_{2}$CuO$_{4}$ and $R_2$CuS$_{2}$O$_{2}$
(see the bottom of Fig.~\ref{fig:weber1}).

\begin{figure}
\includegraphics[width=\columnwidth]{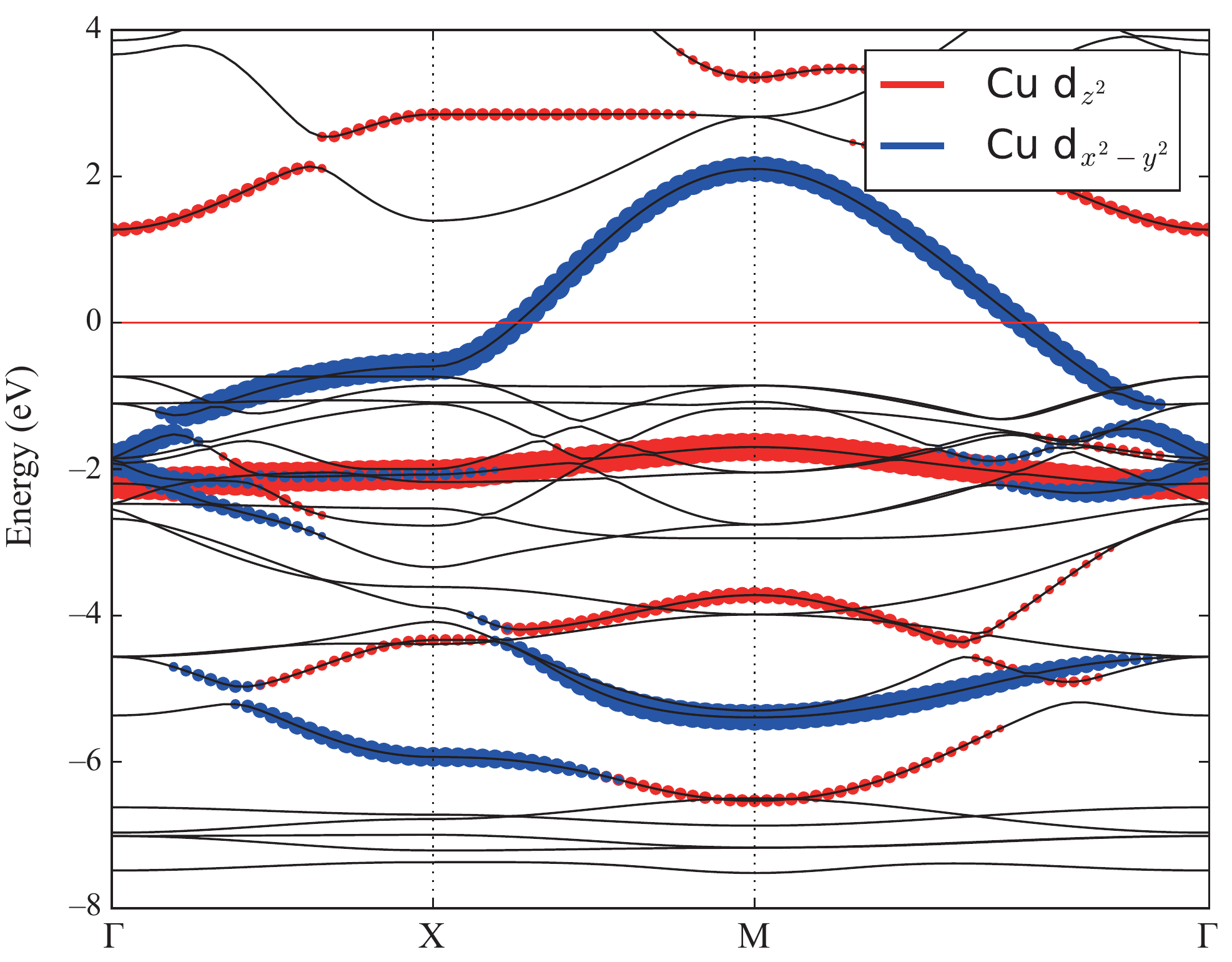}
\caption{Computed band structure of the proposed compound Hg(CaS)$_2$CuO$_2$
using GGA(PBE) nonmagnetic calculation.
Cu $3d_{z^2}$ and $3d_{x^2-y^2}$ orbital characters
are weighted by red and blue filled-circles, respectively.
Similar to the cuprates, a single Cu $3d_{x^2-y^2}$ band
disperses across the Fermi level.}
\label{fig:hcsco-bands}
\end{figure}

It is also useful to plot the orbitally-resolved band structure
(Fig.~\ref{fig:hcsco-bands}). There is indeed a single band crossing the Fermi
level in HCSCO, similar to the other cuprates.

\emph{Structural Stability} --
We tested to check that the desired  structure was
stable  by first point-substituting the proposed elements into the
HgBa$_{2}$CuO$_{4}$ (HBCO) structure and checked for local stability via phonon calculations at the
$\Gamma$, $(\pi,0)$ and $(\pi,\pi)$ points. If the composition was stable in
the HBCO structure, then we used USPEX to perform the roughly week-long
calculations necessary to determine whether the HBCO structure is indeed the
preferred energetic minimum.

We found three BaO substitutions to be stable in the HBCO structure after
phonon screening: CaS, ZrAs, and YbS. USPEX found that only Hg(CaS)$_2$CuO$_2$
(HCSCO) was energetically stable in the layered cuprate structure.

\begin{figure}
  \includegraphics[width=0.9\columnwidth]{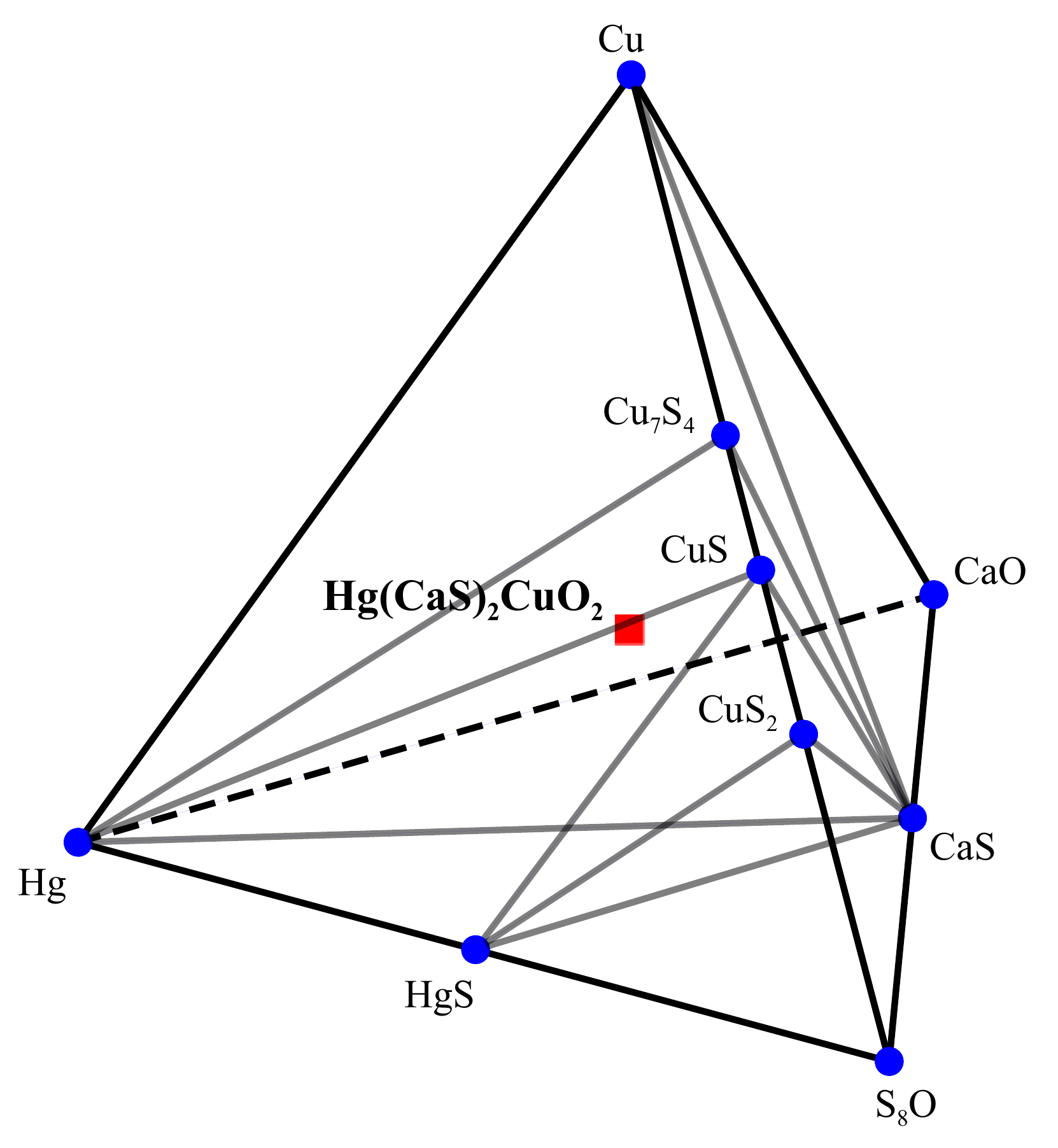}
  \caption{Convex hull for the Hg-Ca-Cu-S chemical system at an oxygen chemical
    potential of $\mu(\text{O}_2) = -16.48$ eV, chosen because HCSCO is least
    unstable at this value. The phase diagram forms a tetrahedron with S$_8$O,
    Hg, Cu and CaO at the vertices (elemental sulfur and calcium are not stable
    under this oxygen environment). HCSCO lies in the interior of the
    tetrahedron, on the triangular face formed by Hg, Cu and CaS.}
  \label{fig:hcsco-hull}
\end{figure}

\emph{Global stability} -- We construct the convex hull by computing the
energies of all known compounds in the Hg-Ca-Cu-S-O system, which would produce
a 4-dimensional tetrahedron. In reality, synthesis is performed in an oxygen
environment parameterized by the chemical potential $\mu(\text{O}_2)$. We find
that for all values of $\mu(\text{O}_2)$, HCSCO is unstable, so we pick the
value for which the compound is closest to the convex hull and plot the results
in Fig.~\ref{fig:hcsco-hull}.

Prior to examining the global stability of HCSCO, we benchmarked the
methodology on HgBa$_2$CuO$_4$ itself. We find that at fixed oxygen
stoichiometry, HBCO lies 74~meV/atom above the hull. Upon modeling the doped
compound using via a $3 \times 3$ supercell calculation (11\% doping), we find
HBCO to be even more unstable at 130~meV/atom above the hull.
As noted above,  La$_{2}$CuO$_{4}$, the metastable structure theoretically predicted to be
65~meV/atom above the hull does indeed exist.

We found that HCSCO lies 170~meV/atom above the hull (existence probability: 0.09) at
fixed oxygen stoichiometry, and 240~meV/atom above the hull (existence probability: 0.05)
for fixed oxygen chemical
potential~\cite{Ong2008},
which means the compound is likely unstable, but not out of the
realm of possibility for successful synthesis.

Revising the phase stability calculation - using a more accurate correlated  method, as suggested in section \ref{sec:Probability-Estimation},
we consider the determinant reaction. This reaction is found by GGA to be HgO + 2 CaS + CuO $\rightarrow$ Hg(CaS)$_{2}$CuO$_{2}$.
Using the GGA(PBE) + Gutzwiller method to calculate the energies of the reactants, we found that HCSCO is stable with 151 meV/atom below the hull
(at fixed oxygen stoichiometry).

\emph{Conclusions} --
This work in modulating the spacer layers in the Hg-based cuprates highlights
the challenges in optimizing properties in highly-explored materials classes.
In the prediction of structures for novel phases, chemical intuition is still
crucial for filtering possible candidates and focusing on the most promising
compositions. Structure prediction is the bottleneck step in the workflow,
consuming the largest fraction of the computational resources required.
Additionally, known materials can be metastable, and new work must explore what
factors select for which metastable compounds form in experiment.  It would therefore be
interesting to pursue the MBE route to approach  the synthesis of this type of compound.
Further searches which  results in similar structures while avoiding the use of toxic chemicals such
as Hg are  also very  important.

Theoretical calculations reproduced the early observation of Raychaudhury
\emph{et al.}~\cite{DeRaychaudhury2007}, that one can correlate $T_c$ with the oxygen-oxygen overlap and the next-nearest-neighbor hopping parameter of the Hubbard model. Other correlations were
pointed out - in particular that decreasing the charge transfer gap in these materials also increases $T_c$ \cite{Weber2012}. Finally, the orbital distillation proposal argues that a large admixture of the apical orbitals, in particular the Cu-$d_{z^2}$ orbital, into the $d_{x^{2} - y^{2}}$ band, suppresses $T_c$~\cite{Sakakibara2010}. It has motivated spectroscopic efforts to determine the degree of orbital distillation
using ARPES~\cite{Matt2017a} and STM measurements to map the charge-transfer gaps \cite{Slezak2008,Ruan_2016}.

Finding materials where the admixture of Cu-$d_{z^2}$ orbital increases
with the smaller charge-transfer gap would greatly advance our understanding
of the mechanism of high temperature superconductivity in the cuprates, and as discussed
in section \ref{sec:tuning}, it remains an outstanding challenge in material design.

Exploring the charge-transfer energy (Refs. ~\onlinecite{Weber2012},~\onlinecite{PhysRevB.82.125107})
dependence of $T_{\text{c}}$ is worth pursuing further .
There are several outstanding problems.
There should be an optimal charge transfer energy to
enhance  $T_{\text{c}}$, as it is clear that a very small charge
transfer energy would result in an uncorrelated metal with very
low $T_{\text{c}}$. It is not known however, where exactly the maximum
of $T_{\text{c}}$ is obtained, as it is clear that a very small charge
transfer energy would result in an uncorrelated material with very
low $T_{\text{c}}$. Furthermore, taking the results of this section together with the previous section,
it seems that the reduction of the charge transfer energy makes the formation of the compound more difficult (i.e. increases its distance from the convex hull).
Once again, what are the limits which can be realized chemically, i.e. how
much can the charge transfer gap be reduced while maintaining a stable compound - is a another interesting question.

\subsection{Valence  Disproportionation in Other High Temperature Superconductors: \texorpdfstring{CsTlCl$_3$}{CsTlCl3}}
\label{sec:tlcscl3}

\begin{figure}
  \includegraphics[width=0.9\columnwidth]{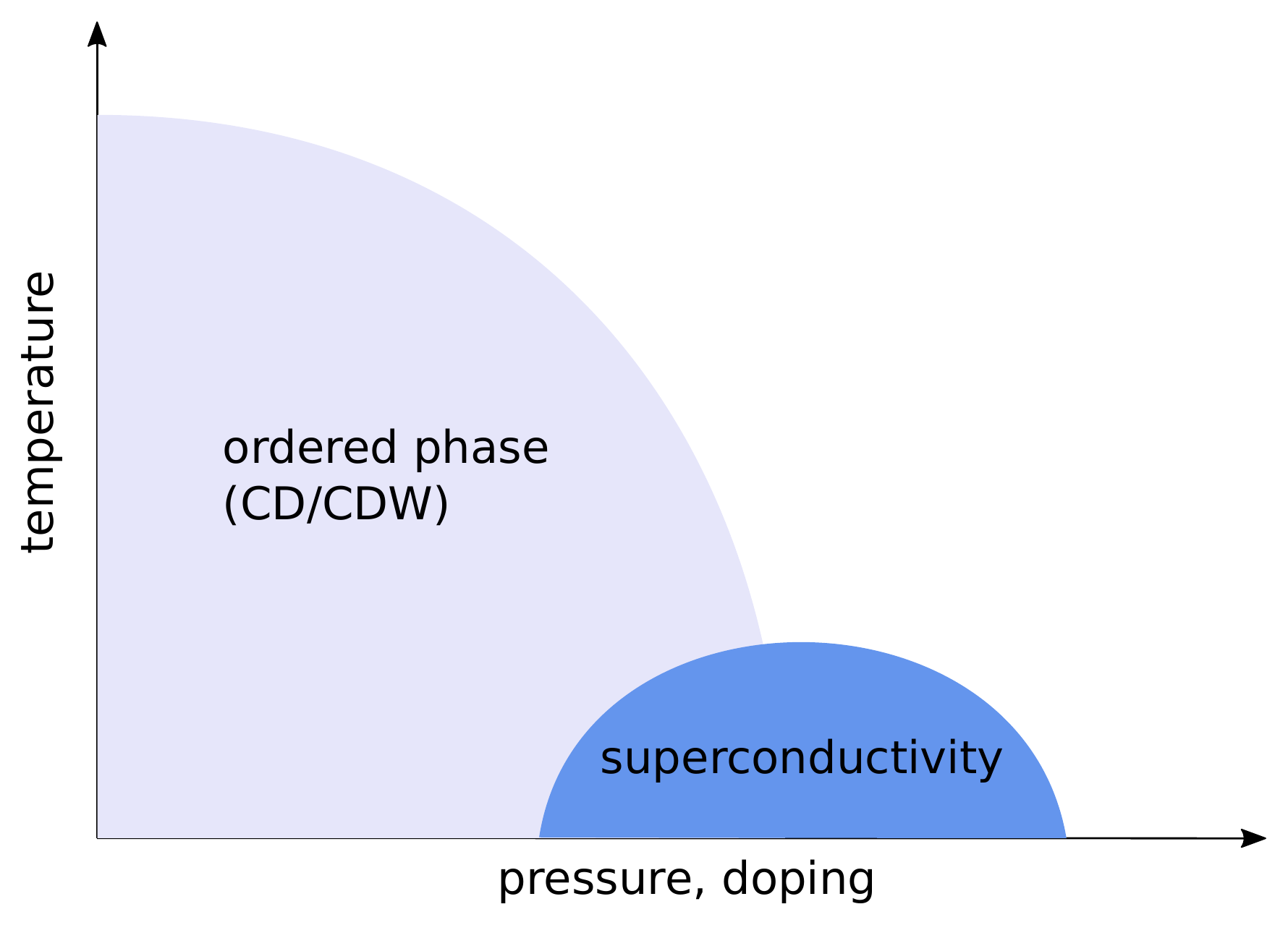}
  \caption{Schematic phase diagram of the potassium-doped BaBiO$_3$ system,
    which is representative of correlated materials as a whole. The phase
    diagram of correlated systems generally contain an ordered ``parent''
    phase. This parent phase can be suppressed as a function of external
    parameters like pressure or doping, and superconductivity can arise near
    the phase transition.}
  \label{fig:sc-phase-diagram}
\end{figure}

\emph{Motivation.} --
In our next example, we seek to design a  parent compound for a new superconductor by creating a
correlated material with valence  disproportionation and strong electron phonon coupling. The hypothesis  guiding the project, is that upon suppression of
the charge{/}valence order by some means (doping, pressure, disorder), pairing will become the dominant instability and  a dome
of superconductivity will emerge. This guiding principle is ubiquitous in
correlated materials: superconductivity generally appears upon suppression of
an ordered ``parent'' phase as a function of experimental tuning parameters
(see the prototypical phase diagram shown in
Fig.~\ref{fig:sc-phase-diagram}).  For a recent realization of this idea
see Ref.~\onlinecite{Li2017}.

There are cases where the  superconducting transition temperatures
turn out to  be well above those expected from  the Migdal Eliashberg theory with the electron phonon coupling evaluated within  LDA.  These materials were  dubbed the ``other" high temperature superconductors~\cite{Pickett2001,yin_kutepov}.

A concrete realization of these principles is provided by Ba$_{1-x}$K$_x$BiO$_3$, a
well-known superconducting system discovered in the 1980s~\cite{Sleight_1975,Cava_1988}. The parent compound is BaBiO$_3$, a $\sim 0.2$~eV band gap
insulator with a distorted perovskite structure~\cite{Sleight_1975}. The order
parameter in this parent state is charge disproportionation, with the bismuth
ions nominally alternating between the $3+$ and $5+$ valence (in reality  the actual  charge valence disproporationation is much smaller \cite{yin_kutepov}). Doping the barium
site with potassium suppresses the structural distortions along with the charge
disproportionation and gives rise to superconductivity with a transition
temperature of nearly 30~K at the optimal doping.

Conventional electronic structure descriptions based on LDA/GGA fail to
describe the insulating character of the parent compound. More importantly, the
DFT estimates of the electron phonon coupling $\lambda$ within Migdal-Eliasberg
theory give a value of 0.34 in the doped compound, too small to account for its
superconductivity~\cite{Meregalli_1998}. Careful  examination
found that $\lambda$ is substantially enhanced relative to its DFT estimate
to a value of nearly $1.0$, and that this enhancement is responsible for
superconductivity in doped BaBiO$_3$~\cite{yin_kutepov}.

It was proposed  that  {\it static}
correlations similarly enhance the electron phonon coupling in other materials
proximate to an insulating state, accounting for superconductivity in systems
such as HfNCl, borocarbides and buckminsterfullerenes. For these materials, the
most important type of correlation that must be captured is the static
contribution, and a GW or hybrid DFT calculation is therefore necessary  to correct the
electronic structure. After these calculations are done, one is left with a
strongly-coupled electron-phonon system with $\lambda \sim 1$. This coupling
induces a large dynamical self energy, which accounts for the observed
anomalous optical properties of doped BaBiO$_3$~\cite{Nourafkan_2012} at
energies below 1~eV. Hence the low energy electron phonon treatment requires DMFT.

\begin{figure}
  \includegraphics[width=0.8\columnwidth]{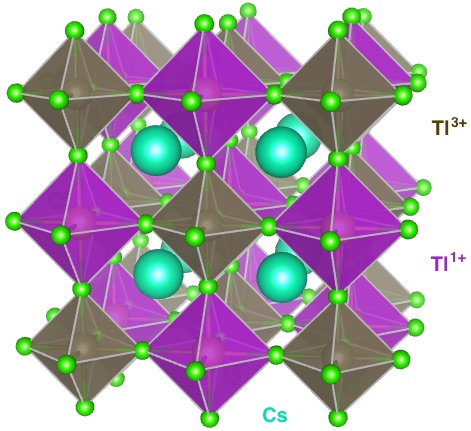}
  \caption{Observed perovskite structure of cubic phase of CsTlCl$_3$. The
    thallium ion disproportionates in to Tl$^{1+}$ and Tl$^{3+}$ nominally, and
    the structure displays an associated alternating expansion and contraction
    of the TlCl$_6$ octahedral cages. Adapted from Ref. \onlinecite{Retuerto_2013}.}
  \label{fig:cstlcl3}
\end{figure}

\emph{Setting the targets and Framing the questions. Heuristic considerations} --
Having identified a  superconducting mechanism, namely   (static) correlation enhanced electron phonon coupling   it is natural to  seek  a realization of this
mechanism in a new material.   This task was undertaken by  Z. Yin \emph{et al.} in Ref.~\onlinecite{Yin2013_epl} .  The parent compound would need to exhibit
charge disproportionation and the  value of the electron-phonon coupling must be underestimated by LDA/GGA. The heuristic reasoning used to select CsTlCl$_3$  follows by analogy with BaBiO$_3$.  It requires  an ion which would charge
disproportionate. Like bismuth, thallium is known to valence skip, preferring
either a $1+$ or $3+$ valence state.  We want also  the same perovskite
structure which requires the  $A$Tl$X_3$ composition. Balancing the
ionic charges in the presence of the average $2+$ charge of the thallium adds
the next constraint. Taking ionic radii into consideration, one arrives at Cs for
the $A$ site, and Cl (or F) for the anion $X$ and lead to the proposal of
CsTlCl$_3$ as a valence disproportionation compound, which could be turned (by application of pressure and doping) into  another ``other" high temperature superconductor~\cite{Yin2013_epl}.

\begin{figure}
  \includegraphics[width=0.85\columnwidth]{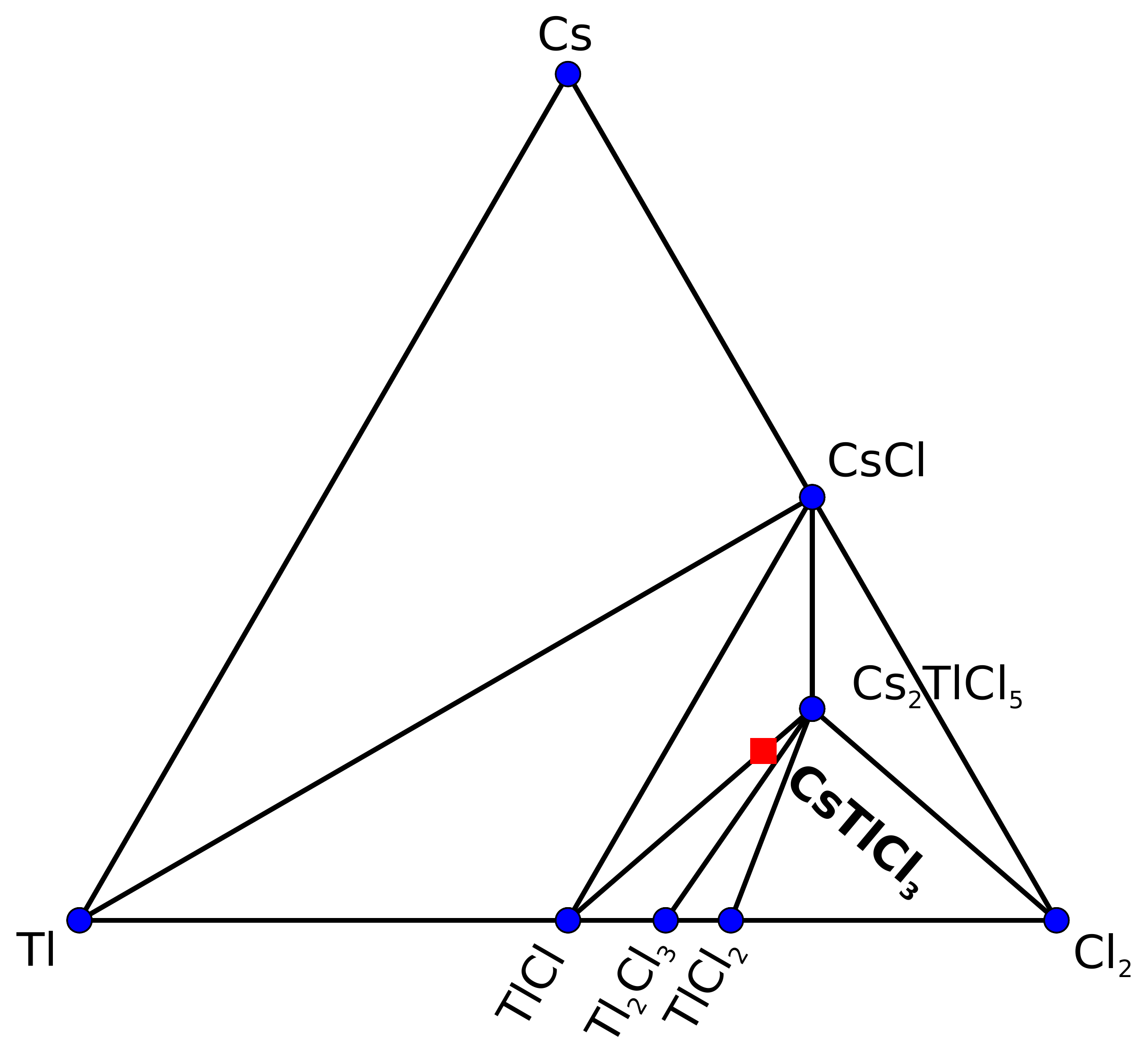}
  \caption{Gibbs phase diagram of the Cs-Tl-Cl chemical system. The target
    compound CsTlCl$_3$ is found to lie 3~meV/atom above the convex hull.
    Although energies above the hull strictly imply instability, its small
    value is indistinguishable from zero given the systematic uncertainties in
    current DFT total energies (up to 50~meV/atom), and many compounds are
    known to be metastable, lying up to 100~meV/atom above the hull. The
    3~meV/atom result should be interpreted as a green light to proceed with
    further investigation as the structure is not obviously unstable.}
  \label{fig:gibbs-cstlcs}
\end{figure}

\emph{Electronic structure} -- Identification of the magnitude and nature of
the correlations are  an  important first step  for a successful electronic structure prediction, which will link the structure of the material to the desired property.
In order to compute the electronic structure, it was  assumed that the
structure of CsTlCl$_3$ would be the desired perovskite crystal structure,
shown in Fig.~\ref{fig:cstlcl3}.

To establish that this material is another potential
``other high temperature superconductor"
the electron-phonon coupling was evaluated
using both LDA and methods that capture
static correlations, namely screened hybrid density functional theory
using the HSE06 functional and GW~\cite{yin_kutepov}.
The calculated electron-phonon couplings at 2.2 GPa with 0.35 hole doping/f.u.
(using the virtual crystal approximation)
are 0.91 and 2.32 for LDA and HSE06 functionals, respectively~\cite{Yin2013_epl}.

The GW method was used as a benchmark to determine the value of the HSE06 screening parameter
that best reproduced the GW band structure.
It was also veritfied  using this methodology that takes into account
{\it static} correlations that charge disproportionation occurs at half filling~\cite{yin_kutepov}.

The phase diagram of the material was then determined as a function of
pressure and doping.
It did resemble the phase diagram of Fig.~\ref{fig:sc-phase-diagram}.
Furthermore, the electron-phonon coupling strength $\lambda$ attains
a value of 2.32 at 2.2~GPa, and the predicted superconducting transition
temperature is 21~K. These results encouraged the  pursuit of  experimental
synthesis.

\emph{Structural prediction} --
Structural  stability   was tested by computing the  most important phonon
modes within DFT and verifying that none were imaginary. It would be
interesting to investigate whether this structure would be correctly predicted
by the methods described in section~\ref{sec:bacoso},
and what additional polymorphs are
possible in this new class of materials.

\emph{Global stability} --
In the original work of Ref.~\onlinecite{Yin2013_epl},  only a few reaction pathways against known binaries were
checked and the authors reported finding exothermic reactions.
We now check the stability of CsTlCl$_3$ for
decomposition against all known elements, binaries and ternaries in the
Cs-Tl-Cl chemical system. Using the computed total energies stored in the
Materials Project database~\cite{Jain_2013} and the phase
diagram~\cite{Ong2008, Ong_2010} functionality provided with
pymatgen~\cite{Ong_2013}, we construct the convex hull plotted in
Fig.~\ref{fig:gibbs-cstlcs}. We find that CsTlCl$_3$ lies 3~meV/atom above the
convex hull, indicating that the material is very likely to form.

\begin{figure}
  \includegraphics[width=1.0\columnwidth]{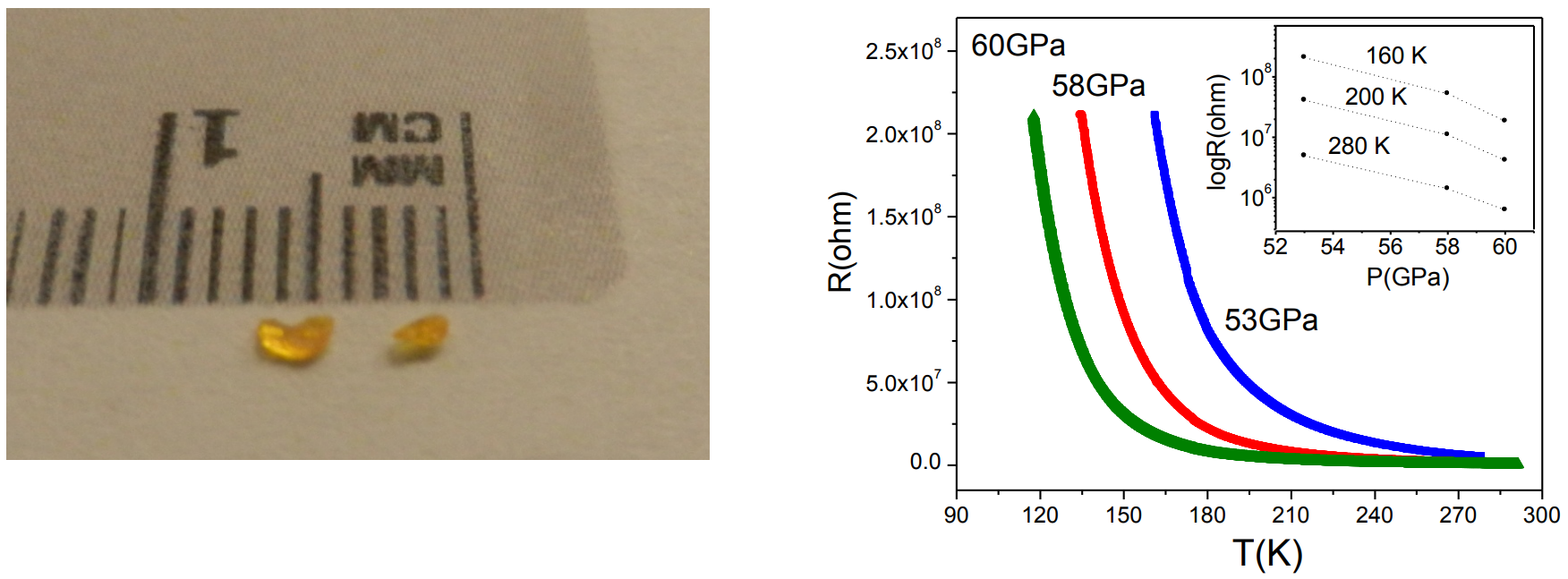}
  \caption{CsTlCl$_{3}$ single crystals resulting from successful material design (left). This material exhibits enhanced electron-phono coupling and is close to metalization with pressure (right), as predicted theoretically (from Ref.~\onlinecite{Retuerto_2013}) }
  \label{fig:csincl3-crystal}
\end{figure}

\emph{Lessons learned} --
Orange crystals of CsTlCl$_3$ were successfully grown, which indeed adopt a
perovskite structure~\cite{Retuerto_2013} exhibiting strong breathing
distortions indicative of charge disproportionation at the thallium sites  as predicted by theory.
The optical gap determined by absortion measurments  was of the order of 2.1 eV, in close agreement with the GW  theoretical  predictions~\cite{Retuerto_2013}.
The crystals are shown in Fig.~\ref{fig:csincl3-crystal}

Pressure studies showed that  as predicted, the compound was very close to metallization.
So clearly in this case, theoretical considerations predicted an interesting  material which was not listed in the ICSD database, and established that the
electron phonon coupling is enhanced over its LDA value due to static correlations.    These halide perovskiets are therefore a  new arena to explore  strong electron phonon coupling.

This  compound turned out to be  challenging to dope: doping via Cs vacancies,
replacement of Cl by O, S, or N, and substitution of Tl by Hg all did not
succeed in pushing the material into a metallic state~\cite{Retuerto_2015}.
This deserves more careful investigation to see if it is the result of phonon
induced self-localization or disorder. The dopablity of a material, and identification of the most probably dopants  was intensely  studied  in weakly correlated semiconducting materials \cite{Zunger2003}  and is  an outstanding open  problem for correlated insulators and semiconductors worth pursuing.

Regardless of whether superconductivity will eventually be discovered in these
thallium halides, we found that when motivated by a guiding principle,
materials design of a correlated material is possible and yields  not only new materials but valuable
insights into the challenges (some unforseen) required to be overcome for
successful end-to-end materials design.
Furthermore, another polymorph  phase  was also found in the experiment, but since an exhaustive structural search had not been performed, it is not possible to ascertain if this phase is among the low lying structures.

\subsection{Fe 112 compounds}
\label{sec:fe112}

\begin{figure}
  \includegraphics[width=\columnwidth]{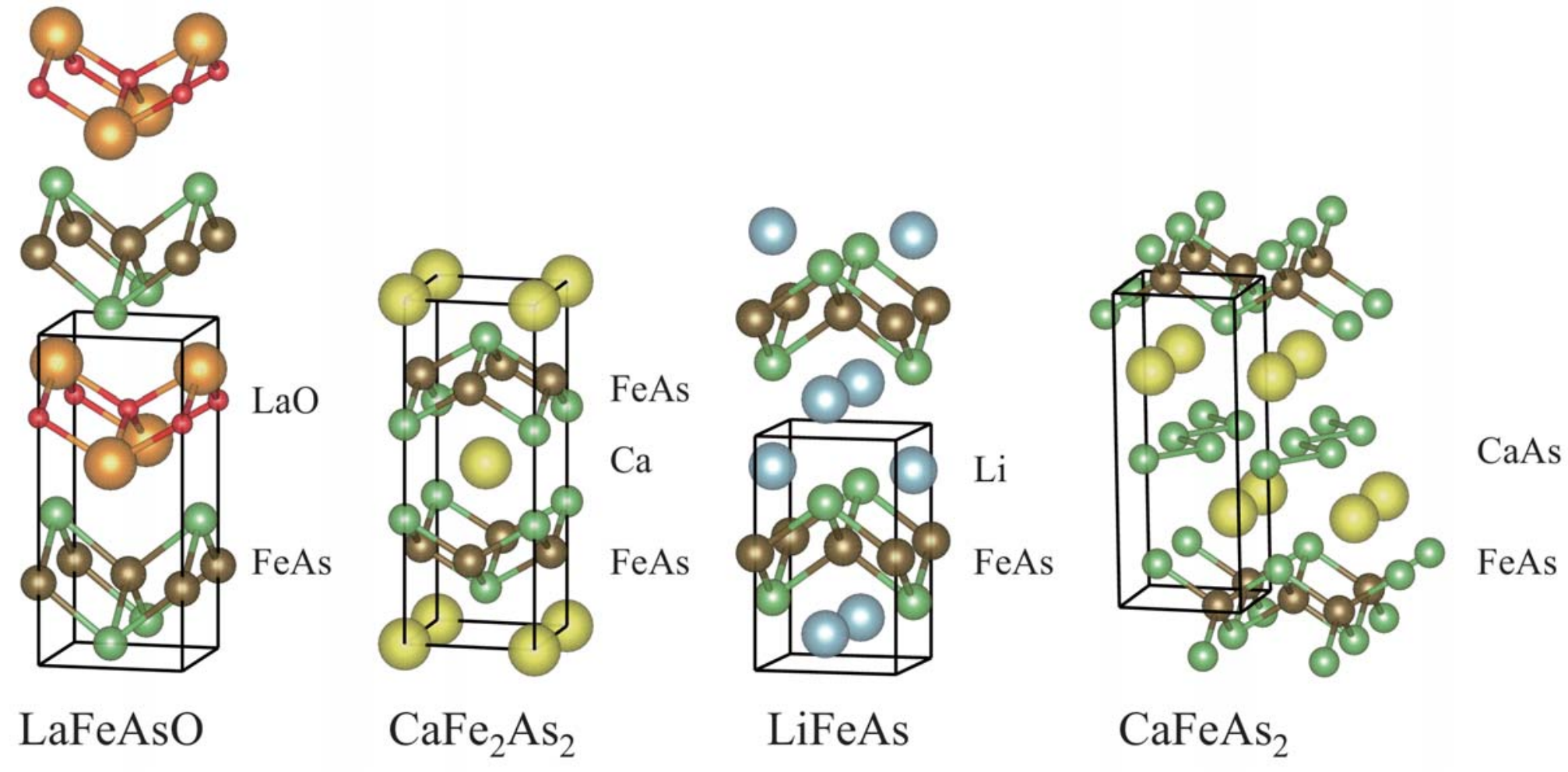}
  \caption{Crystal structures of representative Fe based superconductors.
  They have common FeAs layers but different spacers.}
  \label{fig:Fe-SC-struct}
\end{figure}

\begin{figure*}
\includegraphics[width=1.8\columnwidth]{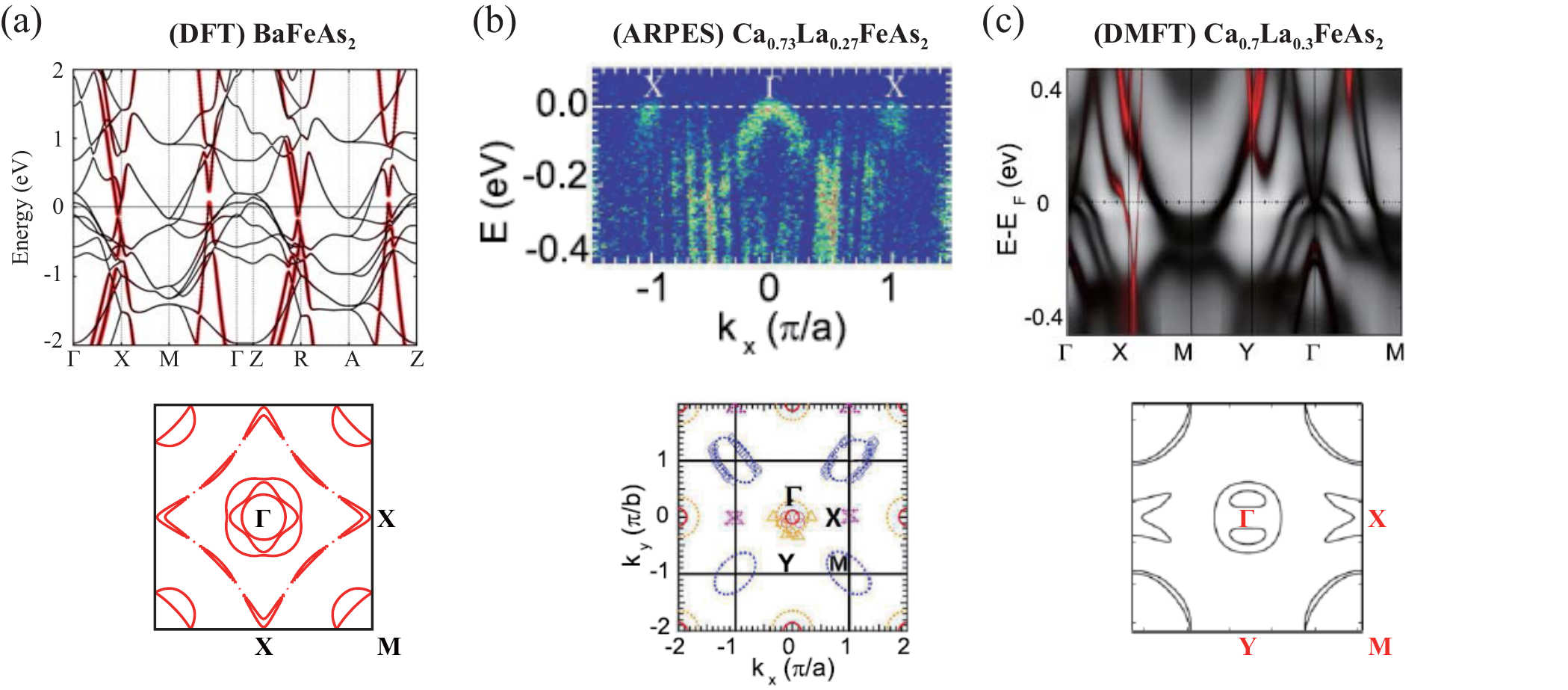}
  \caption{Electronic structures of Fe 112 compounds.
  (a) Band structure (top) and Fermi surface (bottom) of the hypothetical  BaFeAs$_{2}$  material, computed by DFT.
  (b) Angle-resolved photoemission spectroscopy (ARPES) $k-E$ map (top) and the two-dimensional (2D) contour of ARPES Fermi surface (bottom) of Ca$_{0.73}$La$_{0.27}$FeAs$_{2}$.
  (c) Spectral function $A(k,\omega)$ (top) and Fermi surface (bottom) of Ca$_{0.7}$La$_{0.3}$FeAs$_{2}$ computed by DMFT.
  The in-plane $p$ orbital characters ($p_{x}$ and $p_{y}$) of the metallic As spacer
  are highlighted in red in the band dispersions of (a) and (c).
  The 2D Fermi surfaces in (a), (b), and (c) are obtained at $k_{z} \sim \pi/c$.
  (b) and (c) are adapted from Ref.~\onlinecite{jiang_prb16}.
  }
  \label{fig:Fe112-bands}
\end{figure*}

As a final example of  the use of material design to elucidate the mechanism
in correlated  superconducting materials,
we turn our attention into iron-based superconductors.
The discovery of the high temperature superconducting iron-based materials by Hosono \emph{et al.} \cite{hosono}
triggered a heated debate about the superconducting mechanism in these materials.
The original high $T_c$ superconductor was in the 1111 structure,
followed by the 111 and the 122 structure (see Fig.~\ref{fig:Fe-SC-struct}).
In all these cases the spacer layer is inert (LaO for the 1111, Sr++ or Ca++ for the 122 case, Li+ in the 111 case) in the sense that their electrons are far from the Fermi level.

Following the discovery of these materials, multiple ideas where put forward
in order to understand the origin of  their superconductivity.
One set of ideas stresses the importance of the magnetism of the iron ions,
as in spin fluctuation theories~\cite{Mazin2008,Kuroki2008,Graser2008,Chubukov2008}.

A different school of thought posits that what is important
is the high electronic polarizability of the pnictides and chalcogenides.
In the latter case one could envision that presence of additional metallic spacer layers would modify the polarizability
and will strongly modify the superconducting transition temperature
as in the polaronic mechanism of Ref.~\onlinecite{Berciu2009}.
This may also be the case if the superconductivity is mediated by orbital fluctuations ~\cite{Onari2009}.
Designing an iron pnictide superconductor,
with a metallic spacer layer and studying how it affects the critical temperature would help elucidate the mechanism of superconductivity in this important class of compounds.
This task was undertaken by Shim \emph{et al.}~\cite{shim_prb09}.

\emph{Electronic structure} --
A new family of iron based superconductors, the 112 family was proposed theoretically based on chemical analogies and density functional studies \cite{shim_prb09}.
Analogies with known 1111 compounds and the spin fluctuation mechanism lead the authors of Ref.~\onlinecite{shim_prb09} to suggest that
BaFeSb$_{2}$ and BaFeAs$_{2}$ would be high temperature superconductors, but
unlike the known structures at the time the 112 structure would have active spacer layers,
where electrons living in the spacer layers contribute their own Fermi surface
as shown in Fig.~\ref{fig:Fe112-bands}(a).

Attempts to synthesize iron pnictide materials in the  112 structure
proposed in Ref.~\onlinecite{shim_prb09} were not originally successful,
but new Mn-based materials in this structure were found~\cite{wang_prb11,park_prl11}
and both theories~\cite{wang_prb11,lee_prb13}
and experiments~\cite{park_prl11,jo_prl14,jia_prb14,feng_srep14}
show that the spacer layers in the Mn-based materials possess Dirac cones.
New Ag-based materials in this structure were also reported to possess Dirac cones
\cite{Wang12_LaAgSb2,Wang13_LaAgBi2,Shi16_LaAgSb2}.

\emph{Structure prediction} --
The original calculations of Ref.~\onlinecite{shim_prb09}  relaxed lattice parameters but
did not examine  the local stability of the CaAs spacer layer,
and had not considered the stability against phase separation.
Motivated by the earlier theoretical work and experimental developments~\cite{katayama_jpsj13,yakita_jacs14}
these issues were reexamined in Ref.~\onlinecite{kang_prb17}, which investigated the crystal structure of CaFeAs$_{2}$ by using USPEX
in conjunction with VASP as DFT engine.
A  GGA(PBE) functional and a dense Monkhorst-Pack sampling grid
with a resolution of $2\pi \times 0.02 {\AA}^{-1}$ for the $k$-space integrations.
In the structural search in USPEX, two formula-units per unit-cell were considered.
The evolutionary structure prediction preformed by USPEX predicts that
CaFeAs$_{2}$ has monoclinic $P2_{1}/m$ structure
with the distortion of the CaAs layer into zigzag chains.

\begin{figure}
  \includegraphics[width=\columnwidth]{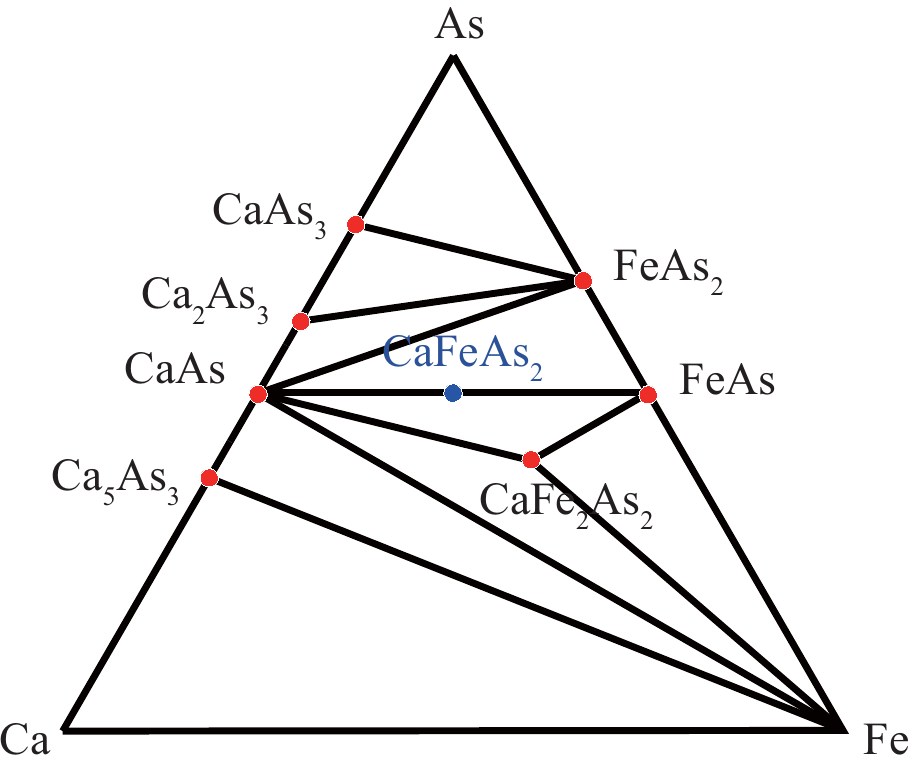}
  \caption{Ternary phase diagram for CaFeAs$_{2}$.
  CaFeAs$_{2}$ is put above the convex hull and the energy above hull is 13~meV/atom.
  Adapted from Ref.~\onlinecite{kang_prb17}.}
  \label{fig:Ca-Fe-As}
\end{figure}

\emph{Global stability} -- A study of the thermodynamic phase stability of CaFeAs$_{2}$
was performed in Ref.~\onlinecite{kang_prb17}.
It is reproduced in Fig.~\ref{fig:Ca-Fe-As}.
Taking into account stripe magnetic order in the predicted
monoclinic $P2_1/m$ structure of CaFeAs$_{2}$
because the additional symmetry breaking due to magnetic order
(for example, antiferromagnetic order) is not considered in the structural prediction
performed by USPEX.
Due to the stripe magnetic order, the monoclinic $P2_1/m$ structure is further
relaxed into a triclinic $P\bar{1}$ structure
and the total energy is lower by 19.50 meV/atom (the magnetic moment is 1.95 $\mu_{B}$/Fe).
CaFeAs$_{2}$ lies 13 meV/atom above the convex hull,
which corresponds to the existence probability of 0.37
as discussed in section~\ref{sec:Probability-Estimation}.
Doping it with rare-earth ions could help its stability
with possible energetic gain or an entropic gain regarding no ordering of the rare-earth atoms~\cite{kang_prb17}.
Similar hypothetical compounds SrFeAs$_{2}$ and BaFeAs$_{2}$ were found to be
24 (existence probability: 0.33) and 17 meV/atom (existence probability: 0.36) above the convex hull,
respectively~\cite{kang_prb17},
which are slightly higher in energy compared to CaFeAs$_{2}$.

Revising the phase stability calculation - using a more accurate correlated  method, as suggested in section \ref{sec:Probability-Estimation},
we consider the determinant reaction. This reaction is found by GGA to be  CaAs + FeAs $\rightarrow$ CaFeAs$_{2}$.
The on-site Coulomb interaction $U$ = 5 eV and Hund's coupling constant $J$ = 0.8 eV
were used in the Gutzwiller calculations~\cite{jiang_prb16,kang_prb17}.
We found that CaFeAs$_{2}$ lies 7 meV/atom above the hull.

\emph{Electronic structure} -- The existence of an extra Fermi surface with CaAs character
in Ca$_{1-x}$La$_{x}$FeAs$_{2}$
was confirmed by angle-resolved photoemission
and the photoemission spectra is in good agreement with DFT+DMFT calculation~\cite{jiang_prb16}
(see Figs.~\ref{fig:Fe112-bands}(b) and (c)).
Since the CaAs layer possesses the conducting zig-zag As chain,
it induces the large electronic anisotropy~\cite{kang_prb17}.

\emph{Conclusion} --
Recently, iron-based superconductors in the 112 structure were synthesized  with rare-earth doping,
Ca$_{1-x}$La$_{x}$FeAs$_{2}$~\cite{katayama_jpsj13} and (Ca,Pr)FeAs$_{2}$~\cite{yakita_jacs14}.
Surprisingly, these materials form in a structure where the As in the CaAs layers
are distorted in zigzag chains (see Fig.~\ref{fig:Fe-SC-struct}).
The space group is monoclinic either $P2_{1}$~\cite{katayama_jpsj13}
or $P2_{1}/m$~\cite{yakita_jacs14}
rather than the originally assumed tetragonal structure.
Second harmonic generation experiments confirmed the space group $P2_{1}$ for La-doped compounds~\cite{harter_prb16},
but similar data are absent for other rare-earth doping compounds
to the best of our knowledge.

\begin{figure}
  \includegraphics[width=\columnwidth]{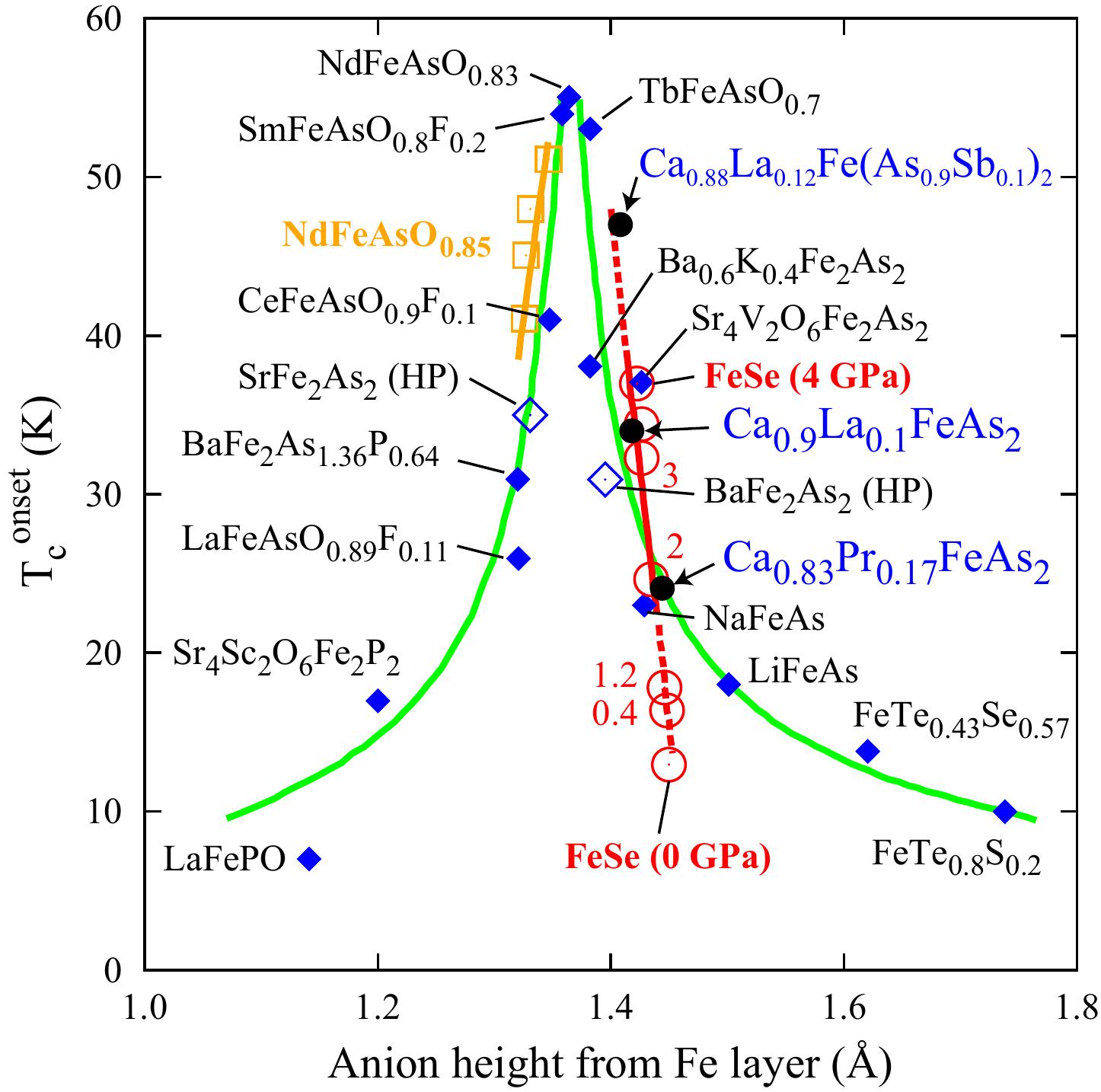}
  \caption{$T_c$ vs anion height in iron pnictide and chalcogenide superconductors. Adapted from Ref.~\onlinecite{kang_prb17}.}
  \label{fig:fesc-Tc-height}
\end{figure}

Armed with this information, the critical temperature of the newly discovered 112 compounds
was replotted on the graph of Mizuguchi \emph{et al.}~\cite{mizuguchi_sst10} which displays the $T_c$ as a function of the pnictogen height
(see Fig.~\ref{fig:fesc-Tc-height}).
The points do not deviate much from their universal plot,
indicating that the fate of the superconductivity resides primarily in the FeAs layers,
and is not affected by the nature of the spacer layers.
This observation helps rule out a charge fluctuation mechanism in favor of spin mediated superconductivity in the iron pnictides.

More recently theoretical studies focused on the spacer layers found that
the As $p_{x}$ and $p_{y}$ orbitals are responsible for the Dirac cones,
and that spin-orbit coupling could not only open a gap,
but also induce topological phases on these layers.
This suggests that the 112 compounds are prime candidates
for proximity induced topological superconductivity~\cite{wu_prb14,wu_prb15}.
More generally the 112 structure provides inspiration for combining  iron pnicitide layers
with layers having non-trivial topological band structure. This area of research is as yet unexplored,
and calls for a new iteration of the material design loop outlined in Fig.~\ref{fig:flow}.

\subsection{BaCoSO}
\label{sec:bacoso}
\begin{figure}
\includegraphics[scale=0.2]{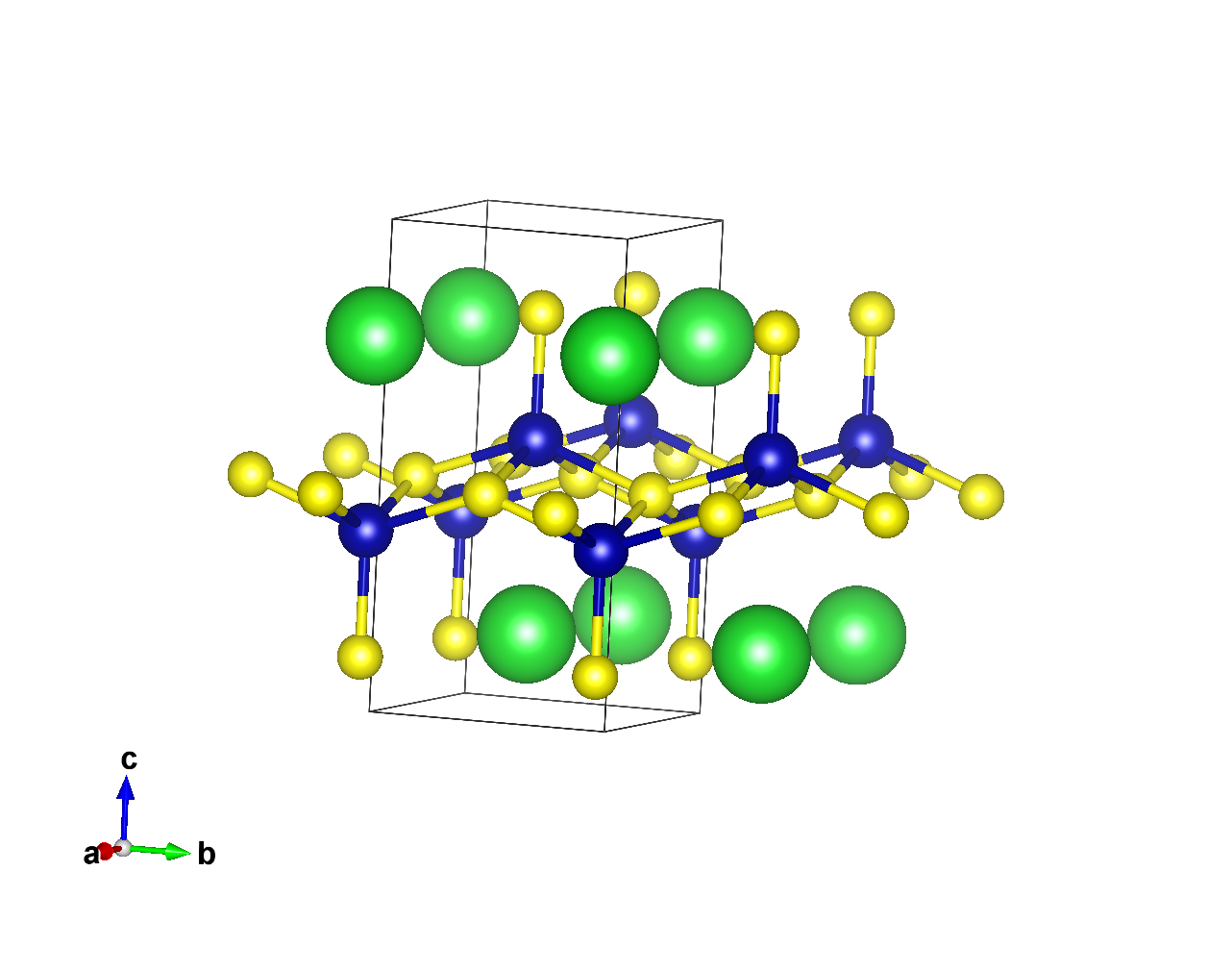}
\caption{Crystal Structure of the 2-dimensional parent material, BaCoS$_{2}$.
The green, blue, and yellow spheres correspond to Ba, Co, and S atoms, respectively.}
\label{fig:BaCoS2}
\end{figure}

\begin{figure}
\includegraphics[scale=1.0]{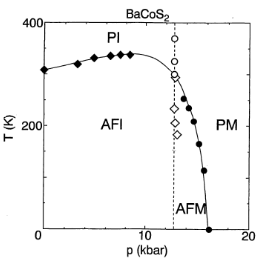}
\caption{BaCoS$_{2}$ phase Diagram from Ref.~\onlinecite{Yasui1999}, (c) (1999) The Physical Society of Japan.}
\label{fig:BaCoS2-pd}
\end{figure}

\emph{Motivation} --
Finally we return to the organizing principle, shown in
Fig.~\ref{fig:sc-phase-diagram}, that exotic phases are often found upon
suppression of a parent ordered phase of layered quasi-two-dimensional compounds. The material BaCoS$_2$ (Fig.~\ref{fig:BaCoS2}) is a layered
antiferromagnetic Mott insulator. The application of pressure does suppress the
magnetic order, but rather than finding a new phase at the critical point, the
material simply becomes metallic down to the lowest temperatures measured as shown in Fig.~\ref{fig:BaCoS2-pd}.

Clearly, the expectation of novel phases near
the suppression of order is not a sure bet! Other factors must be involved.
Still this material is quite interesting, as  it displays a transition or crossover from
a paramagnetic insulating phase to a metallic phase at high temperatures and antiferromagnetically ordered state at low temperatures, features reminiscent of
$V_2 O_3$ and
$Ni Se_x S_{1-x}$, hence a comparative study   of these systems can illuminate the factors that govern the correlation induced metal to insulator transition.

{\it Targets and questions} --
The  question we address is  how the physical properties of BaCoS$_2$  change
if we substitute the
large sulfur ion by the smaller oxygen ion  in  BaCoS$_2$  to
form BaCoSO, in particular how does the gap change as a result.  At first sight one would expect that the larger
size of the sulfur ion would result in a larger gap, but  the answer to this question is not obvious since structural changes can take place.
We approached this problem as a case study of our methodology before this material was reported in the experimental literature~\cite{Salter2016}.  Some of the results were very surprising and are reported in this section.

\begin{figure}
\includegraphics[scale=0.55]{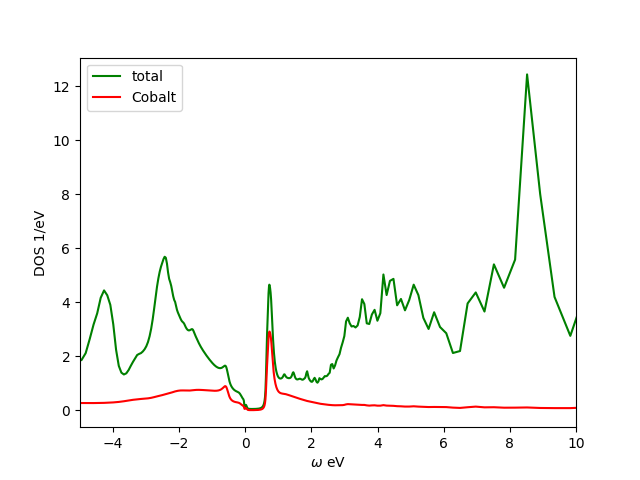}
\caption{Spectral function for BaCoS$_2$ at T=180 K using LDA+DMFT code~\cite{haule_DMFT,Haule-dmft}.}
\label{fig:BaCoS2_DOS}
\end{figure}

\emph{Electronic Structure} --
Previous LDA+U and DMFT studies of BaCoS$_2$ were reported in Refs. \onlinecite{Zainullina2011,Zainullina2012}.
We returned to this problem using LDA+DMFT  using  the implementation of K. Haule ~\cite{haule_DMFT,Haule-dmft}
and display the results in Fig. \ref{fig:BaCoS2_DOS}.  Using the experimental lattice parameters,  BaCoS$_2$  is a small
gap Mott insulator very close to the metal insulator transition.

\emph{Structure prediction} --
We  continue the material design workflow by turning to structural properties of  the Ba-Co-S-O system by first using USPEX, a
structure prediction package based on a genetic-search algorithm, to sample the local minima in the energy landscape. We choose a 1:1:1:1 ratio for the elements and
allowed for two formula units in a unit cell. We use VASP as our DFT engine. We
use spin-polarized PBEsol and do not include $U$ corrections.
In order not to miss crucial seed structures, which
ultimately led to the experimental structure, we found that the randomly
generated initial population of structures must be sufficiently large. An
initial population of size 300 was sufficient with a single generation size of
60. In total $\sim 700$ metastable structures were produced in 8 generations.

With GGA the observed structure does not have the lowest energy:  it is 271~meV above the lowest-energy structure, and there are 23 structures with lower energy.
With spin-polarized GGA the situation is improved, but the structure is still not among the 10 lowest energy results. A naive conclusion would be that it is not the ground state. One would like to distinguish the observed structure within the low-lying set, for example within 0.5~eV/(unit cell) of the final lowest energy structure. However, there are 58 such reported structures (out of the final set of 152 best-fit structure), even after removing similar structures.

\begin{figure}
  \includegraphics[width=\columnwidth]{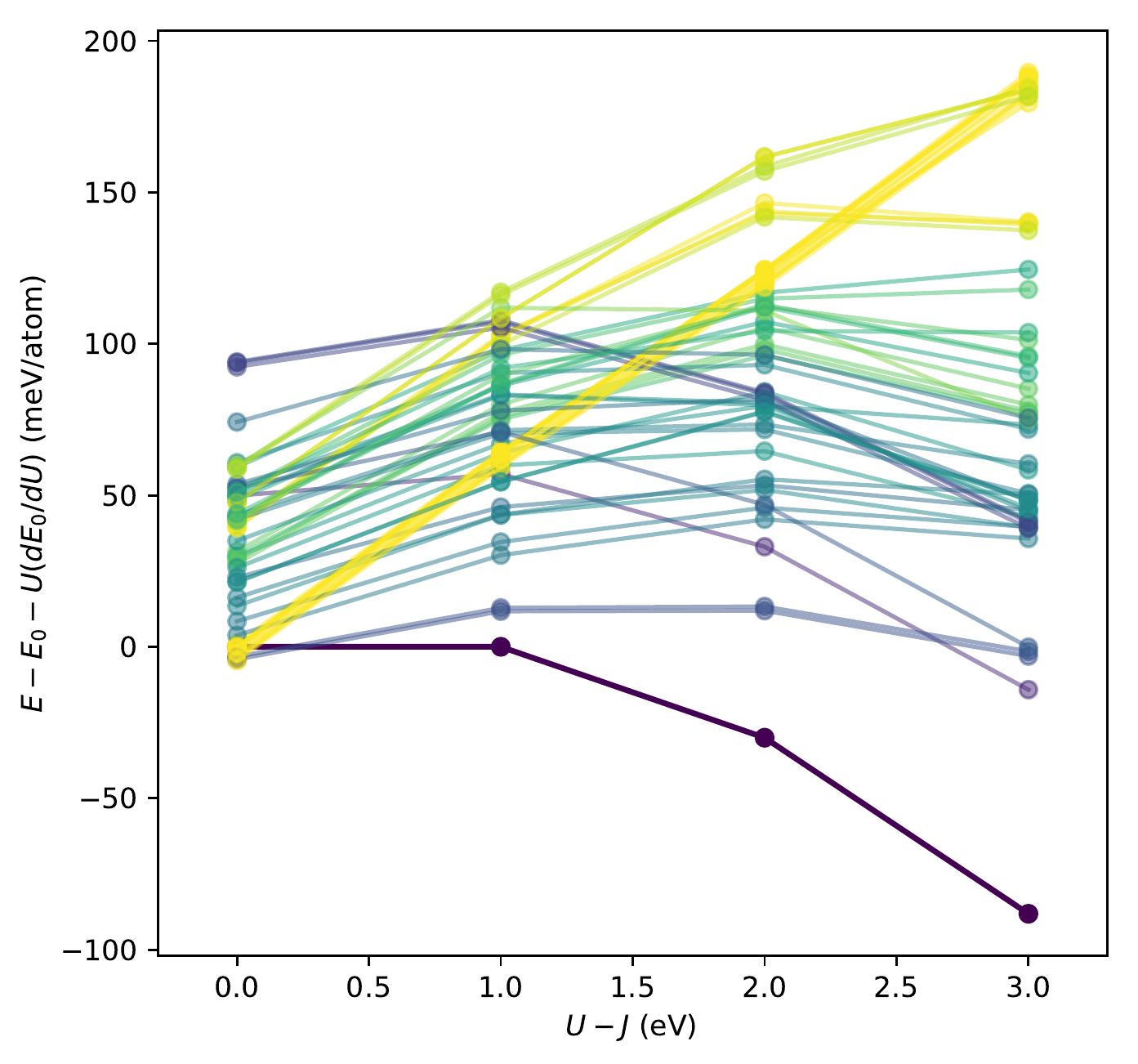}
  \caption{The $U$-dependence of the total energies for 58 of the lowest energy
    structures produced by USPEX during structure prediction for BaCOSO. The
    energies are produced by GGA+U allowing for spin polarization. The
    experimentally observed structure (lowest bold purple line) does not
    initially begin with the lowest energy at $U=0$, but rapidly becomes
    favored energetically with increasing correlation strength. The group of
    structures with the lowest energies at $U=0$ (yellow lines) rapidly rise
    nearly linearly with $U$. Lines colors are graded according to initial
    slopes at $U=0$. The intercept $E_0=-5.92$~eV and slope $dE_0/dU = 0.19$ of
    the energy curve for the experimental structure have been subtracted for
    all curves to improve clarity.}
  \label{fig:bacoso-e-vs-u}
\end{figure}

 We notice however that adding correlations to the Cobalt atom (in the form of a non-zero $U$) resolves this problem.
The energies of the 58 lower-lying structures are examined as a function of $U$ ($J=0$, which we plot in Fig.~\ref{fig:bacoso-e-vs-u}. Crystal structures are kept fixed and only
electronic convergence is performed as $U$ is tuned. The lines are colored by
their slope at $U=0$ which helps visual identification of the different types
of curves.

\begin{figure}
  \includegraphics[width=0.49\columnwidth]{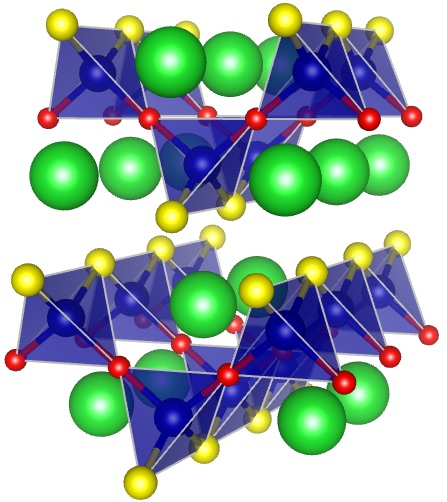}
  \includegraphics[width=0.49\columnwidth]{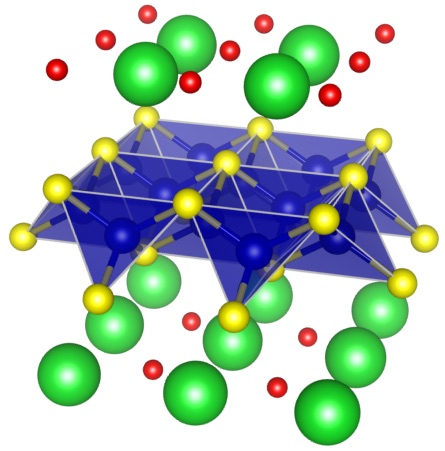}
  \caption{Experimental structure (left),
  and representative of the group of lowest energy structures
  found by GGA (right) for the BaCoSO composition.
  The experimental structure contains corner-shared CoS$_2$O$_2$ tetrahedra
  slightly separated by Ba spacer layers. The GGA structure contains
  edge-sharing CoS$_4$ tetrahedra in a fluorite layer, which are separated by
  BaO rock salt layers.}
\label{fig:bacoso-structs}
\end{figure}

\begin{figure}
  \includegraphics[width=\columnwidth]{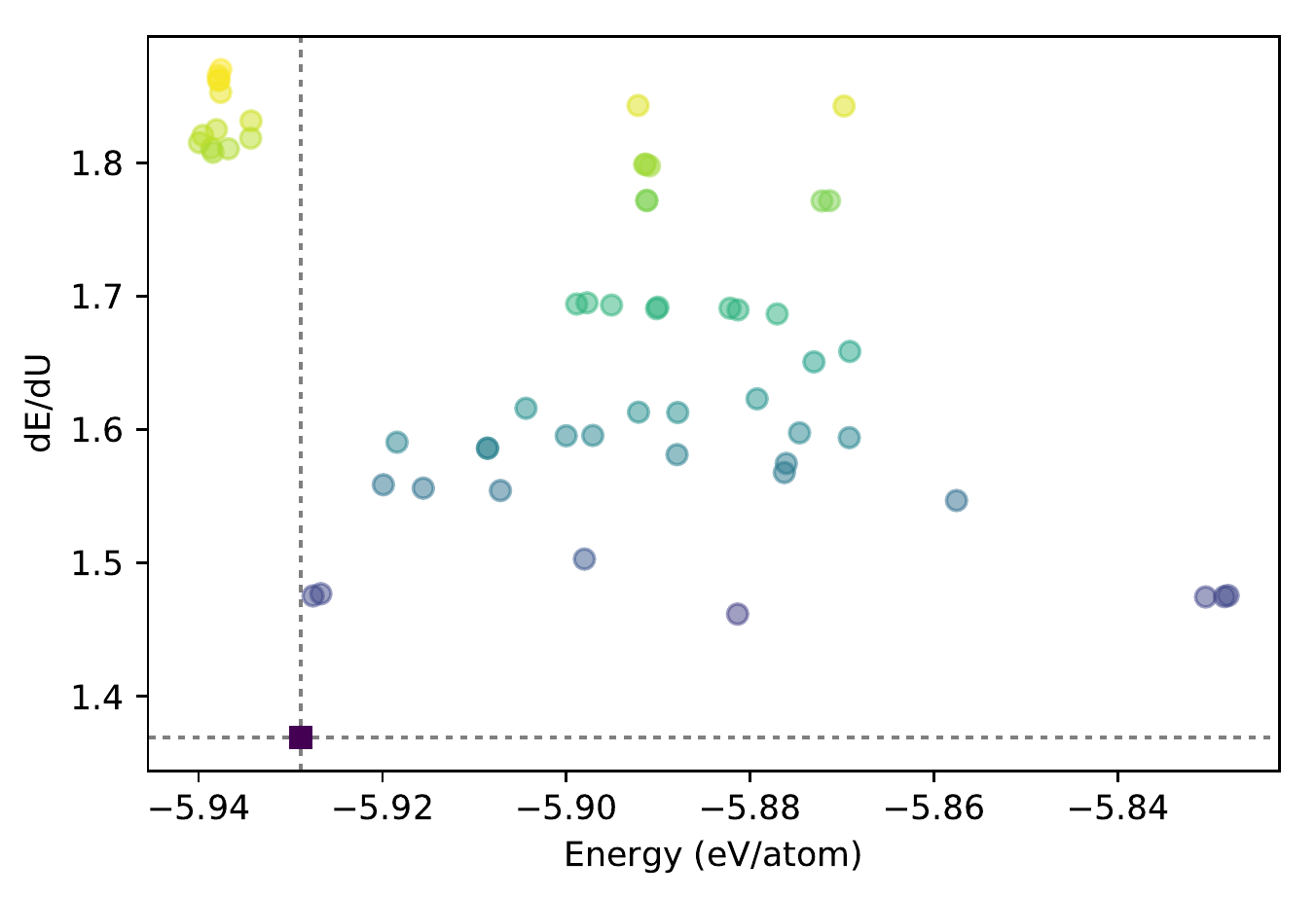}
  \caption{Scatter plot of initial slope of the GGA+U energy curve vs. the
    total energy at $U=0$ for 58 candidate structures produced during the
    search for the crystal structure of BaCoSO. The experimental structure
    (lower left bold purple square) does not have the lowest energy, but has
    the smallest slope, so it is least penalized by increasing $U$. The group
    of structures with the lowest initial energies (upper left cluster of
    yellow dots) have large slopes, and are heavily penalized by correlations.
    Dotted lines centered on the experimental structure are guides to eye. Dots
    are colored by initial slope, matching the convention in
    Fig.~\ref{fig:bacoso-slope}.}
  \label{fig:bacoso-slope}
\end{figure}

We find that the inclusion of even a very small $U \sim 0.5$~eV causes a clear
separation of a single structure from the remaining minima. This curve is the
lowest line plotted in bold purple in Fig.~\ref{fig:bacoso-e-vs-u} and
corresponds to the structure shown in Fig.~\ref{fig:bacoso-structs} (left). The
structure contains corner-linked CoS$_2$O$_2$ tetrahedra forming a corrugated
2$D$ layer, and has been confirmed to be the correct ``ground state'' structure
in experiment. The energy gap between the next-best structure and the ground
state widens significantly as $U$ increases. The group of structures the lowest
energies initially (set of diagonal yellow lines in
Fig.~\ref{fig:bacoso-e-vs-u}) is found to be penalized most rapidly by
correlations, rising nearly linearly. They correspond to slight variations of
structures containing CoS fluorite layers separated by BaO rock salt layers,
and a representative is shown in Fig.~\ref{fig:bacoso-structs} (right). The
remaining structures exhibit intermediate behavior as a function of $U$. This
grouping is most clearly visualized in a plot of the slope vs. the total
energy, as shown in Fig.~\ref{fig:bacoso-slope}. The set of structures favored
by GGA form a clear outlying cluster in the upper left (yellow), the
experimental structure has the smallest slope and a competitive total energy
(bold square, lower left), and the remaining structures are scattered in the upper
right quadrant.

In order to understand this behavior, we examine the density matrix
$n^{\alpha\beta}_\sigma$ of cobalt extracted from the DFT computations. The
indices $\alpha$ and $\beta$ run over the 3$d$ orbitals. The Coulomb $U$ term
in LDA+U has the form $\frac{U-J}{2} \sum_\sigma \tr\{n_\sigma - n_\sigma^2\}$,
where the trace is over the orbital indices. The closer the density matrix
$n_\sigma$ is to idempotency, the less correlations will penalize the state
energetically.

\begin{figure}
  \includegraphics[width=\columnwidth]{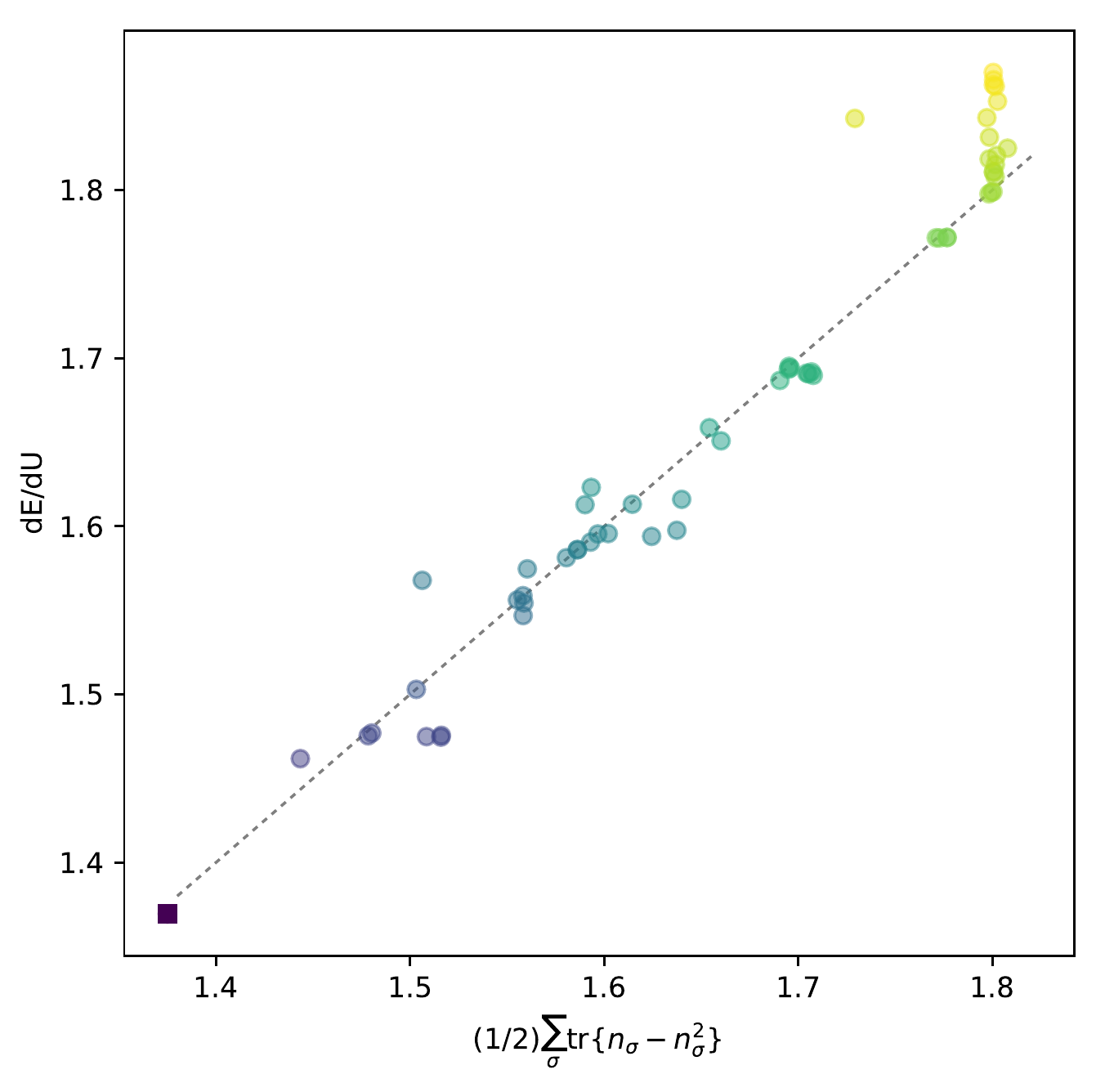}
  \caption{Slope of energy vs. U curve plotted against the LDA+U correlation
    term. Experimental structure is the bold purple square in the lower left.
    Dotted line is the diagonal. Dots are colored by initial slope, matching
    the convention in Fig.~\ref{fig:bacoso-slope}.}
  \label{fig:bacoso-trnn2}
\end{figure}

From a physical viewpoint, LDA+U penalizes structures in which the electrons
are itinerant. The average occupancy of (spin-resolved) orbitals in itinerant
systems cannot be integer: strong charge fluctuations generated by hopping
terms prefer occupancies near 0.5. The LDA+U term penalizes large values of
$n_\sigma-n_\sigma^2$, which is maximal when the eigenvalues of $n_\sigma \sim
0.5$, and thus prefers systems with nearly integer occupancies. We have plotted
the initial slope of the $E(U)$ curve against this LDA+U term in
Fig.~\ref{fig:bacoso-trnn2}. As expected, the slope and LDA+U term track each
other nearly exactly. The experimental structure has orbitals with occupancies
nearest to integral, and is least penalized by correlations. The structures
favored by GGA have the most itinerant orbitals, likely due to the strong
covalent network of CoS$_4$ tetrahedra in its fluorite block, and are strongly
disfavored by $U$.

Our proposed strategy for structural prediction is as follows:
first perform USPEX runs with spin-polarized DFT to generate the list of
structures occupying local minima in the energy landscape. Then, apply an
accurate and likely more computationally expensive method (such as LDA+U or GW)
to the resulting structures to reorder the total energies to determine the true
ground state structure. This is more economical than running USPEX with
hundreds of calls to LDA+U, producing what we estimate to be a factor of 5
speed up or more.

Finally, we verify that our conclusion about the importance of correlation is independent of the method used. Indeed, as can be seen in Tables~\ref{tbl:ecompare} and \ref{tbl:ecompare-gutz}, the ordering between BaCoSO-expt (the observed structure) and BaCoSO-GGA (the lowest energy structure found by GGA) is the same in Wien2K as it is in VASP GGA. Calculating the correlated energy in Gutzwiller~\cite{lanata_prx15} combined with full-potential linearized augmented plane wave Wien2k (W2K)~\cite{wien2k} with $U$ = 10~eV flips the order, and makes BaCoSO-expt favorable.

\begin{figure}
  \includegraphics[width=\columnwidth]{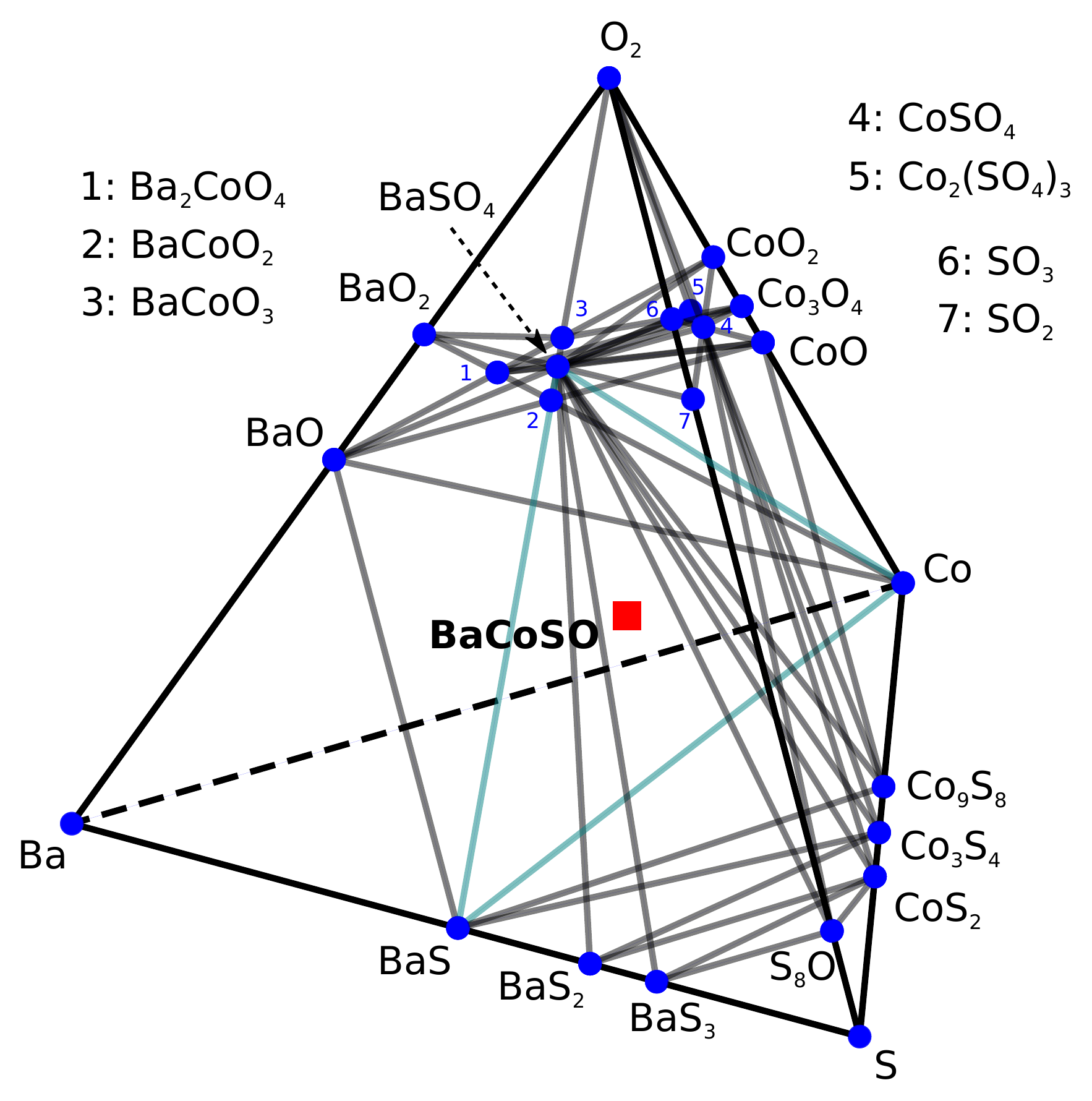}
  \caption{Gibbs phase diagram of Ba-Co-S-O chemical system. The newly
    synthesized compound, BaCoSO is found to be marginally unstable, with an
    energy 102~meV/atom above the hull. The compound lies on the triangular
    facet spanned by Co, BaS and BaSO$_4$.}
  \label{fig:bacoso-hull}
\end{figure}

\emph{Global stability} -- The convex hull for the Ba-Co-S-O system is
shown in Fig.~\ref{fig:bacoso-hull}. We find BaCoSO lies 102~meV/atom above the
hull (in the corrected Materials-Project scheme). Observing the importance of correlations in determining the correct sign or energy differences in this system, we examine the effect of correlations on the convex hull of energies.

\begin{table}
  \begin{tabular}{l|c|cccc}
    \hline
    & GGA & \multicolumn{4}{c}{GGA+U}  \\
    \hline
    Energy & bare & bare & corrected & $U$ & $J$  \\
    \hline
    $E(\text{BaS})$         & -5.106 & -5.106 & -5.438 & 0    & 0  \\
    $E(\text{Co})$          & -6.799 & -7.110 & -7.110 & 0    & 0  \\
    $E(\text{BaSO}_4)$      & -6.554 & -6.554 & -7.133 & 0    & 0  \\
    \hline
    $E(\text{BaCoSO-expt})$ & -5.853 & -5.580 & -6.390 & 3.32 & 0  \\
    $E(\text{BaCoSO-GGA})$  & -5.909 & -5.212 & -6.021 & 3.32 & 0  \\
    \hline
    $\Delta H^f(\text{BaCoSO-expt})$ & 0.219 & 0.570 & 0.102 & - & -  \\
    $\Delta H^f(\text{BaCoSO-GGA})$  & 0.163 & 0.938 & 0.471 & - & -  \\
    \hline
  \end{tabular}
  \vspace{4bp}
  \caption{Comparison of total energies $E$ and energies of formation $H^f$
    (both in eV/atom) for compounds in the Ba-Co-S-O system via VASP pseudopotential code.
    The experimental and GGA-determined structures,
    shown in Fig.~\ref{fig:bacoso-structs}, are labeled BaCoSO-expt and BaCoSO-GGA.
    The bare values of the outputs of non-spin polarized GGA calculations
    are presented in the first column.
    For the uncorrelated calculations, the second column are the
    bare values of the outputs of
    spin-polarized GGA(+U) calculations.
    These energies were modified according to the method used in the
    Materials Project, described in Sec.~\ref{sec:workflow}, to produce the
    corrected energies shown in the next column. The values of $U$ and $J$ (in
    eV) used are listed in the next two columns.
    In the bottom two rows, we display the formation energies, which is the
    result of the reaction $3\text{BaS}+4\text{Co}+\text{BaSO}_4 \rightarrow
    4\text{BaCoSO}$.
    }
  \label{tbl:ecompare}
\end{table}

\begin{table*}
  \begin{tabular}{l|c|cccc}
    \hline
    \multirow{2}{*}{Energy} & W2K & \multicolumn{2}{c}{Gutzwiller} & \multirow{2}{*}{$U$} & \multirow{2}{*}{$J$} \\
    & PM & PM & FM & & \\
    \hline
    $E(\text{BaS})$         & -116,175.236 & -116,175.236 & - &  0 &   0 \\
    $E(\text{Co})$          & -37,917.854 & -37,914.456 & -37,914.634 & 10.0 & 1.0 \\
    $E(\text{BaSO}_4)$      & -40,090.518 &  -40,090.518 & - &  0 &   0 \\
    \hline
    $E(\text{BaCoSO-expt})$ & -68,079.015 & -68,078.217 & -68,078.525 & 10.0 & 1.0 \\
    $E(\text{BaCoSO-GGA})$  & -68,079.077 & -68,078.158 & -68,078.169 & 10.0 & 1.0 \\
    \hline
    $\Delta H^f(\text{BaCoSO-expt})$ & 0.106 & 0.100 & -0.209 & - & - \\
    $\Delta H^f(\text{BaCoSO-GGA})$  & 0.044 & 0.159 & 0.147 & - & - \\
    \hline
  \end{tabular}
  \caption{Comparison of total energies $E$ and energies of formation $H^f$
    (both in eV/atom) for compounds in the Ba-Co-S-O system via Wien2k and Gutzwiller calculations.
    The experimental and GGA-determined structures,
    shown in Fig.~\ref{fig:bacoso-structs}, are labeled BaCoSO-expt and BaCoSO-GGA.
    In the Gutzwiller calculations,
    the reference energies were paramagnetic (non-spin polarized) Wien2K calculations
    with $U = J = 0$. Since Wien2K is an all-electron calculation, the energies
    are extremely large and not directly comparable to those produced using
    VASP in Table~\ref{tbl:ecompare}.
    Shown in the next two columns are the Gutzwiller total energies
    for paramagnetic (PM) and ferromagnetic (FM) orderings, respectively, with the
    Slater-Condon values of $U$ and $J$ shown in the final two columns (in eV).
    In the bottom two rows, we display the formation energies, which is the
    result of the reaction $3 \text{BaS}+4 \text{Co}+\text{BaSO}_4 \rightarrow
    4 \text{BaCoSO}$.
    These energies are directly comparable with VASP results in Table~\ref{tbl:ecompare}.
    Note that the only FM Gutzwiller calculation describes
    the stable phase of BaCoSO-expt,
    suggesting that both correlation and magnetism are important to stabilize BaCoSO.
    }
  \label{tbl:ecompare-gutz}
\end{table*}

The relevant reaction for determining the phase stability of BaCoSO is
\begin{equation}
4~\text{Co} + 3~\text{BaS} + \text{BaSO}_{4} \rightarrow 4~\text{BaCoSO},
\label{eqn:phase-stability}
\end{equation}
where left-hand side compounds are on the corner of the green triangle in
Fig.~\ref{fig:bacoso-hull}.

The W2K and Gutzwiller total energies for materials in Eq.~(\ref{eqn:phase-stability}) are listed in Table~\ref{tbl:ecompare-gutz}. Note that we used the same on-site $U$ and Hund's coupling $J$ in Co $d$ orbitals for BaCoSO and Co materials to avoid the
$GGA/GGA+U$ correction described in Appendix~\ref{sec:Empirical-corrections}.
As can be seen, GGA+U as well as ferromagnetic-Gutzwiller calculations consistently stabilize the observed material BaCoSO-expt more than BaCoSO-GGA.

\begin{figure}
\includegraphics[scale=0.55]{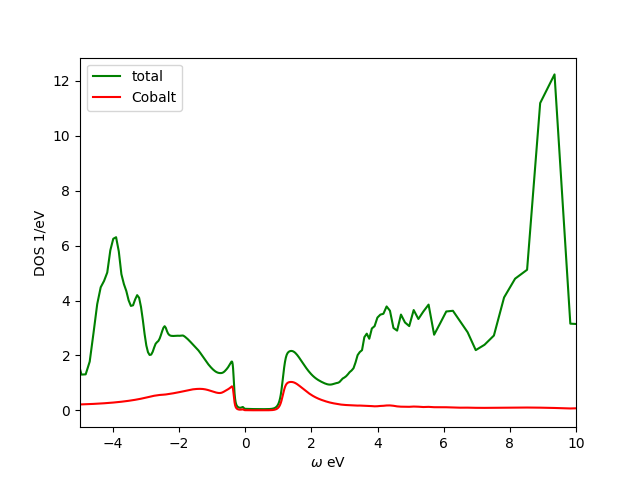}
\caption{Spectral function for BaCoSO at room temperature using LDA+DMFT code~\cite{haule_DMFT,Haule-dmft}.}
\label{fig:BaCoSO_DOS}
\end{figure}

\emph{Electronic Structure} --
Knowing the structure we can now iterate the material design loop and return to the  study the electronic structure. For this purpose we
use  the LDA+DMFT method described in section \ref{sec:How-to-treat}.
We show the calculations of the
spectral functions of  BaCoSO in Fig.~\ref{fig:BaCoSO_DOS} using LDA+DMFT code~\cite{haule_DMFT,Haule-dmft}
and compare it to Fig.~\ref{fig:BaCoS2_DOS}.
The combined effects of the substitution and the structural change result in a Mott insulator
with an increased gap relative to BaCoS$_2$.

\emph{Conclusion} --
As mentioned earlier, BaCoSO has been  very recently synthesized.   It would be very useful to measure its electronic structure by photoemission
and compare with the theoretical predictions.  Ref.~\onlinecite{Qin2017} theorized that it may exhibit high-$T_c$ superconductivity.
Clearly the material design loop can continue to the next iteration.

Application of our  proposed workflow raised several novel
questions. How does $U$ affect the energy landscape? In particular, does $U$
simply shift the local minima relative to one another, or does it create and
destroy minima? Additionally, when is $U$ necessary for correct reordering of
the candidate energies? We expect that $U$ is only necessary for compounds
containing atoms with partially-filled $d$ or $f$ shells or magnetic materials,
and this hypothesis deserves to be investigated.

\subsubsection*{\textbf{Case Studies - wrap-up}}

The examples in sections \ref{sec:tuning}--\ref{sec:bacoso} showcase the question of
``how'' and ``how well'' theory can help guide discovery of new strongly correlated materials.
We found that given the current limitations of existing methods, it is useful to quantify the uncertainties
by thinking in statistical terms about the space of materials. This led to Fig. \ref{fig:probabilities},
which described the probability for the formation of a compound given its calculated energy with various methods.
This should assist materials scientists in the search for new compounds, as it provides a lower bound for the probability
of finding something new in a yet unexplored region of material space. As the methods for evaluation of total energies improve, this
probability distribution will approach a step function centered at zero. While the evaluation of this function with
the LDA+G method is not feasible at this point, we have re-calculated the energetics of the most-relevant reactions of the full LDA convex hull study, and summarized that in Table \ref{tbl:probs}.

Section \ref{sec:tuning} focused on tuning the charge transfer energy via S substitution to increase the superconducting
critical temperature in La$_{2}$CuS$_{2}$O$_{2}$ cuprates.
The probability for this material to exist was estimated to be very small, and indeed in-situ studies showed
that it decomposes into the components predicted by the theory \cite{Hua_2016}.
Pursuing this idea led to Hg(CaS)$_{2}$CuO$_{2}$ (Section \ref{sec:cuprates2}). It is expected to have a higher superconducting transition temperature
than La$_{2}$CuO$_{4}$, but less than HgBa$_{2}$CuO$_{4}$.
Its probability to exist as a ground state structure within DFT is 0.09, but when post-processing
the results with LDA+G, the material is stable, and therefore has a probability larger than 0.5 to exist.

Section \ref{sec:tlcscl3} focused on finding new materials, where the electron-phonon coupling is enhanced by static correlations.
We estimated the probability of CsTlCl$_{3}$ to exist to be close to 0.5 and
indeed the new materials (including the Flourine cousin) were successfully synthesized.
Section \ref{sec:fe112} focused on the iron pnictide, and the idea to create a new family of compounds, the 112 family, to introduce more polarizable metallic spacer layers between the active iron pnictide layers. The probability of CaFeAs$_{2}$ to form was 0.37, and indeed members of this family were synthesized.

Finally in section \ref{sec:bacoso} we studied the Mott insulator BaCoSO. We showed that for this material, considerations based on LDA or GGA would have been misleading both for its ground state structure and stability against phase separation, but treating dynamical correlations gives greatly improved results.
Its probability to exist according to the considerations in section \ref{sec:Probability-Estimation} is 0.17, but when the results are post-processed with LDA+G
the material becomes stable (see Table \ref{tbl:probs}) and its probability to exist is bigger than 0.5.
Indeed this material was reported while this paper was in the process of being completed \cite{Salter2016}, and was also
successfully synthesized in the lab of M. Aronson~\cite{HeUnpublished}.

\begin{table*}
\begin{tabular}{|c|c|c|c|c|}
\hline
Material & Determinant reaction & $\Delta H_{\text{Materials proj.}}$ (eV/atom) & $\mathcal{P}(x)$ & $\Delta H_{\text{Gutzwiller}}$ (eV/atom) \tabularnewline
\hline
\hline
La$_{2}$CuS$_{2}$O$_{2}$
& La$_{2}$SO$_{2}$ + CuS $\rightarrow$ La$_{2}$CuS$_{2}$O$_{2}$
& 0.232 & 0.06 & 0.214
\tabularnewline
\hline
La$_{2}$CuSO$_{3}$
& 3 La$_{2}$SO$_{2}$ + 4 Cu + La$_{2}$SO$_{6}$
$\rightarrow$ 4 La$_{2}$CuSO$_{3}$
& 0.324 & 0.02 & 0.286
\tabularnewline
\hline
Hg(CaS)$_{2}$CuO$_{2}$
& HgO + 2 CaS + CuO $\rightarrow$ Hg(CaS)$_{2}$CuO$_{2}$
& 0.170 & 0.09 & -0.151
\tabularnewline
\hline
CsTlCl$_{3}$
& TlCl + Cs$_{2}$TlCl$_{5}$ $\rightarrow$ 2 CsTlCl$_{3}$
& 0.003 & 0.41 & -
\tabularnewline
\hline
CaFeAs$_{2}$
& CaAs + FeAs $\rightarrow$ CaFeAs$_{2}$
& 0.013 & 0.37 & 0.007
\tabularnewline
\hline
BaCoSO
& 4 Co + 3 BaS + BaSO$_{4}$ $\rightarrow$ 4 BaCoSO
& 0.102 & 0.17 & -0.209
\tabularnewline
\hline
\end{tabular}
\vspace{4bp}
\caption{
The reaction which determines the phase stability
for the materials considered in the paper (determinant reaction).
Given the determinant reaction,
energy above/below the hull is estimated with Materials Project corrections for DFT.
Existence probability is also provided for
the corresponding energy above/below the hull.
Also included in the right-most column is the Gutzwiller energy above/below the hull
for the given determinant reaction.
\label{tbl:probs}
}
\end{table*}

One should stress again the caveats of section \ref{sec:Probability-Estimation}, that the considerations leading
to these probability estimates are crude, as they are based on relatively primitive estimates of energies, and ignore-finite temperature contributions to the entropy - both phononic and electronic. Metastable structures do form, and those are not included in ground state considerations. Estimating the probabilities of metastables states to form is problem which only now is receiving attention, see Ref. \onlinecite{Sun_2016}.

While a compound which is energetically favored will eventually form, the rate of formation
is entirely controlled by kinetics of nucleation and growth. Theory of these phenomena is needed, and the 112 family
illustrates this point. The formation of CaFeAs$_{2}$ requires some La doping, even though this hardly influences
the value of the total energy~\cite{kang_prb17}.

One can use of the bounds on probability of formation discussed in this article
conservatively - as a rejection criterion.
There is little chance for a material to form if its probability to exist is less than
0.08, as shown clearly in the examples of
La$_{2}$CuS$_{2}$O$_{2}$ and La$_{2}$CuSO$_{3}$
in both theory and experiment \cite{Hua_2016}, while all materials with probability of the order of 0.5 did form.
In intermediate cases the experimentalist needs to weigh the interest in properties of the proposed material
against the probability of its formation.

Theory can of course be refined to assist further the synthesis effort.
Knowing that a target compound is unstable against decomposition, one could raise the chemical potential of the products in those reactions to stabilize the desired compounds.
One can also check if pressure would stabilize a desirable composition, and whether the compound would remain metastable once it is synthesized under pressure and the pressure is released~\cite{PhysRevMaterials.2.034604}.

Overall, in the five examples discussed in this article, we see a clear consistency between the
probabilistic estimates and the outcomes of the materials search process. In the process we gained a deeper understanding
of how to control charge transfer energies, and how this can be used to test the mechanism of cuprate superconductivity (sections \ref{sec:tuning}, \ref{sec:cuprates2}).
We found new compounds on which we could study enhanced electron-phonon coupling by static correlations and its connection to valence disproportionation
(section \ref{sec:tlcscl3}), a new family of iron pnictide superconductors which can be studied to understand the mechanism of superconductivity (section \ref{sec:fe112}),
and a new playground to test LDA+DMFT and probe the handles that can be used to move materials around the Mott transition point (section \ref{sec:bacoso}).

The  examples illustrated how the different  methodological aspects of the treatment of correlations  were  used
in practice in the different stages of the  material design workflow:  (i) heuristic and intuitive considerations,
(ii) structure to property relations, (iii) free energy evaluations and electronic structure calculations.

{\it Structure to property}.  It  is  always useful to know whether
the correlations are static or dynamic, and to which extent they are local, as this dictates the methodology used  to 
go from the presumed structure to the  properties. 
Section \ref{sec:tlcscl3} treated compounds where the long range part
of the Coulomb interaction is important and 
static correlations play a decisive role. This class of materials,
which encompass BaBiO$_{3}$, HfNCl and the newly designed CsTlCl$_{3}$. While within LDA the electron-phonon coupling constant $\lambda$
was very small, static correlations make $\lambda$ the order of unity, which results in valence disproportionation as well as high temperature superconductivity
with suitable doping. These were treated using an LAPW implementation of QPGW on the Matsubara axis and hybrid DFT functionals~\cite{Kutepov17}.
Here there  is a clear path for improving the treatment of materials and estimate the size of the corrections  beyond QPGW. 
Vertex corrected treatments increase the accuracy at an increased computational cost~\cite{Shishkin2007,Kutepov17,Kutepov2017}.

Another important direction is to
simplify the $GW$ method so as to accelerate its speed - methods such as the ones proposed in Refs.~\onlinecite{PhysRevB.81.085213,Kang2010a} can be useful in this context.
In Refs.~\onlinecite{yin_kutepov,Wen2018}, which led to the understanding of the electronic structure of the Ba$_{1-x}$K$_{x}$BiO$_3$ system, the  HSE screened hybrid functional was used extensively  with a range
parameter determined by fitting the results for the more accurate QPGW.
This strategy was essential in  the design of the perovskite halides described in section \ref{sec:tlcscl3}.

Sections \ref{sec:bacoso} and \ref{sec:fe112} illustrated the use of photoemission as a theoretical spectroscopy tool to examine the potential properties of a Mott insulator and a Hund's metal material.
These  materials 
are characterized by strong dynamical correlations of quite different nature, charge blocking in \ref{sec:bacoso} and spin blocking in  \ref{sec:fe112}.
Treating these classes of materials  become increasingly difficult
when we probe longer ranges, lower energies and lower temperatures,
as unexpected emergent phenomena, such as superconductivity, appear.
These more complex functionalities at low temperatures -
especially predictions of superconducting transition temperatures and other broken
symmetry states - are very  challenging, in particular in a real materials setting.
LDA + cluster DMFT on a plaquette with
a zero temperature exact diagonalization solver, was the main tool used to guide the project described in
sections \ref{sec:tuning} - \ref{sec:cuprates2}, as it is difficult to reach the very low temperature regime
with Monte-Carlo methods. Work on Monte-Carlo methods to alleviate the minus sign problem
and on alternative accurate finite temperature impurity solvers to cover this important region, are under
active development in many groups around the world, and advances in this area will have substantial
impact in searching for interesting low temperature properties of compounds.

{\it  Structural and Thermodynamical stability  } -- Evaluation of free energies is fundamental to structure prediction and thermodynamic stability. In the projects described in this review article only
the zero temperature electronic energy was estimated. Structural stability requires entropic contributions of both vibrational~\cite{FultzBrent} and electronic origin. For the latter, LDA+DMFT methods will be very valuable~\cite{Haule_2015,Haule2016}. Since LDA+DMFT calculations of free energies are very time consuming, more approximate methods such as LDA+G have been developed~\cite{Deng2009,Yao2011,Lanata2017}. Extension of these methods to finite temperatures is an active area of research \cite{Lanata2015a,Wang2010a,Sandri2013}.

Strong correlations are known to affect structural energy differences. A striking example is its influence on the phase diagram of
elemental Plutonium~\cite{lanata_prx15}. Other examples abound in transition metal oxides, where improvements over
LDA predictions were made using hybrid functionals, LDA+U methods~\cite{Liu2016}, the random phase approximation (RPA)~\cite{Peng2013}, and Monte-Carlo methods~\cite{Schiller2015}, and Gutzwiller methods~\cite{Lee2017}.

In section \ref{sec:bacoso} we showed that correlations are important for structural predictions in a material near the Mott transition - BaCoSO. They can be incorporated first at the LDA+U or the LDA+G level.
While incorporating these into every step of an exhaustive search such as USPEX is expensive, it is enough to consider them on a smaller subset,  by post-processing the results of extensive LDA studies, as we demonstrated in section \ref{sec:bacoso}.

For thermodynamic phase stability, the need to incorporate correlations is more widely
accepted and GGA+U is used in all the Materials Projects \cite{Jain_2013,Kirklin2015,Setyawan2011}. However, this is not enough to achieve the necessary accuracy, and empirical corrections, which can be significant in magnitude, need to be added.

It is important to achieve accuracy
using truly \emph{ab-initio} methods. There is intensive work on using more
elaborate density functionals~\cite{PhysRevMaterials.2.063801}, or RPA~\cite{Kresse2009}. LDA+DMFT with U determined
from first principles and GW+DMFT are promising directions. This is important, as the
values of U currently used for total energy within LDA+U are quite different from the ones needed to describe the photoemission spectra in strongly correlated materials.

We advocated recomputing total energies with more advanced methods as post-processing - after full LDA relaxations and searches. We found that this post-processing to treat correlation, using LDA+U and LDA+G improves bolth structural prediction and the estimation of the probability of formation of this compound. In the future, development of forces for LDA+G, improvement in speed in the forces in LDA+DMFT and automation of user input for electronic methods, could enable a more self-consistent treatment of correlations in structural relaxation and searches.

{\it Developing heuristics and simplified models} --
In addition to theory and computational tools, rapid material design requires
intermediate layers of inference, lying roughly in the space currently occupied
by rules of thumb such as Pauling's rules or the Hume-Rothery rules for structurally complex alloy phases. One
can hope that the development of theory and computational approaches will systematize these rules as well as produce
new ones.

It is important to gain some understanding of the landscape in the
space of materials, and this is one of the main objectives of the materials
genome initiative~\cite{MGI}. This space, as a matter of principle, is infinite as one can synthesize an infinite number of solids with molecular beam epitaxy \cite{JACE:JACE02556}. Mapping a given material onto a model Hamiltonian is akin to providing some set of local coordinates relevant in this space.
The hope in this exercise is that the parametrization captures important physics.

The projects described in sections \ref{sec:tuning}--\ref{sec:cuprates2} illustrated the usefulness of mapping
onto a model Hamiltonian, parametrized by $t_{pd}, t_{pp}, U, \varepsilon_{d}-\varepsilon_{p}$ and how  they can be used  to  explore  the factors that control $T_c$. Just like in the case of the iron pnictides (Fig. \ref{fig:fesc-Tc-height}), further fundamental 
understanding  of the mechanism is needed. We have argued that designing and synthesizing new materials will help in this process.

\section{ Conclusion and Outlook}

\label{sec:conclusion}

Theory and computation have progressed to the point that they are now set to play an important role in material design and exploration, at least for  weakly  correlated electron materials.
For strongly correlated materials,  we argued that the material design process  is intellectually important as it serves as a strong test of physical or chemical intuition, the quality of our methodology and predictive capabilities.
We also showed  through an  admittedly very small number of concrete examples,
that existing methodologies already have a potential to accelerate
material discovery.

The materials considered in the five exemplars were relatively small variations with
respect to known compounds.
In this context LDA+DMFT has substantial predictive power, as one can use a fixed value of the interaction parameters, applied to a localized orbital
supported on a large energy window.
The LDA+DMFT double counting correction can then be determined fully from first principles as proposed in
Ref. \onlinecite{Haule2015}. Improving
the accuracy of the estimates of the remaining parameters such as the strength of the screened Hubbard U and Hund's coupling J and the rest of the Slater parameters
in DMFT from \emph{ab-initio} would be very important for material design as we could contemplate property
prediction in completely unexplored regions in the space of materials.

Combining multiple electronic structure methods to treat correlations constitutes a major software development challenge. Advances in this area will facilitate the development of quantitative measures in the space of materials, and streamline the study of materials. Such tools would enable to scale investigations, such as the ones presented in this review, from five exemplars to a statistically significant sample.
It is also desirable to enable integration between electronic structure tools and the growing databases of materials and their properties. More standardization of data and databases will lead to easier integration between software and databases. Efforts in this direction have been initiated \cite{Nomad}. Finally, machine learning methods will provide alternative routes for exploring structure and property relations.

In solid state physics there is a long tradition of fruitful interactions
between experiment and theory which has resulted in remarkable
advances in understanding condensed matter systems. Until very recently
however, materials discovery was
entirely driven by experiment. This situation is rapidly changing. The highest temperature superconductor, HS$_3$ under pressure, was synthesized \cite{Duan_2014} following a material specific theoretical suggestion~\cite{Li2014,Duan_2014}.

When theory-assisted material design becomes reality,
theory will be able to play new additional roles.
Theory can help guide experiment in selecting which
compositions among the thousands of possibilities would have the highest
probability of forming a new material, and thus be fast-tracked for exploration
over less promising compositions. In return, experiment can provide tests of
theoretical predictions of structure and properties, while new advances in experimental techniques,
where the intermediate products can be monitored in real time within the
reaction vessel, will shed light into how material synthesis actually takes place.
This will be a golden age for experiment - theory interactions, which has
potential to take the field of correlated-electron systems to a new level.

\section{Acknowledgements}
We are grateful to  our theory   (Zhiping Yin, Turan Birol, C. Weber  and Jihoon Shim) and  experimental    (Martha Greenblatt, Meigan Aronson,   Ni Ni,  Cedomir Petrovic and Emilia Morosan) colleagues  for numerous collaborations which shaped our understanding  of the material design challenge.   We are grateful to  Karin Rabe, Kathi Stadler, Piers Coleman and Vladan Stevanovic for useful comments on the manuscript. 
This work was suported by the US Department of Energy, Office of Science Basic Energy Science as a part of the Computational
Materials Science Program.

\begin{appendices}
  
\section{\emph{Empirical corrections}}

\label{sec:Empirical-corrections}

We present a few of the empirical correction schemes used in literature
below. All three schemes that were presented corrected GGA with GGA+U
for certain elements in specific configurations. There is not complete
agreement between the methods on which compounds to apply U to, or
what its value should be. Whereas FERE, building on top of Lany's\cite{Lany2008}
observation, fitted the U to prefer the correct stoichiometry between
oxides with different stoichiometries, Materials Project fitted the
U's on binary oxide formation enthlapies, arriving at different values.
\textcolor{black}{Whereas FERE applies U's universally for the set
of chosen elements} (although it is fitted once on oxides), Materials
Project and OQMD apply GGA+U only for compounds containing oxygen
or flouride. The values of chosen U (or $U_{\textrm{eff}}=U-J$ in
Dudarev's method\cite{Dudarev1998}) in all scheme range between 3-4eV
in all schemes (other than one value). 

In the \textbf{fitted elemental-phase reference energy} (FERE) scheme~\cite{Stevanovic_2012},
experimental formation energies $\Delta H^{\text{exp}}$ are used
to determine the best ``energies'' of the elemental phases. These
fitted elemental-phase reference energies $E^{\text{FERE}}$ are constructed
to minimize the systematic error in the formation energies of the
training set of compounds, and can then be used to predict the formation
energies of new compounds. The procedure requires the tabulation of
the experimental formation energies for 252 binary compounds A$_{m}$B$_{n}$
along with GGA+U calculations for each of the compounds to obtain
the total energies $E^{\text{GGA+U}}$. From this data, the elemental
energies $E^{\text{FERE}}$ of the 50 elements which span the set
of binaries are fitted by solving the linear least-squares problem:
\begin{eqnarray}
\Delta H^{\text{exp}}(\text{A}_{m}\text{B}_{n}) & = & E^{\text{GGA+U}}(\text{A}_{m}\text{B}_{n})\nonumber \\
 & - & mE^{\text{\texttt{FERE}}}(A)-nE^{\text{\texttt{FERE}}}(B).\label{eq:reaction-fit}
\end{eqnarray}
As emphasized by the authors, the energies $E^{\textrm{FERE}}$ are
not meant to improve the absolute total energies for the elemental
phases, but rather constructed to optimize the systematic cancellation
of errors with the GGA+U total energies. A final detail is the choice
of $U$ in the GGA+U calculations. As detailed in Ref.~\onlinecite{Stevanovic_2012},
three values are used: 3~eV for most transition metals, 5~eV for
Cu and Ag, and 0 for the remaining elements. The scheme performs reasonably
well: the mean absolute error of the binaries is 260~meV/atom when
computed using GGA+U, and drops to 54~meV/atom in the FERE scheme.
These improvements carry over when the fitted elemental energies are
used to compute the formation energies of a test set of 55 ternaries,
with the mean absolute error lying at 48~meV/atom.

In the framework of \textbf{Open Quantum Materials Database} (OQMD,
Ref.~\onlinecite{Kirklin2015}), energies of 1,670 reactions were
collected, together with the corresponding GGA or GGA+U results. Unlike
the FERE scheme\cite{Stevanovic_2012}, they applied GGA+U only to
a selected set of ``correlated'' elements when they are in oxides
(V, Cr, Mn, Fe, Co, Ni, Cu, Th, U, Np, Pu). Fitting elemental chemical
potential in experimental formation energies, similarly to above Eq.~(\ref{eq:reaction-fit}),
they compare two different empirically-based correction schemes. In
the \textbf{`fit-partial'} scheme, they fit elemental energies only
for those elements where the DFT energy of the $T$ = 0 K ground state
is not an accurate reference for room-temperature and ambient pressure
(STP) formation energies. These elements include room-temperature
diatomic gases (H, N, O ,F, Cl), room temperature liquids (Br, Hg),
molecular solids (P, S, I), elements with phase transformation between
0 K and room temperature (Na, Ti, Sn), and the ``correlated'' elements listed above. Their \textbf{`fit-all'
}scheme on the other hand, allows all elemental chemical potentials
to be fitted. Interestingly they also find that there is considerable
error within the two different experimental data sets in their database.
It would be interesting to understand the source of this experimental
discrepancy.

In the\textbf{ Materials Project}, an elaborate scheme of physically-motivated
energy shifts are used to respectively correct the formation energy
of gases, compounds containing electronegative anions and energetic
contributions due to correlations at transition metal sites~\cite{Wang_2006,Jain_2011}.
The corrections in this scheme are classified as follows:
\begin{enumerate}
\item \textit{Gas Correction} - formation-energies for gaseous elements
are used as fitting variables for experimental data, similarly to
the other approaches. Therefore the energies for N$_{2}$, F$_{2}$,
Cl$_{2}$ and H$_{2}$ are tabulated.\item \textit{Anion Correction} -- DFT tends to overbind the O$_{2}$ molecule.
It requires too much energy to dissociate O$_{2}$ molecule in oxidation
reactions, which results in underestimating the oxidation energy.
It gives a constant energy shift compared to experimental data \cite{Wang_2006}.
Hence, the DFT total energy for any oxide or sulfide compound is corrected
by the following negative shift, which accounts for DFT overbinding
of anions: 
\begin{equation}
\Delta E=E\cdot N,
\end{equation}
where E is the correction energy, and N is the number of O or S in
the composition. If both oxygen and sulfur are present in the compound,
both corrections are applied. The correction energy E is tabulated
for the ionic state of the anion: oxide(O$^{2-}$), peroxide(O$_{2}^{2-}$),
superoxide(O$_{2}^{-}$), ozonide(O$_{3}^{-}$), and sulfide(S$^{2-}$).
\item \textit{Correction for correlations }-- GGA+U is applied for a selected
set of transition metal compounds containing oxygen or fluorine atoms
(transition metal oxides or transition metal fluorides) ~\cite{hautier_prb12,Jain_2011}.
The value of U was fixed empirically for the following set of elements:
Co, Cr, Fe, Mn, Mo, Ni, V, W.\textcolor{red}{{} }
\item \textit{Correction for GGA+U vs. GGA compatibility} -- The DFT energy
for any oxide or fluoride compound is corrected if the run is a GGA+U
calculation. The correction is applied to any fluoride or oxide containing
one of the following transition metals: \{V, Cr, Mn, Fe, Co, Ni, Mo,
W\}. Note Ti and Cu are absent. The corrections are of the form: 
\begin{equation}
\Delta E=\sum_{i}E_{i}\cdot N_{i},
\end{equation}
where $i$ runs over the transition metals, $N_{i}$ is the number
of atoms of the transition metal present in the compound, and $E_{i}$
is the correction energy is tabulated.
\end{enumerate}
While the corrections are semi-empirical, they perform well from a
practical standpoint, allowing formation energies to be accurately
determined to within 25-50~meV/atom as compared to experimental thermodynamic
benchmarks and successfully reproduce Gibbs phase diagrams~\cite{Stevanovic_2012,Wang_2006,Jain_2011}.
In the paper we used the Materials Project software to construct convex-hulls
and extract energies-above-hull~\cite{Ong2008,Jain_2011}. Therefore
the corrections that we apply are based on their scheme.

\end{appendices}

\bibliography{cormatdes}

\end{document}